\documentclass[aps,prx,twocolumn,amsmath,amssymb,superscriptaddress,hidelinks,author-year,nobalancelastpage,floatfix]{revtex4-2}

\usepackage{bm}
\usepackage{verbatim} 
\usepackage{amsmath}
\usepackage{physics}
\usepackage{amsthm}
\usepackage{amsfonts}
\usepackage{amssymb}
\usepackage{bbold}
\usepackage{mathrsfs}
\usepackage{mathbbol}
\usepackage{braket}
\usepackage{graphicx}
\usepackage[usenames,dvipsnames]{color}
\usepackage[colorlinks]{hyperref}
\hypersetup{
  colorlinks,
  citecolor=blue,
  linkcolor=blue,
  urlcolor=blue}
\usepackage[most]{tcolorbox}
\usepackage{xcolor} 

\usepackage[caption=false]{subfig} 
\usepackage{float}

\usepackage{booktabs}
\usepackage{array}
\usepackage{tabularx}
\usepackage{ragged2e}

\tcbset{
  colback=blue!10!white,
  colframe=white,
  boxrule=0.2mm,
  arc=0.0mm,
}

\newcommand{\Hil}{\mathcal{H}}

\newcommand{\addAA}[1]{\textcolor{magenta}{#1}}

\usepackage[shortlabels]{enumitem}
\usepackage{chemformula}

\usepackage[normalem]{ulem}

\usepackage{siunitx} 
\usepackage[version=4]{mhchem}

\begin{document}
\title{Multi-purpose quantum laboratories from superconducting circuits}
\author{Arpit Arora}
\email{ararora@nvidia.com}
\affiliation{Division of Physical Sciences, College of Letters and Science, University of California, Los Angeles (UCLA), Los Angeles, CA, USA}
\affiliation{Department of Electrical and Computer Engineering, UCLA, Los Angeles, CA, USA}
\affiliation{NVIDIA, Santa Clara, California 95051, USA}
\author{Emily M. Been}
\affiliation{Division of Physical Sciences, College of Letters and Science, University of California, Los Angeles (UCLA), Los Angeles, CA, USA}
\author{William Munizzi}
\affiliation{Division of Physical Sciences, College of Letters and Science, University of California, Los Angeles (UCLA), Los Angeles, CA, USA}
\author{Joel Wang}
\affiliation{Department of Physics, New York University, New York 10003, USA}
\author{Taylor L. Patti}
\affiliation{NVIDIA, Santa Clara, California 95051, USA}
\author{Aaron Chou}
\affiliation{Fermi National Accelerator Laboratory, Batavia, Illinois 60510, USA}
\affiliation{Department of Physics, University of Chicago, Chicago, Illinois 60637, USA}
\author{Prineha Narang}
\email{prineha@ucla.edu}
\affiliation{Division of Physical Sciences, College of Letters and Science, University of California, Los Angeles (UCLA), Los Angeles, CA, USA}
\affiliation{Department of Electrical and Computer Engineering, UCLA, Los Angeles, CA, USA}

\begin{abstract}
Superconducting circuits (SCs) are the cornerstone of modern quantum technology, enabling scalable computing through coherent control of macroscopic quantum states. 
Through a legacy that predates modern quantum computing, SCs have emerged as high-precision instruments for discovery. In this review, we highlight the role of SCs as general-purpose quantum laboratories, outlining the emerging landscape of correlated matter-circuit science. We review and unify the capabilities of superconducting quantum hardware across condensed matter, high energy and quantum information sciences. We trace the technical evolution of these architectures, illustrating how their foundational development has culminated in a toolkit for resolving the complexities of macroscopic quantum states. 
\end{abstract}

\maketitle

\section{Introduction}
The ubiquitous utility of superconducting circuits (SCs) is deeply rooted in their roles in high-precision measurement and control, effectively acting as laboratories of quantum science. Primary examples include defining the voltage standard with AC Josephson circuits, and high-precision metrology standards set by Superconducting Quantum Interference Devices (SQUIDS)~\cite{josephson1964coupled,clarke1988impact,clarke2006squid}. It is a legacy that predates modern quantum computing. Long before the first qubits were realized, SCs served as the testbeds for exploring macroscopic quantum phenomena~\cite{devoret1984resonant,martinis1985energy,devoret1985measurements,clarke1988quantum}, including demonstrations of foundational quantum optics effects, such as squeezed microwave radiation~\cite{yurke1987squeezed,yurke1988observation,yurke1989observation}. 

Fast-forward to current times, one of the most notable examples of coherent macroscopic quantum phenomena is superconducting qubits. Combined with sophisticated fabrication and algorithms, SCs are one of the most mature and adaptable hardware platforms for quantum computation. 
Through sustained progress in coherence and control, SCs now define the technological frontier of quantum engineering. Driven by the quest for fault-tolerance, improvements in coherence times and control handles~\cite{kjaergaard2020superconducting,siddiqi2021engineering,somoroff2023millisecond} have served to amplify and mature the foundational discovery capabilities of SCs. In this review, we chart
a unified approach to reilluminate SCs as multi-purpose quantum laboratories: devices equally capable of acting as high-precision probes for fundamental discoveries and managing quantum resources. 
 
\begin{figure*}
    \centering
    \includegraphics[width=0.7\linewidth, page=1, trim={0 0 240 0}, clip]{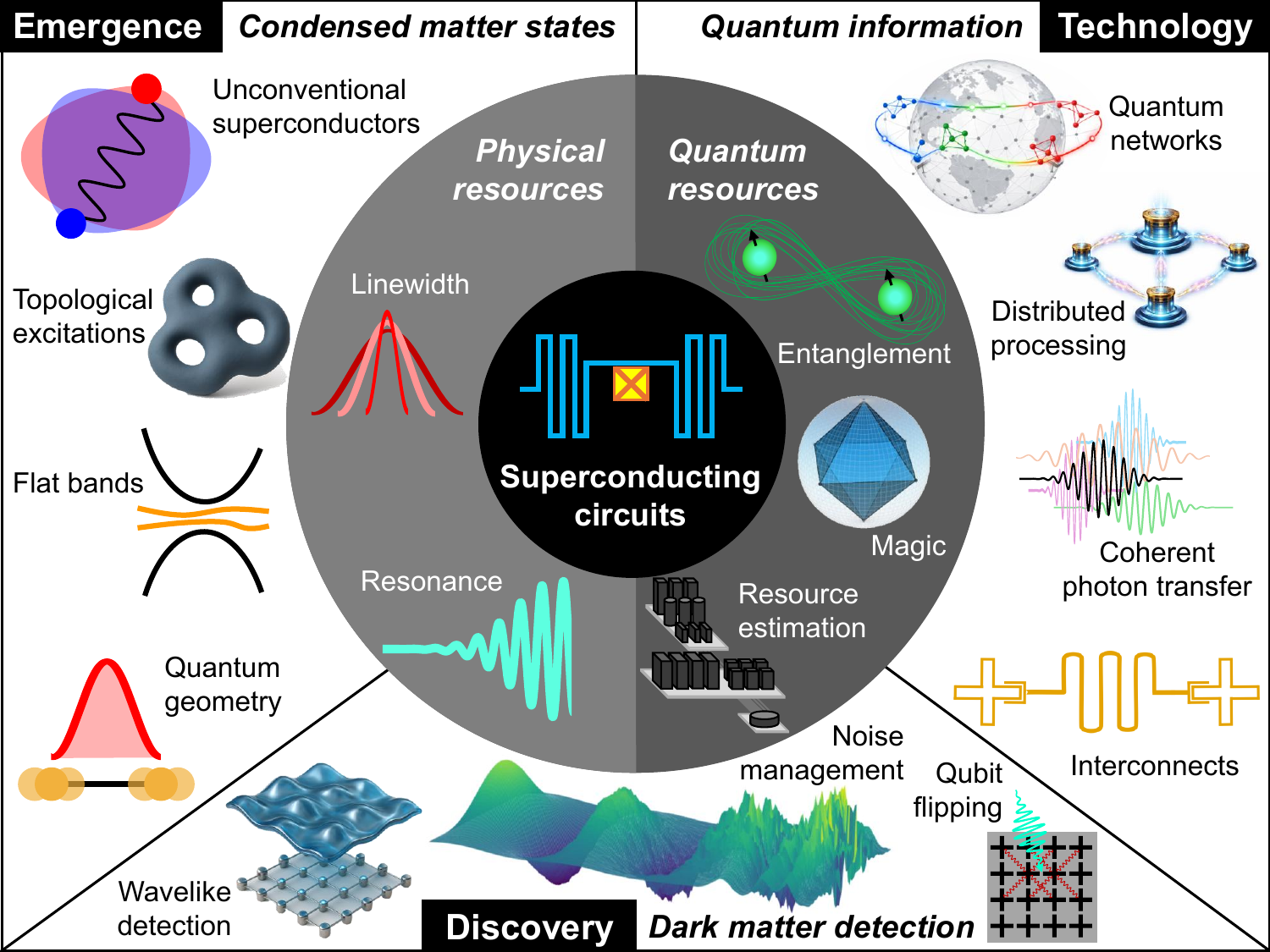}
    \caption{Conceptual map illustrating how superconducting circuit (SC) platforms link physical resources and quantum resources for use in a vast array of applications. The benefits of SCs  span from sensing fundamental physics, such as discovering emergent condensed-matter phenomena and detecting elusive dark matter, to serving as quantum-information primitives and enabling technologies at the forefront of quantum science.}
    \label{fig:overarching}
\end{figure*}

We review the dual capacity of superconducting circuits as both a fundamental subject of quantum research and a probing tool across condensed matter, particle physics, and quantum information. We consider SCs in the microwave photonics regime including their capabilities as resonant circuits, circuit-quantum electrodynamics (circuit-QED), and waveguide-QED platforms.
We highlight superconducting electronics' established use in dark matter (DM) searches, recently developed application as probes for correlated phases in quantum materials, and ongoing development as conduits for engineering complex quantum resources. We further consider AI as a cross-cutting layer that can accelerate circuit design, automate calibration and control, and extract weak physical signals from high-dimensional measurements. Figure~\ref{fig:overarching} shows a multi-directional, multi-level stack for superconducting hardware, highlighting its role as a diverse platform for fundamental physics research and realizing new maps for compiling abstraction.

\subsection{Superconducting circuits toolbox}
Microwave pulses and high quality resonators are central to the SC toolbox: their low photon energies prevent heating fragile quantum states~\cite{romero2009microwave}, their long wavelengths ensure compatibility with mesoscale devices~\cite{schoelkopf1998radio}, and their integration with lithographic fabrication techniques facilitates scalability~\cite{blais_circuit_2021}. Together, these attributes have positioned SCs as a unifying framework across computing, sensing, and fundamental physics. The quantum photonics principles of SCs, established for off-resonant dispersive readout and on-resonant transmission spectroscopy~\cite{blais_circuit_2021}, now provide a mature foundation upon which new functionalities can be built. Table~\ref{tab:cross_domain_toolbox} summarizes how the resulting
toolbox maps onto the applications, experimental maturity, and
limitations discussed throughout this review.

\begin{table*}[t]
\caption{From superconducting-circuit primitives to
cross-domain applications: same hardware elements support both
the measurement of external physical systems and the generation,
distribution, and characterization of quantum states.}
\label{tab:cross_domain_toolbox}

\centering
\footnotesize
\renewcommand{\arraystretch}{1.25}
\setlength{\tabcolsep}{5pt}
\setlength{\arrayrulewidth}{0.6pt}

\begin{tabularx}{\textwidth}{
    |>{\RaggedRight\arraybackslash}p{0.17\textwidth}
    |>{\RaggedRight\arraybackslash}p{0.24\textwidth}
    |>{\RaggedRight\arraybackslash}p{0.27\textwidth}
    |>{\RaggedRight\arraybackslash}X|
}
\hline

\textbf{Toolbox element} &
\textbf{Operational function} &
\textbf{Applications developed} &
\textbf{Present status and limitations} \\
\hline

High-$Q$ microwave and LC resonators &
Convert changes in inductance, capacitance, and dissipation
into frequency and linewidth shifts &
Quantum material spectroscopy, including superconducting, stiffness, topological excitations
(Sec~\ref{sec:cqed-cm}) &
Demonstrated; limited by material
loading, dielectric loss, and magnetic field compatibility. \\
\hline

Dispersive readout and single photon detectors &
Detect photon absorption, quasiparticles and cavity occupation &
Far-infrared and cosmic-ray detection, and transmon-based searches for
low-mass dark matter (Sec~\ref{sec:cqed-dm}) &
Demonstrated; limited by dark counts, stray photons and magnetic shielding. \\
\hline

Josephson nonlinearities and parametric devices &
Amplify/squeeze microwave fields, produce non-Gaussian states &
Quantum-enhanced dark-matter sensing and prospective quantum-limited
materials measurement (Sec~\ref{sec:cqed-cm}, \ref{sec:cqed-dm}) &
Metrological gain demonstrated with squeezing and with non-Gaussian states, limited by loss and measurement inefficiency. \\
\hline

Circuit-QED and waveguide-QED interfaces &
Exchange and route photons; distribute entanglement &
Quantum interconnects, detection and emulation of highly correlated matter (Sec~\ref{sec:cqed-cm}, \ref{sec:cqed-dm}, \ref{quantumresource}) &
Remote photon transfer and entanglement are demonstrated; limited by loss, thermal photons and decoherence. \\
\hline

Engineered nonlinear interactions &
Prepare and control entanglement and magic using Clifford and non-Clifford gates. &
Many-body and effective field-theory emulation, including QED, QCD and holographic models (Sec.~\ref{quantumresource}), entanglement-enhanced sensing. &
Resource generation demonstrated; limited by gate errors, decoherence and measurement overhead. \\
\hline

Electromagnetic and Hamiltonian modelling, circuit databases, and AI &
Accelerate inverse design, calibration, control, and measurement
analysis &
Component design, noise prediction, pulse and gate optimization,
readout, and resource characterization (Sec. \ref{sec:cqed-cm}, \ref{quantumresource}, and \ref{sec:outlook}) &
Individual uses demonstrated; autonomous operation remains limited by model fidelity and experimental validation. \\
\hline

\end{tabularx}
\end{table*}

For most experimentally accessible protocols, SC applications are built on the simple principle that the electromagnetic environment of the circuit carries information about the system under study. An LC oscillator, or a microwave resonator, has a resonance frequency $f = 1/2\pi\sqrt{LC}$, where \textit{L} and \textit{C} denote the effective inductance and capacitance of the resonator. SCs are based on microwave resonators and transmission lines fabricated from aluminum, niobium, or tantalum, and can achieve quality factors exceeding $10^6$~\cite{kjaergaard2020superconducting, bland2025millisecond}. Consequently, any changes to the effective inductance or capacitance, whether through charge distribution, spin alignment, density of states, or quasiparticle distribution, will produce a measurable shift in frequency. Because of the high quality factors, even subtle modifications become detectable. 

This sensitivity underpins SCs as multi-disciplinary probes, which is discussed in Sec.~\ref{fundamentalsearch}. In condensed matter systems, the electromagnetic response can be investigated by integrating the material into a resonator and comparing the resulting response to that of the bare resonator~\cite{clerk2020hybrid}. 
In photon and dark matter detection, quantum capacitance detectors based on single Cooper-pair box superconducting circuits have demonstrated single-photon sensitivity at terahertz frequencies and have been proposed as sensors for low-mass dark matter signals~\cite{echternach_large_2021, golwala_novel_2022}. In these devices, incident energy breaks Cooper pairs in a superconducting absorber, and quasiparticle tunneling onto a nearby superconducting island modifies the quantum capacitance of the single Cooper-pair box, producing a measurable shift in the dispersive response of the coupled resonator~\cite{echternach_large_2021}.
This highlights that SCs can simultaneously serve as fast burst detectors for high-energy cosmic rays and far infrared photons, and as we shall see below, as stable, frequency-agile sensors for low-mass dark matter candidates that interact as a continuous field. 
Furthermore, Josephson nonlinearities enable active signal management, for instance, Kerr shifts allow for parametric amplification and noise squeezing~\cite{rebic2009giant,moon2005theory,aumentado2020superconducting}, while nonreciprocal responses~\cite{arora2025chiral,dirnegger2025nonreciprocal} ensure signals are routed without back-action.

SCs also provide a robust platform for generating key quantum resources: entanglement for distributed computing, squeezed states for sub-vacuum sensing, and coherent superposition for quantum-enhanced simulation. While there is a deep theoretical understanding of these resources~\cite{chitambar_quantum_2019}, the quest to achieve long-range coherence and optimize quantum resources is still on-going. In Sec.~\ref{quantumresource}, we transition from discussing SCs as static hardware to exploring their role as active refineries for quantum states, specifically focusing on the hardware-native generation of entanglement and magic. This capability allows the platform to simulate complex many-body dynamics and search for elusive particles, such as dark matter.

Furthermore, we combine the potential for fundamental discoveries and managing quantum resources with the growing role of AI in the design and operation of quantum information technology, including SCs~\cite{alexeev2025artificial}. While efforts to automate the design of SCs hardware have been underway for more than a half decade \cite{menke2021automated,nugraha2023machine}, this ambitious goal is becoming more realizable with the ongoing diversification and acceleration of classical simulations at both the electromagnetic \cite{ye2025electromagnetic,saslow2025analysis,yao2022massively,sommers2025open} and Hamiltonian \cite{bayraktar2023cuquantum,de2025universal,chakraborty2025gpu} level of abstraction. AI efforts can be further accelerated by SC databases \cite{shanto2024squadds}, which provide users with pre-computed training data that links SC designs to system properties. Using such data, a recent tandem neural-network workflow mapped target qubit frequency and anharmonicity to component-level parameters of a planar transmon, achieving electromagnetic-solver-validated mean errors below 2\% and a roughly 2,000-fold reduction in single-query runtime relative to conventional capacitance extraction~\cite{seidel2026component}.



\section{Fundamental physics searches with quantum laboratories}\label{fundamentalsearch}
Superconducting circuits can be used as task-specific processor and universal high-precision sensor~\cite{degen2017quantum}. This is driven by the platform's unique ability to interface with various energy scales, ranging from few GHz for correlated low-energy excitations~\cite{cottet2017cavity}, to the elusive, weak-coupling regimes of axion-like dark matter~\cite{day2003broadband}. By leveraging the same hardware-native resources used in quantum computing, such as squeezed states and non-linear resonators, we can now project these capabilities outward. In what follows we review how this inherent flexibility enables the detection of emergent phases in quantum materials and the search for dark matter.

\subsection{Superconducting circuits for probing emergent phenomena in condensed matter} \label{sec:cqed-cm}

The study of emergent phenomena in quantum materials relies on a diverse suite of experimental probes, each offering unique insights into different aspects of macroscopic and microscopic properties. Transport measurements serve as a bedrock of condensed matter physics, providing definitive signatures of large-scale collective responses and macroscopic phases in bulk systems~\cite{basov2017towards}. For high-energy dynamics, optical and THz spectroscopies offer powerful techniques to interrogate quasiparticle excitations and correlated orders across the eV to meV range~\cite{de2021colloquium}. When atomic-scale precision is required, Scanning Tunneling Microscopy (STM) provides a view of the local density of states~\cite{yin2021probing}. Complementing these established methodologies, superconducting circuits introduce a distinct capability by operating in the microwave regime. By utilizing low-energy ($\mu$eV) photons that couple coherently to the system's collective modes, SC platforms act as high-sensitivity mesoscale probes. This enables thermodynamic quantities such as superfluid stiffness to be extracted from the measured microwave response and fragile superconducting states to be characterized, particularly in atomically thin materials~\cite{bottcher2024circuit,tanaka2025superfluid,banerjee2025superfluid,zheng2026encapsulation}.


\begin{figure}
  \centering
  \includegraphics[width=\linewidth]{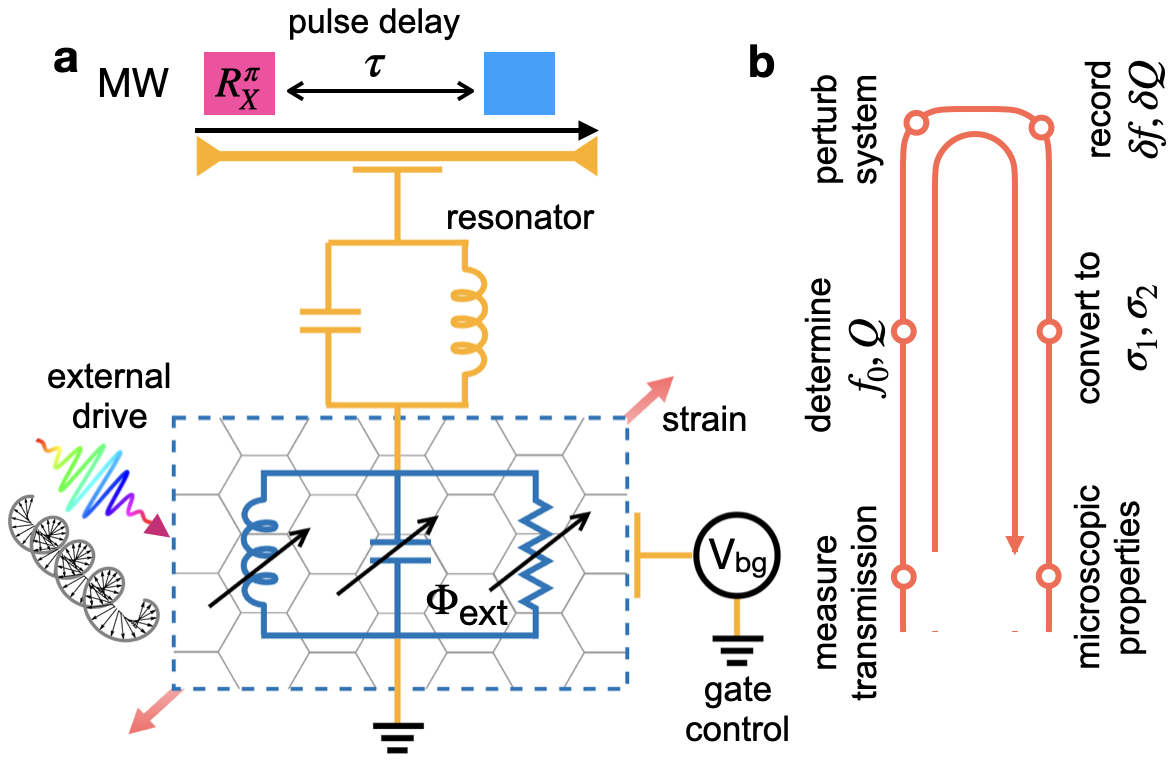}
  \caption{Overview of SCs for probing quantum materials. (a) The sample under investigation can couple capacitively, inductively or galvanically to the microwave resonator, which in turn can be externally controlled and measured via a feedline. The sample can thus be controlled via external fields, mechanical modifications and driving. (b) Workflow of SC measurement, which includes measurement of the transmission spectrum of the hybrid resonator and records the changes in frequency and linewidth under external perturbation. These changes are converted to complex conductivity, $\sigma = \sigma_1 + i\sigma_2$, analyzing which yields the microscopic properties of interest.}
  \label{fig:cm-schematic}
\end{figure}

This sensitivity has long been exploited in quantum computing to benchmark the loss tangent of component materials which impose fundamental limits on coherence times~\cite{blais_circuit_2021}. Extending these ideas to correlated systems reveals much richer opportunities. A straightforward approach is to probe the resonance properties of a microwave resonator coupled to novel superconductors via inductive, capacitive, or galvanic coupling. The frequency shift can then be precisely extracted from complex scattering coefficients via the standard circular fitting~\cite{probst2015efficient}. The frequency shift reflects the change in the effective inductance or capacitance of the loaded resonator, which in turn encodes the sample's response to applied excitations such as gate voltage, bias current, or magnetic field, providing insights into corresponding microscopic properties, see Fig.~\ref{fig:cm-schematic}. Specifically, recent experiments employed this technique to probe superconducting order parameters, e.g., superconductivity in hybrid superconductor–ferromagnet bilayer~\cite{bottcher2024circuit} and twisted graphene multilayers~\cite{tanaka2025superfluid,banerjee2025superfluid}.

For instance, the variation of resonance shifts with temperature has been developed as a probe for unconventional pairing symmetries. Bottcher, et al.~\cite{bottcher2024circuit} used power law fitting of resonance shift ($\delta f$) to temperature ($T$), $\delta f/f_0 \propto T^n$, of the loaded resonator in the presence of external magnetic field to conclude $p$-wave nodal order in Nb-Py superconductor–ferromagnet bilayer; $n$ is the fitting parameter. Note that $\delta f/f_0$ is proportional to changes in kinetic inductance and thus captures the change in superfluid stiffness which follows an exponential temperature dependence for isotropic fully gapped superconductors and a power law temperature dependence for anisotropic gapped and nodal superconductors~\cite{prozorov2006magnetic}. 
More examples include detection of induced $p\pm ip$ pairing at the Al-InAs interface~\cite{phan2022detecting}, superconductivity in cuprates~\cite{jin2025exploring} and Nb-doped STO~\cite{thiemann2018single}. Power law fitting was also used by Tanaka, et. al.~\cite{tanaka2025superfluid} to probe superconducting order in twisted bilayer graphene and demonstrated an anisotropic superconducting gap. Strikingly, in the same experiment, the superfluid stiffness was extracted and provided direct verification of its quantum geometric origin in flat bands. In a parallel effort, the nonlinear Meissner effect was probed with SCs for extraction of superfluid stiffness to probe the nodal superconductivity in twisted trilayer graphene~\cite{banerjee2025superfluid}. Here, the quadratic behavior of superfluid stiffness was tracked down to 30 mK in variation with the applied current; diverging nonlinear behavior was observed, a strong signature of nodal superconductivity consistent with STM measurements~\cite{kim2022evidence}.

In the context of superconductivity, SCs  have been employed to study the voltage-tunable Josephson inductance in graphene that acquired its superconducting order through the proximity effect~\cite{schmidt2018ballistic}. Recent studies have characterized few-layer NbSe$_2$, a transition-metal dichalcogenide (TMD) superconductor, revealing a large kinetic inductance reaching up to 1.2 nH/$\bm{\Box}$ in the monolayer limit and a thickness-dependent crossover from clean- to dirty-limit superconducting behavior~\cite{zaman2025kinetic} (see also Ref.~\cite{kreidel2024measuring}). Beyond using superconducting circuits to characterize quantum materials, recent work has inverted this relationship by integrating large-area, air-stable monolayer NbSe$_2$ into superconducting circuits through encapsulation epitaxy and oxidation-free contacts, yielding a sheet kinetic inductance of approximately 0.7 nH/$\bm{\Box}$ and providing a route toward scalable van der Waals circuit elements~\cite{zheng2026encapsulation}.

Additionally, SCs present a unique opportunity to directly probe chiral superconductivity, a superconducting phase that spontaneously breaks time reversal symmetry. Reliable realization of chiral superconductivity is crucial to harness the benefits of superconductors and topology for applications in fault-tolerant quantum computing. Transport measurements do not suffice for this purpose as the DC anomalous Hall effect vanishes in the superconducting phase~\cite{lutchyn2009frequency}. Current inferences for chiral superconductivity~\cite{han2025signatures} have three necessary conditions: magnetic hysteresis in longitudinal transport, robustness of superconductivity against in-plane magnetic fields, and an anomalous Hall effect in the normal phase. SCs present a unique opportunity to probe the finite frequency Hall response in the superconducting phase, without significant quasiparticle poisoning arising from Cooper-pair breaking by high-energy photons. Other multimode resonators (such as cross~\cite{petrides2025probing,petrides2025passive} and ring geometries~\cite{dirnegger2025nonlinear}) have also been proposed to probe chiral superconductivity based on the coupling of resonator modes mediated by time reversal symmetry broken order of ground states. Additionally, microwave sensing techniques can be extended for other topological excitations such as vortex states~\cite{ren2024microwave,dmytruk2024hybrid,shulga2025observation}. Another potential application is the detection of quantum spin liquids, however such experiments will require new SC device designs that can sustain substantial magnetic fields of at least a few Tesla. Existing SC devices operating in the range of few mT are already in-use for electron and nuclear spin spectroscopy~\cite{wang2023single,budakian2024roadmap,travesedo2025all}.

While investigating superconductivity is a natural path for exploring new functionalities of SCs, owing to the direct change in inductance of the LC resonator, the approach of extracting microscopic properties is far more general. Capacitance offers complementary sensitivity to dielectric properties of a sample~\cite{gorgi2025high}, as can be seen in circuit-QED based dielectric loss characterization of hBN~\cite{wang2022hexagonal, antony2021miniaturizing}, and the probing of 2D material band gaps~\cite{maji2024superconducting}. Correlated insulating phases, such as those of anomalous Hall crystals~\cite{cai2023signatures, lu2024fractional} or charge-density waves~\cite{lee1979electric,littlewood1987screened}, are expected to produce sharp dielectric signatures, resulting in detectable shifts in resonance frequency. Moreover, circuit-QED noise spectroscopy around a quantum critical point can resolve multi-partite entanglement structures~\cite{mazza2024quantum,fang2025amplified}.

In another application, superconducting resonators form an on-chip photonic platform to control electronic correlations. This is particularly potent for flat band systems~\cite{torma2022superconductivity,regnault2022catalogue,checkelsky2024flat} where electronic kinetic energy is quenched and cavity interactions dominate the system's topological and collective behavior. 
Particularly, moir\'e materials hosting these flat bands are a playground for sensitive photonic phenomena due to their large unit cell as light-matter interaction scales with the size of unit cell. For instance, the quantum geometric nature of superconductivity in flat bands~\cite{peotta2015superfluidity,yu2025quantum} is ideal to track quantum state alteration, and the resultant changes in superconducting order~\cite{arora2025chiral,arora2025quantum} which can be harnessed for probing the quantum metric (encodes wavefunction overlap in materials), a quantity difficult to measure in solids~\cite{verma2025quantum, kim2025direct}. More broadly, integrating SCs with flat-band systems reveals a new understanding of wavefunction controlled phenomena, e.g., emulation of Chern and fractional Chern insulator phases, similar to recent demonstration of optical control in twisted MoTe$_2$~\cite{cai2025optical,huber2025optical,holtzmann2025optical}, enabling on-demand control of fractional excitations and non-Abelian statistics.

Current architectures typically treat integrated materials as passive elements within a resonator. However, there is an opportunity to access quantum-limited measurements of correlated phenomena through nonlinearities to enhance measurement sensitivity, for instance in hybrid Josephson junction arrays~\cite{bottcher2024berezinskii,sasmal2025voltage,wang2026long} and ring resonators~\cite{dirnegger2025nonlinear}. Achieving quantum-limited measurements in SCs requires balancing extreme sensitivity with the inherent noise of the quantum vacuum. Machine learning methods function as high-dimensional inference engines that isolate signal patterns within stochastic datasets. For instance, convolutional neural networks~\cite{sarker2026machine}, multilayer preceptrons~\cite{gupta2025expedited} and long short-term memory (LSTM) neural networks have been used to train surrogate models~\cite{koolstra2022monitoring} for the characterization and predictive mitigation of environmental noise in qubits.

The same SC toolkit can also be used towards development of quantum computing devices. For instance, nonreciprocity is a bottleneck in state-of-the-art superconducting circuits~\cite{barzanjeh2025nonreciprocity,arora2025chiral,dirnegger2025nonreciprocal,hovhannisyan2025demonstration}, which require new nonreciprocal elements that are on-chip compatible and reduce device complexity. The measurement of the diode response with circuit-QED devices was recently proposed~\cite{arora2025chiral,dirnegger2025nonreciprocal}, and various applications in context of nonreciprocal information processing~\cite{dirnegger2025nonreciprocal} and microwave rectification~\cite{hovhannisyan2025demonstration} were demonstrated. The SC based discoveries can in-turn supplement the AI protocols of SC device modeling which currently rely on blackbox models~\cite{genois2021quantum}. More broadly, this feedback represents an opportunity for surrogate modeling of SC devices, an underexplored domain, as such AI models can accelerate the collection of quantum training data for downstream AI tasks by many orders of magnitude~\cite{shah2026fourier,waugh2026toward,pipi2025inverse}.

\subsection{Dark-matter detection leveraging superconducting circuits}\label{sec:cqed-dm}



\begin{tcolorbox}[breakable]
\textbf{Low-mass dark matter primer}\\
As the effort to illuminate cosmic dark matter (DM) continues, detection mechanisms have diversified well beyond traditional nuclear-recoil particle searches. Signatures of DM in direct detection experiments may extend from scattering off targets to absorption into targets, depending on the unknown mass of DM particles. Astrophysical data bounds the coupling to normal matter and the self-interaction of potential DM candidates, however their mass is largely unrestricted. In addition to their possible particle-like or wave-like behavior, the microscopic identity of DM remains unknown. 


While DM could be fermionic or bosonic, SC techniques are particularly well-suited to search for low-mass, bosonic DM, which is expected to have high mode-occupation number and therefore behave as classical waves. This signal can be modeled as a weak, coherent drive, often at microwave frequencies accessible to circuit-QED hardware. This makes superconducting resonators, tunable circuits, and quantum-limited microwave readout especially natural tools for detection. Well-studied bosonic DM candidates relevant to such searches include scalar DM, pseudoscalar DM (quantum chromodynamics (QCD) axions and axion-like particles), and vector DM (dark photons) \cite{mitridate_dark_2021}.


To design DM searches, the central challenge is that both the DM mass and its coupling to the target are unknown. 
The broadest experiments are therefore motivated to obtain logarithmic coverage in coupling–mass parameter space \cite{AxionLimits}, while sharper targets can be set by focusing on well-motivated candidates: particles postulated elsewhere in physics.
Of the many possible dark matter candidates, the QCD axion is particularly well-motivated as it resolves the strong-charge-parity problem in QCD \cite{peccei_mathrmcp_1977}, and also has a plausible cosmological origin story, being naturally and inevitably produced in early universe phase transitions \cite{preskill_cosmology_1983,abbott_cosmological_1983,dine_not-so-harmless_1983,sikivie_axion_2008}.

\end{tcolorbox}

\begin{figure}
    \centering
    \includegraphics[width=\linewidth,trim={0 100 0 0},clip]{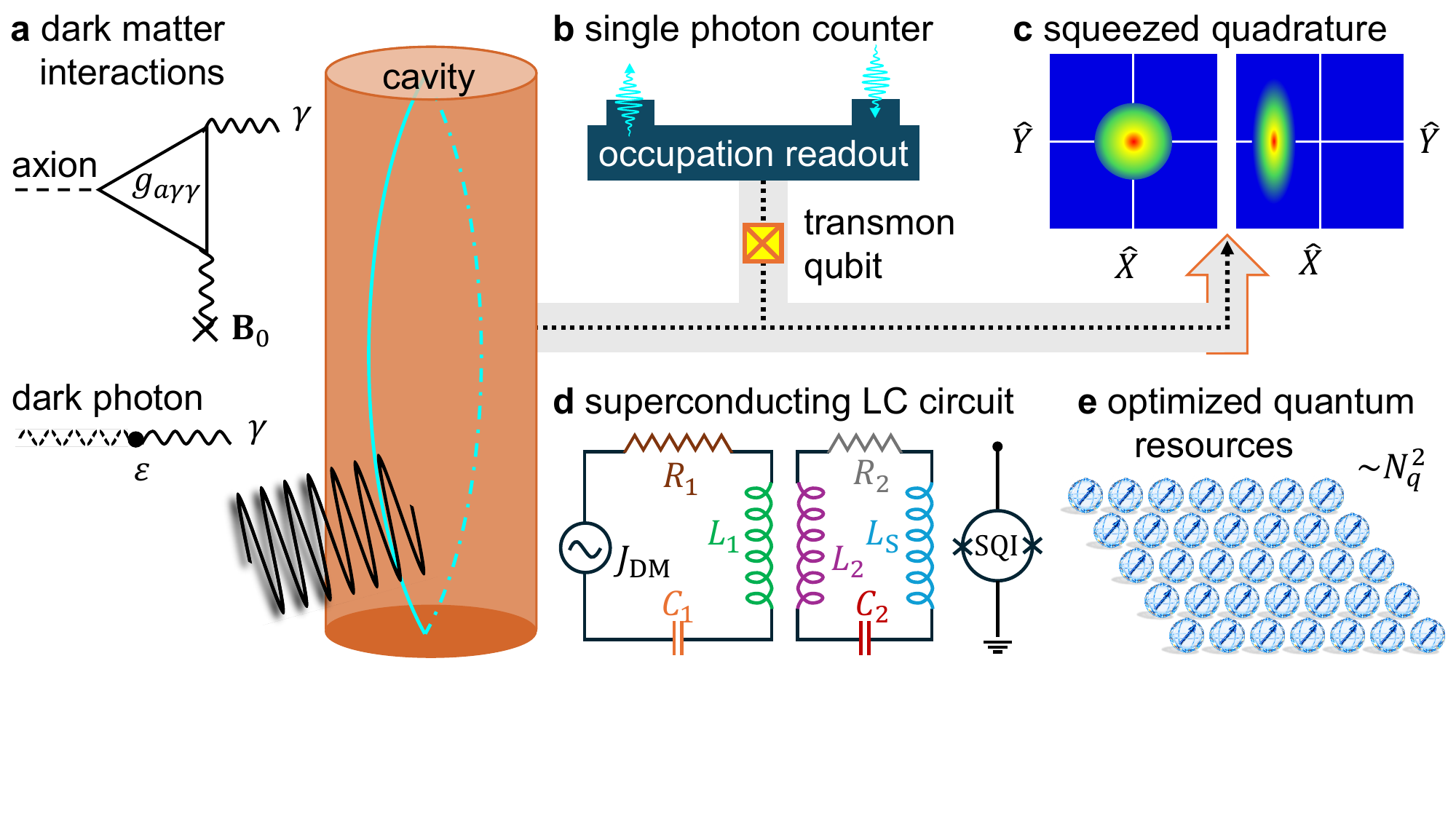}
    \caption{Cavity-based dark matter detection schemes using SC concepts. (a) Axion-like particle interaction with electromagnetism via the Primikoff effect or topological magnetoelectric effect in an external magnetic field $B_0$, and dark photon kinetic mixing with electromagnetic photon. (b) Using transmon qubits to circumvent the standard-quantum-limit by measuring photon number instead of signal quadratures. (c) Squeezing quantum vacuum to improve scan rates by increasing single-quadrature detection bandwidth while maintaining signal-to-noise ratio. (d) Describing the DM detection methods via lumped-element circuits, maximizing signal-to-noise across wide frequency ranges. (e) Incorporating quantum resource allocation into detection protocols. }
    \label{fig:DM-section}
\end{figure}

Conventional wave-like DM searches rely on cavities coupled to linear amplifiers, providing coherent and direct electromagnetic field information in a framework amenable to SC enhancement.
Since the amplified signal provides information on the non-commuting amplitude and phase quadratures of the original signal,
the linear amplifier is also subject to quantum fluctuations, imposing a quantum-noise floor at the standard quantum limit (SQL). 
These limits can be surpassed by exchanging linear amplifiers for single photon detectors which measure only amplitude without phase \cite{caves_quantum_1982,lamoreaux_analysis_2013,Baudis_QROCODILE_2025}.
 %
Despite the SQL noise, record-setting results came from the first deployment of a quantum-limited amplifier in the Axion Dark Matter eXperiment (ADMX). Located in a remote, magnetically-shielded region well away from the high $B$ field needed to induce axion-photon conversion via the 3-wave mixing from the topological magnetoelectric effect (see \autoref{fig:DM-section}a), this near-SQL amplifier demonstrated its utility by reaching sensitivity to the most pessimistic QCD axion couplings in the $\mu$eV mass range \cite{ADMX-MSA}.
Subsequently, multiple promising proposals recognized that circuit-QED could supply the non-classical resources needed to beat SQL noise for cavity searches  \cite{zheng_accelerating_2016,dixit_detecting_2018,braggio_quantum-enhanced_2025}.  

Early concrete DM--SC proposals leveraged superconducting qubits, 
operating in magnetic field-free regions to avoid uncontrolled Aharonov-Bohm phases in the circuit and to avoid quenching the superconducting state \cite{dixit_detecting_2018}.
%
While not sensitive to axions, magnetic-field-free devices could still be used to search for 
dark photons which, as vector particles,  can directly kinetically mix with standard visible photons without requiring a high field magnet to provide a third wave (see \autoref{fig:DM-section}a) 
\cite{Fabbrichesi:2020wbt}. 
%
A qubit in a resonant cavity was shown, in principle, to be able to detect single cavity photons via number-selective dispersive readout, a quantum-non-demolition measurement that circumvents the SQL
\cite{dixit_searching_2021}.
%
These ideas were inspired by rapid progress in quantum information, where qubit-cavity systems already achieved single-photon resolution \cite{blais_circuit_2021}, while the field was increasingly being framed by quantum resource-theory language for quantifying non-classicality, entanglement, and quantum advantages \cite{chitambar_quantum_2019}.


Concurrent theoretical work recast axion haloscope searches as an energy-constrained additive-noise estimation problem, identifying the quantum-limited scan-rate and optimal quantum resources via nulling, i.e., anti-squeezing followed by photon counting ~\cite{shi_ultimate_2023}.
Building on this metrological picture, in-cavity state preparation combined with in-situ transient control of  axion accumulation can mitigate Gaussian-state limitations under loss, helping broaden the usable bandwidth without sacrificing on-resonance sensitivity~\cite{shi_quantum-enhanced_2025}.
More generally, in the experimentally relevant regime where losses dominate on all modes, achieving the fundamental precision limit (i.e. saturating the extended-channel quantum Cram{\'e}r-Rao bound) requires highly non-Gaussian state preparation and measurements, such as Fock states, cat states, or Gottesman-Kitaev-Preskill (GKP) states ~\cite{gardner_stochastic_2025,qmss-lc5x}.  Enhanced dark matter-induced transition rates using Fock states have been demonstrated in Ref.~\cite{Agrawal_2024}, and number-squeezed $N~500$ Fock states with much larger metrological gain have also been achieved using lensing techniques in Fock space~\cite{hua2026quantumconfocalmicroscopyfock}. 
A GKP state with average photon number $\sim 9$ was created and stored using a long-lived $Q\sim10^7$ resonator; a huge triumph for bosonic quantum computing, enabled by the long resonator lifetime which can be viewed as the critical quantum resource
\cite{campagne-ibarcq_quantum_2020}.
Intuitively, loss washes out the Gaussian advantages used in linear sensing and drives Gaussian protocols into a regime where the information about a weak stochastic signal becomes parametrically suppressed; non-Gaussian encoding and readout restores sensitivity and can saturate the extended-channel limit.
These considerations highlight the importance of the long coherence times of superconducting resonators and devices as a vital quantum resource for both quantum sensing and long-lived quantum memories.  For example, the extremely long 34 ms lifetime of a niobium cavity was used to create a Schrodinger's cat state of $N=1024$ photons which persists despite decohering $N$ times faster than a single photon bosonic qubit state \cite{milul_superconducting_2023}.

Three directions define the current landscape of quantum-enhanced DM searches 
using superconducting-circuit technologies, as shown in \autoref{fig:DM-section}:

\begin{enumerate}[i)]

    \item {\bf Transmon-based single-photon counters.}  \\
    In addition to earlier proposals \cite{lamoreaux_analysis_2013,zheng_accelerating_2016,dixit_detecting_2018,dixit_searching_2021},
    recent experiments \cite{braggio_quantum-enhanced_2025, balembois_cyclically_2024, pallegoix_enhancing_2025, may2025noisemitigationsinglemicrowave} 
    utilize a transmon-based single-microwave photon counter, which is both remotely operated and shielded from magnetic field regions required for axion searches.  
    Some experiments \cite{balembois_cyclically_2024,pallegoix_enhancing_2025} reached power sensitivities of $10^{-22}$ and $3\times10^{-23}$ W$/\sqrt{\rm Hz}$, respectively. Along a similar vein, single-microwave photon detectors were able to achieve one of the lowest dark-count rates of $\sim 6\,\mathrm{s}^{-1}$%
    , dominated by thermal and leakage photons, by leveraging cascaded quantum measurements encoded on several superconducting qubits \cite{may2025noisemitigationsinglemicrowave}.  The intrinsic dark count rate of this device is estimated to be $\sim  0.1 s^{-1}$ which should be sufficient to reach QCD axion sensitivity at frequencies up to $\sim10$ GHz \cite{kuo_maximizing_2025}.

    \item \textbf{State squeezing optimization.} \\
    In circuit-QED haloscopes, Josephson parametric devices generate and route quadrature-squeezed vacuum through high-$Q$ readout cavities, enabling sub-quantum-limited noise in the measured quadrature \cite{andersen_30_2016,haystac_collaboration_new_2023}. Taken together, quantum-efficient amplification, impedance-matched coupling, and calibrated squeezing injection, turn quantum noise from a hard limit into an engineering resource \cite{sushkov_quantum_2023}. This SC toolset increases the sensitivity bandwidth while keeping signal-to-noise ratio fixed and, critically, accelerates scan rate and usable bandwidth for dark-matter searches.
    \cite{Shokair_2014,backes_quantum_2021,zheng_accelerating_2016,shi_quantum-enhanced_2025}.

\item \textbf{Superconducting resonant circuits.} \\
A third, well-established direction employs superconducting resonant circuits themselves as the dark-matter sensor, such as the architectures of DMRadio \cite{silva-feaver_design_2017} and ABRACADABRA-style detectors
\cite{ouellet_first_2019}. 
Tunable lumped-element LC resonators and pickup-loop circuits are engineered for ultralow loss and are read out with superconducting quantum interference device (SQUID)-based magnetometry or other quantum-limited amplification chains \cite{salemi_search_2021}. 
While these devices are typically described within classical circuit electrodynamics, they share fabrication, cryogenic, and readout infrastructure with circuit-QED platforms and operate in parameter regimes where quantum-limited measurement becomes essential.

\end{enumerate}


A less established but increasingly active direction is \textbf{quantum resource driven detection}, where non-classical state preparation, entanglement, and control are treated as explicit levers for improving sensitivity and/or scan rate beyond standard quantum limits. 
A quantum circuit can be used to coherently combine the DM-induced phase evolution across many sensor qubits, yielding an enhanced signal rate that can scale as $N_q^{2}$ under suitable error models (rather than linearly with $N_q$, the number of qubits) \cite{chen_quantum_2024,chen_search_2024,bodas2025speedupwavelikedarkmatter}.
This produces a Greenberger-Horne-Zeilinger (GHZ) state with many qubits, where one gets the usual Heisenberg scaling of sensitivity with an extra $\sqrt{N_q}$, and in axion searches this quantum enhancement can be combined with cavity-resonant amplification in transmon-based architectures \cite{sushkov_quantum_2023}.
Closely related work by Shi et al., \cite{shi_quantum-enhanced_2025}, emphasizes that genuinely superlinear scan-rate gains require phase-based protocols 
and Bodas et al., \cite{bodas2025speedupwavelikedarkmatter}, goes on to elucidate the regimes where such entanglement-assisted advantages survive realistic errors and decoherence. 
Recent work by Chen, et al. has pointed out that in the limit of low signals in a common mode detector like a radio telescope array, the quantum state is a n-qubit W state where the single photon is equally distributed in each individual sensor mode \cite{chen_background_2025}.  Quantum-error-correction (QEC) methods can then be used to select this particular common mode signal state and reject noise that is not common to all qubits \cite{fukuda_quantum_2025}.
Framing these approaches in terms of quantum resources and information-theoretic figures of merit (e.g. Fisher-information rates) provides a common language connecting entanglement-based qubit protocols, in-cavity control of squeezed/entangled probe states, and broader quantum-sensing perspectives on DM searches.

\section{Superconducting circuits for Quantum Resources}\label{quantumresource}


Superconducting circuits access many quantum resources; squeezing, for example, already enhances microwave sensing and dark-matter searches~(see Sec.~\ref{sec:cqed-dm}). As summarized in
Table~\ref{tab:cross_domain_toolbox}, here we focus on how these circuits generate and harness entanglement and magic. Entanglement enables the preparation and distribution of correlations central to many-body physics and emergent geometry, whereas magic supports classically difficult non-stabilizer dynamics with connections to quantum matter and holographic models. Their use for cross-domain discovery remains underdeveloped, leaving opportunities for new sensing and simulation advantages.

The ability to generate precise and controllable entanglement structures on quantum hardware remains a central requirement for the development of scalable, fault-tolerant quantum systems~\cite{Yan:2025etq,PhysRevA.107.042412,PhysRevA.88.042324}. In circuit-QED, entanglement is realized through light–matter interactions between superconducting qubits and microwave photons, often confined to on-chip resonators~\cite{Blais_2021,Wallraff_2004, Devoret_2013}. These interactions are defined by fabricated circuit elements, tunable couplers, and calibrated microwave control lines.

Large-scale superconducting circuit (SC) processors integrate many qubits within engineered circuit layouts, where shared resonator modes mediate high-fidelity two-qubit interactions. Dispersive coupling to a common resonator enables selective entangling operations between qubit pairs, supporting the controlled generation of multipartite entanglement patterns~\cite{Barends_2014,Chow_2013,Reagor_2016}. Tunable qubit–resonator couplings and shaped microwave drive pulses can then be used to prepare a broad and configurable family of entangled states~\cite{Krantz_2019,Kjaergaard_2020}. As a result, SC platforms function as programmable laboratories for the preparation, manipulation, and characterization of entanglement-based resources. Another exciting prospect is to extend the same characterization and control techniques to quantum sensing and transduction, in order to identify and harness quantum resources that may be already present in nature, e.g. in the highly non-classical states of quantum materials.


Extensive theoretical work has demonstrated that specific entanglement structures both enable and constrain quantum processes, reflecting fundamental information-theoretic restrictions encoded in entropy inequalities~\cite{Araki:1970,Lieb:1973,Hayden:2013,Ingleton:1971,Fong2008Ingleton,Dougherty2009,ZhangYeung1998,Matus2007,Fuentes:2025kwg,Khumalo:2025xfv}. These constraints organize states according to their entanglement structure, as formalized through entropy-vector and entropy-cone classification schemes~\cite{Linden:2013kal, Bao:2015bfa, Bao:2020mqq, Hern_ndez_Cuenca_2024, Czech:2022fzb, He:2020xuo, Keeler:2022ajf, Keeler:2023xcx, Keeler:2023shl, Munizzi:2023ihc}. Complementary approaches based on graph states further underscore the importance of efficiently distributing and manipulating entanglement, with applications ranging from quantum communication to quantum sensing~\cite{Walter2016,Hein2003,WalterGrossEisert2016,HeinDurEisertBriegel2006,Burchardt:2023odi,Fuentes:2025kwg}. In superconducting architectures, the interplay of tunable couplings, resonator-mediated interactions, and quantum non-demolition measurements provides a flexible platform for engineering precise and custom entanglement structures, as well as enabling direct experimental tests of their role in information processing and emerging distributed quantum technologies~\cite{gong2021quantum}. In the present context, these classifications provide targets and benchmarks for the correlation structures prepared in circuit-based many-body simulators; extracting the same information directly from spectroscopic measurements of quantum materials remains largely prospective.


A particularly powerful realization of distributed entanglement in SC platforms is remote entanglement generation mediated by microwave photons~\cite{Zhong2019}. Remote entanglement generation schemes require physical quantum interconnects linking distant qubits. In SC systems, shared microwave resonators can function as such an interconnect, coherently linking spatially separated qubits coupled to either end, and enabling entanglement through controlled photon exchange~\cite{Zhong2021}. These resonator-based interconnects facilitate multipartite entanglement between several qubits, on a length scale constrained by the free spectral range of the resonator, providing a scalable hardware realization of distributed entanglement architectures.

Propagating, or itinerant, microwave photons in waveguide interconnects can mediate entanglement generation between qubits separated by macroscopic distances~\cite{Kurpiers2018, kannan2023, Negrin:2024tyj,Almanakly2025}. The routing, emission and absorption processes underlying these protocols are governed by waveguide-QED, a particular regime of circuit-QED in which long-range, multi-qubit interactions arise from coupling to a continuum of photonic modes in a shared waveguide~\cite{Lalumiere2013}. Quantum interconnects based on waveguide-QED provide all-to-all connectivity between distant qubits, greatly extending the scale of multi-partite entanglement generation. Such configurable entanglement networks provide a useful platform for simulating many-body physics and exploring emergent quantum states of light and matter~\cite{Pichler2015, Lodahl2017}. At the same time, these tunable light-matter interfaces serve as the foundation for superconducting quantum networks, where microwave-photon interconnects support distributed entanglement~\cite{Hughes:2025jwa}, gate teleportation, and scalable quantum communication~\cite{DiAdamo:2023tqg}. This functionality establishes SC as a unified experimental framework for engineering and dynamically controlling quantum resources across both local processors and distributed architectures.\\

\begin{tcolorbox}[breakable]
\textbf{Entanglement and Magic}\\
Entanglement and magic are complementary resources underlying quantum advantage. Entanglement quantifies non-classical correlations, while magic quantifies the non-stabilizerness needed for classical intractability. For a state $\ket{\psi}$ on a factorizable Hilbert space, entanglement monotones such as the von Neumann entropy~\cite{vonNeumann1927,nielsen2010quantum} 
\begin{equation}
    S_I = -\Tr(\rho_I \log_d \rho_I), \, \text{with} \, \, \rho_I = \Tr_{\bar I}(\ket{\psi}\bra{\psi}),
\end{equation} 
measure correlations between subsystems. A generalization is the R\'{e}nyi entropy~\cite{Rnyi1961OnMO}
\begin{equation}
S_I^{(\alpha)} = \frac{1}{1-\alpha}\log_d \Tr\!\left(\rho_I^\alpha\right), \, \text{for} \, \, \alpha \geq 0, \alpha \neq 1,
\end{equation}
which interpolates between the log of the Schmidt rank ($\alpha \to 0$), von Neumann entropy ($\alpha \to 1$), and collision entropy ($\alpha = 2$). Since $S_I^{(\alpha)}$ is monotonically non-increasing in $\alpha$, the family characterizes the full entanglement spectrum, which no single measure captures.

Entanglement alone is insufficient for advantage, as many maximally entangled states are efficiently classically simulable~\cite{gottesman1998heisenberg,aaronson2004improved}. Quantum advantage requires another resource, \emph{quantum magic}~\cite{bravyi2005universal,howard2017application}, measured as  the trace distance to stabilizer state set, which is intractable at large $n$ given the exponential scaling of the stabilizer polytope. For pure states an efficient measure exists, the stabilizer $\alpha$-R\'{e}nyi entropy (SRE)~\cite{Leone:2021rzd},
\begin{equation*}
\begin{split}
    M_\alpha(\psi)
        &= \frac{1}{1-\alpha}\log P_\alpha(\psi),\\
\textnormal{where}& \quad P_\alpha(\psi)
= \frac{1}{d_n}\sum_{P\in\Pi_n}
|\langle\psi|P|\psi\rangle|^{2\alpha},
\end{split}
\end{equation*}
with $\Pi_n$ the $n$-qubit Pauli group. Magic can further exhibit a genuinely non-local character~\cite{Bao_2022,cao2025gravitational,Andreadakis:2025mfw,Qian:2025oit,Munizzi:2025suf,Munizzi:2026qkv}. For $\Hil=\Hil_A\otimes\Hil_B$, the \textit{non-local magic}
\begin{equation}
\mathcal{M}^{NL}(\ket{\psi}) = \min_{U_A\otimes U_B} \mathcal{M}(U_A\otimes U_B\ket{\psi}),
\end{equation}
is the residual magic after optimizing over all local unitaries $U_A \in \mathcal{L}(\Hil_A)$, $U_B \in \mathcal{L}(\Hil_B)$. Non-local magic depends on the entanglement spectrum~\cite{cao2025gravitational}. Maximally entangled states, which posses flat spectra by definition, have no non-local magic, while states with highly-skewed spectra have significant non-local magic.
\end{tcolorbox}

Circuit-QED provides a hardware platform for the controlled generation of magic in superconducting architectures. Since magic quantifies separation from what is achievable within the stabilizer formalism, its reliable production requires the high-fidelity implementation of non-Clifford gates, such as the $T$ gate. In SC systems, non-Clifford operations can be engineered using precisely shaped microwave drives, parametric modulation, and tunable qubit-resonator interactions, followed by magic state distillation to improve fidelity, enabling the controlled injection of magic into otherwise Clifford circuits. These hardware-native capabilities position superconducting circuits as an efficient testbed for optimizing magic generation under realistic noise and coherence constraints, while also allowing for the systematic preparation of non-stabilizer states and direct measurement of their resource content. Techniques such as Wigner tomography~\cite{devra2024wigner} for small qubit number, and more generally stabilizer-based witnesses~\cite{andersen2019entanglement} for larger systems, further allow the dynamics of magic to be tracked under coherent control, dissipation, and hybrid qubit-bosonic interactions.

Since real quantum devices are fragile, inherently noisy, and subject to parameter drift, reliable magic generation demands adaptive control and real-time optimization. Data-driven and AI-assisted strategies are therefore well-suited to this setting, where learning-based protocols can be used to mitigate noise, refine pulse sequences, and dynamically optimize resource injection under partial and destructive measurements~\cite{alexeev2025artificial}. Beyond pulse-level optimization, recent advances in magic state cultivation further demonstrate how hardware-aware approaches can reduce physical qubit overhead in hybrid architectures~\cite{rosenfeld2025}, particularly through the use of code switching and surface-code grafting.

Generating states with substantial non-local magic $\mathcal{M}^{NL}$ requires not only the ability to implement local non-Clifford operations, but also the precise control over multi-qubit entangling operations, such that the state's entanglement spectrum can be driven away from flat configurations. Circuit-QED architectures are particularly well-suited to this task. Resonator-mediated interactions, tunable couplers, engineered nonlinearities~\cite{hashim2025efficient}, and mid-circuit measurement in SC platforms provide a fine-grained control over how correlations are distributed across different qubits, and allow the direct observation of non-local magic generation, redistribution, and decay. This hardware-level control allows for a customization of entanglement spectra, which in turn enables the preparation of states spanning maximally-entangled (flat spectrum with $\mathcal{M}^{NL} = 0$) to highly-skewed spectra that can support large non-local magic, thereby providing powerful quantum resource states for both entanglement and magic-driven tasks. The combination of strong dispersive coupling, fast entangling gates, and programmable pulse control allows these powerful resource states to be prepared directly at the physical layer, reducing overhead compared to purely circuit-level synthesis.

In the context of fundamental physics, recent analyses of QED and QCD scattering processes have shown that certain interactions generate (non-local) magic, while others produce none~\cite{Robin:2025ymq,White:2024bjp,PhysRevD.110.116016}. The first experimental observation of non-local magic on superconducting hardware~\cite{ahmad2025experimental} further demonstrates that SC platforms can simulate interaction-driven resource dynamics in a controlled manner. The combination of these results establish superconducting architectures as a testbed for probing the generation and redistribution of quantum resources under effective field-theoretic evolution. Beyond characterization, SC platforms also enable the exploration of related phenomena  such as Bell inequality violations in states possessing non-local magic~\cite{cusumano2025non}, toy models of gravitational fields~\cite{cao2025gravitational}, and signatures of quantum critical behavior. In this way, SC systems function as laboratories for investigating how fundamental interactions generate structured, distinguishable quantum resources in emergent physical regimes.

For quantum resource theory on SC platforms, AI-assisted approaches offer several advantages. At the control layer, reinforcement learning and differentiable programming can shape pulses to maximize entanglement generation or non-Clifford gate fidelity under realistic noise models~\cite{bukov2018reinforcement,niu2019universal,baum2021experimental}, improving the production of magic and entangled resource states at the physical-layer. For characterization, neural-network tomography and generative models offer scalable alternatives to full state tomography~\cite{torlai2018neural,carrasquilla2019reconstructing}, enabling estimation of entanglement spectra, stabilizer Rényi entropies, and related magic monotones in regimes where exact reconstruction is intractable. Classical shadow protocols, which likewise admit natural machine-learning post-processing, have been used to estimate nonlinear functionals of density matrices, including magic measures and subsystem entropies~\cite{huang2020predicting,elben2023randomized}, and are a  natural fit for mid-circuit measurement on SC hardware.

Tensor network and neural-quantum-state ansatze can efficiently represent states with structured entanglement spectra~\cite{carleo2017solving,sharir2020deep}, making them well-suited to the preparation of states with tunable non-local magic. Machine-learning has been used to detect phase transitions and resource-theoretic structure from measurement data~\cite{carrasquilla2017machine,vannieuwenburg2017learning,Khumalo:2025xfv}, suggesting techniques for diagnosing measurement-induced transitions in entanglement and magic on SC hardware. Furthermore, AI-driven optimization is a strong candidate for non-local magic estimation, since computing $\mathcal{M}^{NL}$ requires minimizing a magic monotone over a manifold of local unitaries, typically employing non-convex optimization~\cite{Lipardi2025SREMachineLearning,Busoni2026NonLocalMagicQudits} over a structured parameter space for which gradient-based and evolutionary machine learning is well-suited.

\section{Conclusions \& Outlook}\label{sec:outlook}


Superconducting circuits occupy a distinctive place in modern physics because their development has long intertwined precision measurement, macroscopic quantum control, and, more recently, quantum computation. Long before the first qubits were realized, superconducting technologies had already established themselves as tools for metrology and high-precision sensing. Today, advances driven by the quest for scalable quantum processors, including improvements in coherence, control, tunability, and fabrication, have fed back into this broader legacy, sharpening superconducting circuits into platforms for discovery. As emphasized throughout this review, their reach now spans established roles in fundamental detection, rapidly developing applications in correlated and low-dimensional quantum matter, and the controlled generation and interrogation of quantum resources.

SCs when integrated with AI enable adaptive learning to handle dynamic and unpredictable environments. Much of the AI-based innovation in SC technologies has taken a modular approach, focusing on pulse optimization \cite{nam2024reinforcement,genois2025quantum}, gate design \cite{daraeizadeh2020machine,wright2023fast}, and readout \cite{chatterjee2025enhanced} for SC systems. AI-assisted readout protocols have even been deployed on custom field-programmable gate arrays, realizing low-latency single-shot fidelity \cite{di2025end}. Deep reinforcement learning has played a particularly prominent role in qubit optimization \cite{liu2025superconducting}, and remains a favored modality for automated state-preparation protocols \cite{porotti2022deep,reuer2023realizing}. Moreover, large language models have also been deployed for full automation of experimental process~\cite{cao2025automating}. The synergy of SCs and AI is ideally suited to address currently intractable challenges refining abilities for fundamental discoveries and managing quantum information resources alike, projecting a next phase towards autonomous quantum laboratories.


Viewed in this way, superconducting circuits are best understood as a maturing hardware ecosystem in which discovery and information processing increasingly reinforce one another. Quantum computation has supplied powerful new benchmarks and engineering discipline, while sensing and hybrid measurement applications continue to expand the scientific scope of the platform beyond computation alone. The broader outlook is therefore one of convergence: superconducting circuits as multi-purpose quantum laboratories that connect precision detection, quantum materials, and quantum information science through a shared physical toolkit. Their future importance lies not only in enabling better processors, but in serving as adaptable experimental interfaces for both fundamental discovery and the next generation of quantum technologies.

\section{Acknowledgments}

Authors acknowledge discussions with William Oliver (MIT), Tanuj Khattar (Google Quantum AI), Murat Can Sarihan (Google Quantum AI), Aziza Almanakly (NYU), Nicolas Dirnegger (UCLA), Abhishek Banerjee (Amazon), Sadman Ahmed Shanto (USC), Elaine Taylor (Stanford). This work is supported by the Quantum Science Center (QSC), a National Quantum Information Science Research Center of the U.S. Department of Energy (DOE). This work is also supported by the Department of Energy (DOE) Office of Science (SC) Grant No DOE DE-FOA-0003432. This work is also supported by Grant No GBMF12976 of the Gordon and Betty Moore Foundation.
This work was produced by Fermi Forward Discovery Group, LLC under Contract No. 89243024CSC000002 with the U.S. Department of Energy, Office of Science, Office of High Energy Physics. The United States Government retains and the publisher, by accepting the work for publication, acknowledges that the United States Government retains a non-exclusive, paid-up, irrevocable, world-wide license to publish or reproduce the published form of this work, or allow others to do so, for United States Government purposes. The Department of Energy will provide public access to these results of federally sponsored research in accordance with the 
\href{http://energy.gov/downloads/doe-public-access-plan}{DOE Public Access Plan}
(\url{http://energy.gov/downloads/doe-public-access-plan})
.

\clearpage

\bibliography{references_clean}

\begin{thebibliography}{231}%
\makeatletter
\providecommand \@ifxundefined [1]{%
 \@ifx{#1\undefined}
}%
\providecommand \@ifnum [1]{%
 \ifnum #1\expandafter \@firstoftwo
 \else \expandafter \@secondoftwo
 \fi
}%
\providecommand \@ifx [1]{%
 \ifx #1\expandafter \@firstoftwo
 \else \expandafter \@secondoftwo
 \fi
}%
\providecommand \natexlab [1]{#1}%
\providecommand \enquote  [1]{``#1''}%
\providecommand \bibnamefont  [1]{#1}%
\providecommand \bibfnamefont [1]{#1}%
\providecommand \citenamefont [1]{#1}%
\providecommand \href@noop [0]{\@secondoftwo}%
\providecommand \href [0]{\begingroup \@sanitize@url \@href}%
\providecommand \@href[1]{\@@startlink{#1}\@@href}%
\providecommand \@@href[1]{\endgroup#1\@@endlink}%
\providecommand \@sanitize@url [0]{\catcode `\\12\catcode `\$12\catcode `\&12\catcode `\#12\catcode `\^12\catcode `\_12\catcode `\%12\relax}%
\providecommand \@@startlink[1]{}%
\providecommand \@@endlink[0]{}%
\providecommand \url  [0]{\begingroup\@sanitize@url \@url }%
\providecommand \@url [1]{\endgroup\@href {#1}{\urlprefix }}%
\providecommand \urlprefix  [0]{URL }%
\providecommand \Eprint [0]{\href }%
\providecommand \doibase [0]{https://doi.org/}%
\providecommand \selectlanguage [0]{\@gobble}%
\providecommand \bibinfo  [0]{\@secondoftwo}%
\providecommand \bibfield  [0]{\@secondoftwo}%
\providecommand \translation [1]{[#1]}%
\providecommand \BibitemOpen [0]{}%
\providecommand \bibitemStop [0]{}%
\providecommand \bibitemNoStop [0]{.\EOS\space}%
\providecommand \EOS [0]{\spacefactor3000\relax}%
\providecommand \BibitemShut  [1]{\csname bibitem#1\endcsname}%
\let\auto@bib@innerbib\@empty
\bibitem [{\citenamefont {Josephson}(1964)}]{josephson1964coupled}%
  \BibitemOpen
  \bibfield  {author} {\bibinfo {author} {\bibfnamefont {B.}~\bibnamefont {Josephson}},\ }\bibfield  {title} {\bibinfo {title} {Coupled superconductors},\ }\href@noop {} {\bibfield  {journal} {\bibinfo  {journal} {Reviews of Modern Physics}\ }\textbf {\bibinfo {volume} {36}},\ \bibinfo {pages} {216} (\bibinfo {year} {1964})}\BibitemShut {NoStop}%
\bibitem [{\citenamefont {Clarke}\ and\ \citenamefont {Koch}(1988)}]{clarke1988impact}%
  \BibitemOpen
  \bibfield  {author} {\bibinfo {author} {\bibfnamefont {J.}~\bibnamefont {Clarke}}\ and\ \bibinfo {author} {\bibfnamefont {R.~H.}\ \bibnamefont {Koch}},\ }\bibfield  {title} {\bibinfo {title} {The impact of high-temperature superconductivity on squid magnetometers},\ }\href@noop {} {\bibfield  {journal} {\bibinfo  {journal} {Science}\ }\textbf {\bibinfo {volume} {242}},\ \bibinfo {pages} {217} (\bibinfo {year} {1988})}\BibitemShut {NoStop}%
\bibitem [{\citenamefont {Clarke}\ and\ \citenamefont {Braginski}(2006)}]{clarke2006squid}%
  \BibitemOpen
  \bibfield  {author} {\bibinfo {author} {\bibfnamefont {J.}~\bibnamefont {Clarke}}\ and\ \bibinfo {author} {\bibfnamefont {A.~I.}\ \bibnamefont {Braginski}},\ }\href@noop {} {\emph {\bibinfo {title} {The SQUID handbook: Applications of SQUIDs and SQUID systems}}}\ (\bibinfo  {publisher} {John Wiley \& Sons},\ \bibinfo {year} {2006})\BibitemShut {NoStop}%
\bibitem [{\citenamefont {Devoret}\ \emph {et~al.}(1984)\citenamefont {Devoret}, \citenamefont {Martinis}, \citenamefont {Esteve},\ and\ \citenamefont {Clarke}}]{devoret1984resonant}%
  \BibitemOpen
  \bibfield  {author} {\bibinfo {author} {\bibfnamefont {M.~H.}\ \bibnamefont {Devoret}}, \bibinfo {author} {\bibfnamefont {J.~M.}\ \bibnamefont {Martinis}}, \bibinfo {author} {\bibfnamefont {D.}~\bibnamefont {Esteve}},\ and\ \bibinfo {author} {\bibfnamefont {J.}~\bibnamefont {Clarke}},\ }\bibfield  {title} {\bibinfo {title} {Resonant activation from the zero-voltage state of a current-biased josephson junction},\ }\href@noop {} {\bibfield  {journal} {\bibinfo  {journal} {Physical review letters}\ }\textbf {\bibinfo {volume} {53}},\ \bibinfo {pages} {1260} (\bibinfo {year} {1984})}\BibitemShut {NoStop}%
\bibitem [{\citenamefont {Martinis}\ \emph {et~al.}(1985)\citenamefont {Martinis}, \citenamefont {Devoret},\ and\ \citenamefont {Clarke}}]{martinis1985energy}%
  \BibitemOpen
  \bibfield  {author} {\bibinfo {author} {\bibfnamefont {J.~M.}\ \bibnamefont {Martinis}}, \bibinfo {author} {\bibfnamefont {M.~H.}\ \bibnamefont {Devoret}},\ and\ \bibinfo {author} {\bibfnamefont {J.}~\bibnamefont {Clarke}},\ }\bibfield  {title} {\bibinfo {title} {Energy-level quantization in the zero-voltage state of a current-biased josephson junction},\ }\href@noop {} {\bibfield  {journal} {\bibinfo  {journal} {Physical review letters}\ }\textbf {\bibinfo {volume} {55}},\ \bibinfo {pages} {1543} (\bibinfo {year} {1985})}\BibitemShut {NoStop}%
\bibitem [{\citenamefont {Devoret}\ \emph {et~al.}(1985)\citenamefont {Devoret}, \citenamefont {Martinis},\ and\ \citenamefont {Clarke}}]{devoret1985measurements}%
  \BibitemOpen
  \bibfield  {author} {\bibinfo {author} {\bibfnamefont {M.~H.}\ \bibnamefont {Devoret}}, \bibinfo {author} {\bibfnamefont {J.~M.}\ \bibnamefont {Martinis}},\ and\ \bibinfo {author} {\bibfnamefont {J.}~\bibnamefont {Clarke}},\ }\bibfield  {title} {\bibinfo {title} {Measurements of macroscopic quantum tunneling out of the zero-voltage state of a current-biased josephson junction},\ }\href@noop {} {\bibfield  {journal} {\bibinfo  {journal} {Physical review letters}\ }\textbf {\bibinfo {volume} {55}},\ \bibinfo {pages} {1908} (\bibinfo {year} {1985})}\BibitemShut {NoStop}%
\bibitem [{\citenamefont {Clarke}\ \emph {et~al.}(1988)\citenamefont {Clarke}, \citenamefont {Cleland}, \citenamefont {Devoret}, \citenamefont {Esteve},\ and\ \citenamefont {Martinis}}]{clarke1988quantum}%
  \BibitemOpen
  \bibfield  {author} {\bibinfo {author} {\bibfnamefont {J.}~\bibnamefont {Clarke}}, \bibinfo {author} {\bibfnamefont {A.~N.}\ \bibnamefont {Cleland}}, \bibinfo {author} {\bibfnamefont {M.~H.}\ \bibnamefont {Devoret}}, \bibinfo {author} {\bibfnamefont {D.}~\bibnamefont {Esteve}},\ and\ \bibinfo {author} {\bibfnamefont {J.~M.}\ \bibnamefont {Martinis}},\ }\bibfield  {title} {\bibinfo {title} {Quantum mechanics of a macroscopic variable: the phase difference of a josephson junction},\ }\href@noop {} {\bibfield  {journal} {\bibinfo  {journal} {Science}\ }\textbf {\bibinfo {volume} {239}},\ \bibinfo {pages} {992} (\bibinfo {year} {1988})}\BibitemShut {NoStop}%
\bibitem [{\citenamefont {Yurke}(1987)}]{yurke1987squeezed}%
  \BibitemOpen
  \bibfield  {author} {\bibinfo {author} {\bibfnamefont {B.}~\bibnamefont {Yurke}},\ }\bibfield  {title} {\bibinfo {title} {Squeezed-state generation using a josephson parametric amplifier},\ }\href@noop {} {\bibfield  {journal} {\bibinfo  {journal} {Journal of the Optical Society of America B}\ }\textbf {\bibinfo {volume} {4}},\ \bibinfo {pages} {1551} (\bibinfo {year} {1987})}\BibitemShut {NoStop}%
\bibitem [{\citenamefont {Yurke}\ \emph {et~al.}(1988)\citenamefont {Yurke}, \citenamefont {Kaminsky}, \citenamefont {Miller}, \citenamefont {Whittaker}, \citenamefont {Smith}, \citenamefont {Silver},\ and\ \citenamefont {Simon}}]{yurke1988observation}%
  \BibitemOpen
  \bibfield  {author} {\bibinfo {author} {\bibfnamefont {B.}~\bibnamefont {Yurke}}, \bibinfo {author} {\bibfnamefont {P.}~\bibnamefont {Kaminsky}}, \bibinfo {author} {\bibfnamefont {R.}~\bibnamefont {Miller}}, \bibinfo {author} {\bibfnamefont {E.}~\bibnamefont {Whittaker}}, \bibinfo {author} {\bibfnamefont {A.}~\bibnamefont {Smith}}, \bibinfo {author} {\bibfnamefont {A.}~\bibnamefont {Silver}},\ and\ \bibinfo {author} {\bibfnamefont {R.}~\bibnamefont {Simon}},\ }\bibfield  {title} {\bibinfo {title} {Observation of 4.2-k equilibrium-noise squeezing via a josephson-parametric amplifier},\ }\href@noop {} {\bibfield  {journal} {\bibinfo  {journal} {Physical review letters}\ }\textbf {\bibinfo {volume} {60}},\ \bibinfo {pages} {764} (\bibinfo {year} {1988})}\BibitemShut {NoStop}%
\bibitem [{\citenamefont {Yurke}\ \emph {et~al.}(1989)\citenamefont {Yurke}, \citenamefont {Corruccini}, \citenamefont {Kaminsky}, \citenamefont {Rupp}, \citenamefont {Smith}, \citenamefont {Silver}, \citenamefont {Simon},\ and\ \citenamefont {Whittaker}}]{yurke1989observation}%
  \BibitemOpen
  \bibfield  {author} {\bibinfo {author} {\bibfnamefont {B.}~\bibnamefont {Yurke}}, \bibinfo {author} {\bibfnamefont {L.}~\bibnamefont {Corruccini}}, \bibinfo {author} {\bibfnamefont {P.}~\bibnamefont {Kaminsky}}, \bibinfo {author} {\bibfnamefont {L.}~\bibnamefont {Rupp}}, \bibinfo {author} {\bibfnamefont {A.}~\bibnamefont {Smith}}, \bibinfo {author} {\bibfnamefont {A.}~\bibnamefont {Silver}}, \bibinfo {author} {\bibfnamefont {R.}~\bibnamefont {Simon}},\ and\ \bibinfo {author} {\bibfnamefont {E.}~\bibnamefont {Whittaker}},\ }\bibfield  {title} {\bibinfo {title} {Observation of parametric amplification and deamplification in a josephson parametric amplifier},\ }\href@noop {} {\bibfield  {journal} {\bibinfo  {journal} {Physical Review A}\ }\textbf {\bibinfo {volume} {39}},\ \bibinfo {pages} {2519} (\bibinfo {year} {1989})}\BibitemShut {NoStop}%
\bibitem [{\citenamefont {Kjaergaard}\ \emph {et~al.}(2020)\citenamefont {Kjaergaard}, \citenamefont {Schwartz}, \citenamefont {Braum{\"u}ller}, \citenamefont {Krantz}, \citenamefont {Wang}, \citenamefont {Gustavsson},\ and\ \citenamefont {Oliver}}]{kjaergaard2020superconducting}%
  \BibitemOpen
  \bibfield  {author} {\bibinfo {author} {\bibfnamefont {M.}~\bibnamefont {Kjaergaard}}, \bibinfo {author} {\bibfnamefont {M.~E.}\ \bibnamefont {Schwartz}}, \bibinfo {author} {\bibfnamefont {J.}~\bibnamefont {Braum{\"u}ller}}, \bibinfo {author} {\bibfnamefont {P.}~\bibnamefont {Krantz}}, \bibinfo {author} {\bibfnamefont {J.~I.-J.}\ \bibnamefont {Wang}}, \bibinfo {author} {\bibfnamefont {S.}~\bibnamefont {Gustavsson}},\ and\ \bibinfo {author} {\bibfnamefont {W.~D.}\ \bibnamefont {Oliver}},\ }\bibfield  {title} {\bibinfo {title} {Superconducting qubits: Current state of play},\ }\href@noop {} {\bibfield  {journal} {\bibinfo  {journal} {Annual Review of Condensed Matter Physics}\ }\textbf {\bibinfo {volume} {11}},\ \bibinfo {pages} {369} (\bibinfo {year} {2020})}\BibitemShut {NoStop}%
\bibitem [{\citenamefont {Siddiqi}(2021)}]{siddiqi2021engineering}%
  \BibitemOpen
  \bibfield  {author} {\bibinfo {author} {\bibfnamefont {I.}~\bibnamefont {Siddiqi}},\ }\bibfield  {title} {\bibinfo {title} {Engineering high-coherence superconducting qubits},\ }\href@noop {} {\bibfield  {journal} {\bibinfo  {journal} {Nature Reviews Materials}\ }\textbf {\bibinfo {volume} {6}},\ \bibinfo {pages} {875} (\bibinfo {year} {2021})}\BibitemShut {NoStop}%
\bibitem [{\citenamefont {Somoroff}\ \emph {et~al.}(2023)\citenamefont {Somoroff}, \citenamefont {Ficheux}, \citenamefont {Mencia}, \citenamefont {Xiong}, \citenamefont {Kuzmin},\ and\ \citenamefont {Manucharyan}}]{somoroff2023millisecond}%
  \BibitemOpen
  \bibfield  {author} {\bibinfo {author} {\bibfnamefont {A.}~\bibnamefont {Somoroff}}, \bibinfo {author} {\bibfnamefont {Q.}~\bibnamefont {Ficheux}}, \bibinfo {author} {\bibfnamefont {R.~A.}\ \bibnamefont {Mencia}}, \bibinfo {author} {\bibfnamefont {H.}~\bibnamefont {Xiong}}, \bibinfo {author} {\bibfnamefont {R.}~\bibnamefont {Kuzmin}},\ and\ \bibinfo {author} {\bibfnamefont {V.~E.}\ \bibnamefont {Manucharyan}},\ }\bibfield  {title} {\bibinfo {title} {Millisecond coherence in a superconducting qubit},\ }\href@noop {} {\bibfield  {journal} {\bibinfo  {journal} {Physical Review Letters}\ }\textbf {\bibinfo {volume} {130}},\ \bibinfo {pages} {267001} (\bibinfo {year} {2023})}\BibitemShut {NoStop}%
\bibitem [{\citenamefont {Romero}\ \emph {et~al.}(2009)\citenamefont {Romero}, \citenamefont {Garc{\'\i}a-Ripoll},\ and\ \citenamefont {Solano}}]{romero2009microwave}%
  \BibitemOpen
  \bibfield  {author} {\bibinfo {author} {\bibfnamefont {G.}~\bibnamefont {Romero}}, \bibinfo {author} {\bibfnamefont {J.~J.}\ \bibnamefont {Garc{\'\i}a-Ripoll}},\ and\ \bibinfo {author} {\bibfnamefont {E.}~\bibnamefont {Solano}},\ }\bibfield  {title} {\bibinfo {title} {Microwave photon detector in circuit qed},\ }\href@noop {} {\bibfield  {journal} {\bibinfo  {journal} {Physical review letters}\ }\textbf {\bibinfo {volume} {102}},\ \bibinfo {pages} {173602} (\bibinfo {year} {2009})}\BibitemShut {NoStop}%
\bibitem [{\citenamefont {Schoelkopf}\ \emph {et~al.}(1998)\citenamefont {Schoelkopf}, \citenamefont {Wahlgren}, \citenamefont {Kozhevnikov}, \citenamefont {Delsing},\ and\ \citenamefont {Prober}}]{schoelkopf1998radio}%
  \BibitemOpen
  \bibfield  {author} {\bibinfo {author} {\bibfnamefont {R.~J.}\ \bibnamefont {Schoelkopf}}, \bibinfo {author} {\bibfnamefont {P.}~\bibnamefont {Wahlgren}}, \bibinfo {author} {\bibfnamefont {A.}~\bibnamefont {Kozhevnikov}}, \bibinfo {author} {\bibfnamefont {P.}~\bibnamefont {Delsing}},\ and\ \bibinfo {author} {\bibfnamefont {D.}~\bibnamefont {Prober}},\ }\bibfield  {title} {\bibinfo {title} {The radio-frequency single-electron transistor (rf-set): A fast and ultrasensitive electrometer},\ }\href@noop {} {\bibfield  {journal} {\bibinfo  {journal} {science}\ }\textbf {\bibinfo {volume} {280}},\ \bibinfo {pages} {1238} (\bibinfo {year} {1998})}\BibitemShut {NoStop}%
\bibitem [{\citenamefont {Blais}\ \emph {et~al.}(2021)\citenamefont {Blais}, \citenamefont {Grimsmo}, \citenamefont {Girvin},\ and\ \citenamefont {Wallraff}}]{blais_circuit_2021}%
  \BibitemOpen
  \bibfield  {author} {\bibinfo {author} {\bibfnamefont {A.}~\bibnamefont {Blais}}, \bibinfo {author} {\bibfnamefont {A.~L.}\ \bibnamefont {Grimsmo}}, \bibinfo {author} {\bibfnamefont {S.~M.}\ \bibnamefont {Girvin}},\ and\ \bibinfo {author} {\bibfnamefont {A.}~\bibnamefont {Wallraff}},\ }\bibfield  {title} {\bibinfo {title} {Circuit quantum electrodynamics},\ }\href {https://doi.org/10.1103/RevModPhys.93.025005} {\bibfield  {journal} {\bibinfo  {journal} {Reviews of Modern Physics}\ }\textbf {\bibinfo {volume} {93}},\ \bibinfo {pages} {025005} (\bibinfo {year} {2021})},\ \bibinfo {note} {publisher: American Physical Society}\BibitemShut {NoStop}%
\bibitem [{\citenamefont {Bland}\ \emph {et~al.}(2025)\citenamefont {Bland}, \citenamefont {Bahrami}, \citenamefont {Martinez}, \citenamefont {Prestegaard}, \citenamefont {Smitham}, \citenamefont {Joshi}, \citenamefont {Hedrick}, \citenamefont {Kumar}, \citenamefont {Yang}, \citenamefont {Pakpour-Tabrizi} \emph {et~al.}}]{bland2025millisecond}%
  \BibitemOpen
  \bibfield  {author} {\bibinfo {author} {\bibfnamefont {M.~P.}\ \bibnamefont {Bland}}, \bibinfo {author} {\bibfnamefont {F.}~\bibnamefont {Bahrami}}, \bibinfo {author} {\bibfnamefont {J.~G.}\ \bibnamefont {Martinez}}, \bibinfo {author} {\bibfnamefont {P.~H.}\ \bibnamefont {Prestegaard}}, \bibinfo {author} {\bibfnamefont {B.~M.}\ \bibnamefont {Smitham}}, \bibinfo {author} {\bibfnamefont {A.}~\bibnamefont {Joshi}}, \bibinfo {author} {\bibfnamefont {E.}~\bibnamefont {Hedrick}}, \bibinfo {author} {\bibfnamefont {S.}~\bibnamefont {Kumar}}, \bibinfo {author} {\bibfnamefont {A.}~\bibnamefont {Yang}}, \bibinfo {author} {\bibfnamefont {A.~C.}\ \bibnamefont {Pakpour-Tabrizi}}, \emph {et~al.},\ }\bibfield  {title} {\bibinfo {title} {Millisecond lifetimes and coherence times in 2d transmon qubits},\ }\href@noop {} {\bibfield  {journal} {\bibinfo  {journal} {Nature}\ ,\ \bibinfo {pages} {1}} (\bibinfo {year} {2025})}\BibitemShut {NoStop}%
\bibitem [{\citenamefont {Clerk}\ \emph {et~al.}(2020)\citenamefont {Clerk}, \citenamefont {Lehnert}, \citenamefont {Bertet}, \citenamefont {Petta},\ and\ \citenamefont {Nakamura}}]{clerk2020hybrid}%
  \BibitemOpen
  \bibfield  {author} {\bibinfo {author} {\bibfnamefont {A.}~\bibnamefont {Clerk}}, \bibinfo {author} {\bibfnamefont {K.}~\bibnamefont {Lehnert}}, \bibinfo {author} {\bibfnamefont {P.}~\bibnamefont {Bertet}}, \bibinfo {author} {\bibfnamefont {J.}~\bibnamefont {Petta}},\ and\ \bibinfo {author} {\bibfnamefont {Y.}~\bibnamefont {Nakamura}},\ }\bibfield  {title} {\bibinfo {title} {Hybrid quantum systems with circuit quantum electrodynamics},\ }\href@noop {} {\bibfield  {journal} {\bibinfo  {journal} {Nature Physics}\ }\textbf {\bibinfo {volume} {16}},\ \bibinfo {pages} {257} (\bibinfo {year} {2020})}\BibitemShut {NoStop}%
\bibitem [{\citenamefont {Echternach}\ \emph {et~al.}(2021)\citenamefont {Echternach}, \citenamefont {Beyer},\ and\ \citenamefont {Bradford}}]{echternach_large_2021}%
  \BibitemOpen
  \bibfield  {author} {\bibinfo {author} {\bibfnamefont {P.~M.}\ \bibnamefont {Echternach}}, \bibinfo {author} {\bibfnamefont {A.~D.}\ \bibnamefont {Beyer}},\ and\ \bibinfo {author} {\bibfnamefont {C.~M.}\ \bibnamefont {Bradford}},\ }\bibfield  {title} {\bibinfo {title} {Large array of low-frequency readout quantum capacitance detectors},\ }\href {https://doi.org/10.1117/1.JATIS.7.1.011003} {\bibfield  {journal} {\bibinfo  {journal} {Journal of Astronomical Telescopes, Instruments, and Systems}\ }\textbf {\bibinfo {volume} {7}},\ \bibinfo {pages} {011003} (\bibinfo {year} {2021})}\BibitemShut {NoStop}%
\bibitem [{\citenamefont {Golwala}\ and\ \citenamefont {Figueroa-Feliciano}(2022)}]{golwala_novel_2022}%
  \BibitemOpen
  \bibfield  {author} {\bibinfo {author} {\bibfnamefont {S.~R.}\ \bibnamefont {Golwala}}\ and\ \bibinfo {author} {\bibfnamefont {E.}~\bibnamefont {Figueroa-Feliciano}},\ }\bibfield  {title} {\bibinfo {title} {Novel {Quantum} {Sensors} for {Light} {Dark} {Matter} and {Neutrino} {Detection}},\ }\href {https://doi.org/10.1146/annurev-nucl-102020-112133} {\bibfield  {journal} {\bibinfo  {journal} {Annual Review of Nuclear and Particle Science}\ }\textbf {\bibinfo {volume} {72}},\ \bibinfo {pages} {419} (\bibinfo {year} {2022})}\BibitemShut {NoStop}%
\bibitem [{\citenamefont {Rebi{\'c}}\ \emph {et~al.}(2009)\citenamefont {Rebi{\'c}}, \citenamefont {Twamley},\ and\ \citenamefont {Milburn}}]{rebic2009giant}%
  \BibitemOpen
  \bibfield  {author} {\bibinfo {author} {\bibfnamefont {S.}~\bibnamefont {Rebi{\'c}}}, \bibinfo {author} {\bibfnamefont {J.}~\bibnamefont {Twamley}},\ and\ \bibinfo {author} {\bibfnamefont {G.~J.}\ \bibnamefont {Milburn}},\ }\bibfield  {title} {\bibinfo {title} {Giant kerr nonlinearities in circuit quantum electrodynamics},\ }\href@noop {} {\bibfield  {journal} {\bibinfo  {journal} {Physical review letters}\ }\textbf {\bibinfo {volume} {103}},\ \bibinfo {pages} {150503} (\bibinfo {year} {2009})}\BibitemShut {NoStop}%
\bibitem [{\citenamefont {Moon}\ and\ \citenamefont {Girvin}(2005)}]{moon2005theory}%
  \BibitemOpen
  \bibfield  {author} {\bibinfo {author} {\bibfnamefont {K.}~\bibnamefont {Moon}}\ and\ \bibinfo {author} {\bibfnamefont {S.}~\bibnamefont {Girvin}},\ }\bibfield  {title} {\bibinfo {title} {Theory of microwave parametric down-conversion and squeezing using circuit qed},\ }\href@noop {} {\bibfield  {journal} {\bibinfo  {journal} {Physical review letters}\ }\textbf {\bibinfo {volume} {95}},\ \bibinfo {pages} {140504} (\bibinfo {year} {2005})}\BibitemShut {NoStop}%
\bibitem [{\citenamefont {Aumentado}(2020)}]{aumentado2020superconducting}%
  \BibitemOpen
  \bibfield  {author} {\bibinfo {author} {\bibfnamefont {J.}~\bibnamefont {Aumentado}},\ }\bibfield  {title} {\bibinfo {title} {Superconducting parametric amplifiers: The state of the art in josephson parametric amplifiers},\ }\href@noop {} {\bibfield  {journal} {\bibinfo  {journal} {IEEE Microwave magazine}\ }\textbf {\bibinfo {volume} {21}},\ \bibinfo {pages} {45} (\bibinfo {year} {2020})}\BibitemShut {NoStop}%
\bibitem [{\citenamefont {Arora}\ and\ \citenamefont {Narang}(2025)}]{arora2025chiral}%
  \BibitemOpen
  \bibfield  {author} {\bibinfo {author} {\bibfnamefont {A.}~\bibnamefont {Arora}}\ and\ \bibinfo {author} {\bibfnamefont {P.}~\bibnamefont {Narang}},\ }\bibfield  {title} {\bibinfo {title} {Chiral cavity control of superconducting diode-like nonlinearities},\ }\href@noop {} {\bibfield  {journal} {\bibinfo  {journal} {arXiv preprint arXiv:2501.17924}\ } (\bibinfo {year} {2025})}\BibitemShut {NoStop}%
\bibitem [{\citenamefont {Dirnegger}\ \emph {et~al.}(2025{\natexlab{a}})\citenamefont {Dirnegger}, \citenamefont {Narang},\ and\ \citenamefont {Arora}}]{dirnegger2025nonreciprocal}%
  \BibitemOpen
  \bibfield  {author} {\bibinfo {author} {\bibfnamefont {N.}~\bibnamefont {Dirnegger}}, \bibinfo {author} {\bibfnamefont {P.}~\bibnamefont {Narang}},\ and\ \bibinfo {author} {\bibfnamefont {A.}~\bibnamefont {Arora}},\ }\bibfield  {title} {\bibinfo {title} {Nonreciprocal quantum information processing with superconducting diodes in circuit quantum electrodynamics},\ }\href@noop {} {\bibfield  {journal} {\bibinfo  {journal} {arXiv preprint arXiv:2511.20758}\ } (\bibinfo {year} {2025}{\natexlab{a}})}\BibitemShut {NoStop}%
\bibitem [{\citenamefont {Chitambar}\ and\ \citenamefont {Gour}(2019)}]{chitambar_quantum_2019}%
  \BibitemOpen
  \bibfield  {author} {\bibinfo {author} {\bibfnamefont {E.}~\bibnamefont {Chitambar}}\ and\ \bibinfo {author} {\bibfnamefont {G.}~\bibnamefont {Gour}},\ }\bibfield  {title} {\bibinfo {title} {Quantum resource theories},\ }\href {https://doi.org/10.1103/RevModPhys.91.025001} {\bibfield  {journal} {\bibinfo  {journal} {Reviews of Modern Physics}\ }\textbf {\bibinfo {volume} {91}},\ \bibinfo {pages} {025001} (\bibinfo {year} {2019})},\ \bibinfo {note} {publisher: American Physical Society}\BibitemShut {NoStop}%
\bibitem [{\citenamefont {Alexeev}\ \emph {et~al.}(2025)\citenamefont {Alexeev}, \citenamefont {Farag}, \citenamefont {Patti}, \citenamefont {Wolf}, \citenamefont {Ares}, \citenamefont {Aspuru-Guzik}, \citenamefont {Benjamin}, \citenamefont {Cai}, \citenamefont {Cao}, \citenamefont {Chamberland} \emph {et~al.}}]{alexeev2025artificial}%
  \BibitemOpen
  \bibfield  {author} {\bibinfo {author} {\bibfnamefont {Y.}~\bibnamefont {Alexeev}}, \bibinfo {author} {\bibfnamefont {M.~H.}\ \bibnamefont {Farag}}, \bibinfo {author} {\bibfnamefont {T.~L.}\ \bibnamefont {Patti}}, \bibinfo {author} {\bibfnamefont {M.~E.}\ \bibnamefont {Wolf}}, \bibinfo {author} {\bibfnamefont {N.}~\bibnamefont {Ares}}, \bibinfo {author} {\bibfnamefont {A.}~\bibnamefont {Aspuru-Guzik}}, \bibinfo {author} {\bibfnamefont {S.~C.}\ \bibnamefont {Benjamin}}, \bibinfo {author} {\bibfnamefont {Z.}~\bibnamefont {Cai}}, \bibinfo {author} {\bibfnamefont {S.}~\bibnamefont {Cao}}, \bibinfo {author} {\bibfnamefont {C.}~\bibnamefont {Chamberland}}, \emph {et~al.},\ }\bibfield  {title} {\bibinfo {title} {Artificial intelligence for quantum computing},\ }\href@noop {} {\bibfield  {journal} {\bibinfo  {journal} {Nature Communications}\ }\textbf {\bibinfo {volume} {16}},\ \bibinfo {pages} {10829} (\bibinfo {year} {2025})}\BibitemShut {NoStop}%
\bibitem [{\citenamefont {Menke}\ \emph {et~al.}(2021)\citenamefont {Menke}, \citenamefont {H{\"a}se}, \citenamefont {Gustavsson}, \citenamefont {Kerman}, \citenamefont {Oliver},\ and\ \citenamefont {Aspuru-Guzik}}]{menke2021automated}%
  \BibitemOpen
  \bibfield  {author} {\bibinfo {author} {\bibfnamefont {T.}~\bibnamefont {Menke}}, \bibinfo {author} {\bibfnamefont {F.}~\bibnamefont {H{\"a}se}}, \bibinfo {author} {\bibfnamefont {S.}~\bibnamefont {Gustavsson}}, \bibinfo {author} {\bibfnamefont {A.~J.}\ \bibnamefont {Kerman}}, \bibinfo {author} {\bibfnamefont {W.~D.}\ \bibnamefont {Oliver}},\ and\ \bibinfo {author} {\bibfnamefont {A.}~\bibnamefont {Aspuru-Guzik}},\ }\bibfield  {title} {\bibinfo {title} {Automated design of superconducting circuits and its application to 4-local couplers},\ }\href@noop {} {\bibfield  {journal} {\bibinfo  {journal} {npj Quantum Information}\ }\textbf {\bibinfo {volume} {7}},\ \bibinfo {pages} {49} (\bibinfo {year} {2021})}\BibitemShut {NoStop}%
\bibitem [{\citenamefont {Nugraha}\ and\ \citenamefont {Shao}(2023)}]{nugraha2023machine}%
  \BibitemOpen
  \bibfield  {author} {\bibinfo {author} {\bibfnamefont {F.~P.}\ \bibnamefont {Nugraha}}\ and\ \bibinfo {author} {\bibfnamefont {Q.}~\bibnamefont {Shao}},\ }\bibfield  {title} {\bibinfo {title} {Machine learning-based predictive modeling for designing transmon superconducting qubits},\ }in\ \href@noop {} {\emph {\bibinfo {booktitle} {2023 IEEE International Conference on Quantum Computing and Engineering (QCE)}}},\ Vol.~\bibinfo {volume} {1}\ (\bibinfo {organization} {IEEE},\ \bibinfo {year} {2023})\ pp.\ \bibinfo {pages} {1360--1368}\BibitemShut {NoStop}%
\bibitem [{\citenamefont {Ye}\ \emph {et~al.}(2025)\citenamefont {Ye}, \citenamefont {Wang},\ and\ \citenamefont {Liu}}]{ye2025electromagnetic}%
  \BibitemOpen
  \bibfield  {author} {\bibinfo {author} {\bibfnamefont {J.}~\bibnamefont {Ye}}, \bibinfo {author} {\bibfnamefont {J.}~\bibnamefont {Wang}},\ and\ \bibinfo {author} {\bibfnamefont {Y.-X.}\ \bibnamefont {Liu}},\ }\bibfield  {title} {\bibinfo {title} {Electromagnetic feature extraction in superconducting quantum circuits: An open-source finite-element workflow using palace},\ }in\ \href@noop {} {\emph {\bibinfo {booktitle} {2025 International Applied Computational Electromagnetics Society Symposium (ACES-China)}}}\ (\bibinfo {organization} {IEEE},\ \bibinfo {year} {2025})\ pp.\ \bibinfo {pages} {1--3}\BibitemShut {NoStop}%
\bibitem [{\citenamefont {Saslow}\ and\ \citenamefont {Wong}(2025)}]{saslow2025analysis}%
  \BibitemOpen
  \bibfield  {author} {\bibinfo {author} {\bibfnamefont {J.}~\bibnamefont {Saslow}}\ and\ \bibinfo {author} {\bibfnamefont {H.~Y.}\ \bibnamefont {Wong}},\ }\bibfield  {title} {\bibinfo {title} {Analysis of a 3d integrated superconducting quantum chip structure},\ }in\ \href@noop {} {\emph {\bibinfo {booktitle} {2025 IEEE International Conference on Quantum Computing and Engineering (QCE)}}},\ Vol.~\bibinfo {volume} {1}\ (\bibinfo {organization} {IEEE},\ \bibinfo {year} {2025})\ pp.\ \bibinfo {pages} {1410--1418}\BibitemShut {NoStop}%
\bibitem [{\citenamefont {Yao}\ \emph {et~al.}(2022)\citenamefont {Yao}, \citenamefont {Jambunathan}, \citenamefont {Zeng},\ and\ \citenamefont {Nonaka}}]{yao2022massively}%
  \BibitemOpen
  \bibfield  {author} {\bibinfo {author} {\bibfnamefont {Z.}~\bibnamefont {Yao}}, \bibinfo {author} {\bibfnamefont {R.}~\bibnamefont {Jambunathan}}, \bibinfo {author} {\bibfnamefont {Y.}~\bibnamefont {Zeng}},\ and\ \bibinfo {author} {\bibfnamefont {A.}~\bibnamefont {Nonaka}},\ }\bibfield  {title} {\bibinfo {title} {A massively parallel time-domain coupled electrodynamics--micromagnetics solver},\ }\href@noop {} {\bibfield  {journal} {\bibinfo  {journal} {The International Journal of High Performance Computing Applications}\ }\textbf {\bibinfo {volume} {36}},\ \bibinfo {pages} {167} (\bibinfo {year} {2022})}\BibitemShut {NoStop}%
\bibitem [{\citenamefont {Sommers}\ \emph {et~al.}(2025)\citenamefont {Sommers}, \citenamefont {Degnan}, \citenamefont {Gautam}, \citenamefont {Chen}, \citenamefont {Chiu}, \citenamefont {Fedorov},\ and\ \citenamefont {Pakkiam}}]{sommers2025open}%
  \BibitemOpen
  \bibfield  {author} {\bibinfo {author} {\bibfnamefont {D.}~\bibnamefont {Sommers}}, \bibinfo {author} {\bibfnamefont {Z.}~\bibnamefont {Degnan}}, \bibinfo {author} {\bibfnamefont {D.}~\bibnamefont {Gautam}}, \bibinfo {author} {\bibfnamefont {Y.-H.}\ \bibnamefont {Chen}}, \bibinfo {author} {\bibfnamefont {C.-C.}\ \bibnamefont {Chiu}}, \bibinfo {author} {\bibfnamefont {A.}~\bibnamefont {Fedorov}},\ and\ \bibinfo {author} {\bibfnamefont {P.}~\bibnamefont {Pakkiam}},\ }\bibfield  {title} {\bibinfo {title} {Open-source highly parallel electromagnetic simulations for superconducting circuits},\ }\href@noop {} {\bibfield  {journal} {\bibinfo  {journal} {arXiv preprint arXiv:2511.01220}\ } (\bibinfo {year} {2025})}\BibitemShut {NoStop}%
\bibitem [{\citenamefont {Bayraktar}\ \emph {et~al.}(2023)\citenamefont {Bayraktar}, \citenamefont {Charara}, \citenamefont {Clark}, \citenamefont {Cohen}, \citenamefont {Costa}, \citenamefont {Fang}, \citenamefont {Gao}, \citenamefont {Guan}, \citenamefont {Gunnels}, \citenamefont {Haidar} \emph {et~al.}}]{bayraktar2023cuquantum}%
  \BibitemOpen
  \bibfield  {author} {\bibinfo {author} {\bibfnamefont {H.}~\bibnamefont {Bayraktar}}, \bibinfo {author} {\bibfnamefont {A.}~\bibnamefont {Charara}}, \bibinfo {author} {\bibfnamefont {D.}~\bibnamefont {Clark}}, \bibinfo {author} {\bibfnamefont {S.}~\bibnamefont {Cohen}}, \bibinfo {author} {\bibfnamefont {T.}~\bibnamefont {Costa}}, \bibinfo {author} {\bibfnamefont {Y.-L.~L.}\ \bibnamefont {Fang}}, \bibinfo {author} {\bibfnamefont {Y.}~\bibnamefont {Gao}}, \bibinfo {author} {\bibfnamefont {J.}~\bibnamefont {Guan}}, \bibinfo {author} {\bibfnamefont {J.}~\bibnamefont {Gunnels}}, \bibinfo {author} {\bibfnamefont {A.}~\bibnamefont {Haidar}}, \emph {et~al.},\ }\bibfield  {title} {\bibinfo {title} {cuquantum sdk: A high-performance library for accelerating quantum science},\ }in\ \href@noop {} {\emph {\bibinfo {booktitle} {2023 IEEE International Conference on Quantum Computing and Engineering (QCE)}}},\ Vol.~\bibinfo {volume} {1}\ (\bibinfo {organization} {IEEE},\ \bibinfo {year} {2023})\ pp.\ \bibinfo {pages}
  {1050--1061}\BibitemShut {NoStop}%
\bibitem [{\citenamefont {De~Raedt}\ \emph {et~al.}(2025)\citenamefont {De~Raedt}, \citenamefont {Kraus}, \citenamefont {Herten}, \citenamefont {Mehta}, \citenamefont {Bode}, \citenamefont {Hrywniak}, \citenamefont {Michielsen},\ and\ \citenamefont {Lippert}}]{de2025universal}%
  \BibitemOpen
  \bibfield  {author} {\bibinfo {author} {\bibfnamefont {H.}~\bibnamefont {De~Raedt}}, \bibinfo {author} {\bibfnamefont {J.}~\bibnamefont {Kraus}}, \bibinfo {author} {\bibfnamefont {A.}~\bibnamefont {Herten}}, \bibinfo {author} {\bibfnamefont {V.}~\bibnamefont {Mehta}}, \bibinfo {author} {\bibfnamefont {M.}~\bibnamefont {Bode}}, \bibinfo {author} {\bibfnamefont {M.}~\bibnamefont {Hrywniak}}, \bibinfo {author} {\bibfnamefont {K.}~\bibnamefont {Michielsen}},\ and\ \bibinfo {author} {\bibfnamefont {T.}~\bibnamefont {Lippert}},\ }\bibfield  {title} {\bibinfo {title} {Universal quantum simulation of 50 qubits on europe's first exascale supercomputer harnessing its heterogeneous cpu-gpu architecture},\ }\href@noop {} {\bibfield  {journal} {\bibinfo  {journal} {arXiv preprint arXiv:2511.03359}\ } (\bibinfo {year} {2025})}\BibitemShut {NoStop}%
\bibitem [{\citenamefont {Chakraborty}\ \emph {et~al.}(2025)\citenamefont {Chakraborty}, \citenamefont {Patti}, \citenamefont {Khailany}, \citenamefont {Jordan},\ and\ \citenamefont {Anandkumar}}]{chakraborty2025gpu}%
  \BibitemOpen
  \bibfield  {author} {\bibinfo {author} {\bibfnamefont {A.}~\bibnamefont {Chakraborty}}, \bibinfo {author} {\bibfnamefont {T.~L.}\ \bibnamefont {Patti}}, \bibinfo {author} {\bibfnamefont {B.}~\bibnamefont {Khailany}}, \bibinfo {author} {\bibfnamefont {A.~N.}\ \bibnamefont {Jordan}},\ and\ \bibinfo {author} {\bibfnamefont {A.}~\bibnamefont {Anandkumar}},\ }\bibfield  {title} {\bibinfo {title} {Gpu-accelerated effective hamiltonian calculator},\ }\href@noop {} {\bibfield  {journal} {\bibinfo  {journal} {Quantum}\ }\textbf {\bibinfo {volume} {9}},\ \bibinfo {pages} {1946} (\bibinfo {year} {2025})}\BibitemShut {NoStop}%
\bibitem [{\citenamefont {Shanto}\ \emph {et~al.}(2024)\citenamefont {Shanto}, \citenamefont {Kuo}, \citenamefont {Miyamoto}, \citenamefont {Zhang}, \citenamefont {Maurya}, \citenamefont {Vlachos}, \citenamefont {Hecht}, \citenamefont {Shum},\ and\ \citenamefont {Levenson-Falk}}]{shanto2024squadds}%
  \BibitemOpen
  \bibfield  {author} {\bibinfo {author} {\bibfnamefont {S.}~\bibnamefont {Shanto}}, \bibinfo {author} {\bibfnamefont {A.}~\bibnamefont {Kuo}}, \bibinfo {author} {\bibfnamefont {C.}~\bibnamefont {Miyamoto}}, \bibinfo {author} {\bibfnamefont {H.}~\bibnamefont {Zhang}}, \bibinfo {author} {\bibfnamefont {V.}~\bibnamefont {Maurya}}, \bibinfo {author} {\bibfnamefont {E.}~\bibnamefont {Vlachos}}, \bibinfo {author} {\bibfnamefont {M.}~\bibnamefont {Hecht}}, \bibinfo {author} {\bibfnamefont {C.~W.}\ \bibnamefont {Shum}},\ and\ \bibinfo {author} {\bibfnamefont {E.}~\bibnamefont {Levenson-Falk}},\ }\bibfield  {title} {\bibinfo {title} {Squadds: A validated design database and simulation workflow for superconducting qubit design},\ }\href@noop {} {\bibfield  {journal} {\bibinfo  {journal} {Quantum}\ }\textbf {\bibinfo {volume} {8}},\ \bibinfo {pages} {1465} (\bibinfo {year} {2024})}\BibitemShut {NoStop}%
\bibitem [{\citenamefont {Seidel}\ \emph {et~al.}(2026)\citenamefont {Seidel}, \citenamefont {Abouzahr}, \citenamefont {Chakraborty}, \citenamefont {Shanto}, \citenamefont {Das}, \citenamefont {Baxter}, \citenamefont {Asaadi}, \citenamefont {Pancotti}, \citenamefont {Yang}, \citenamefont {Khailany} \emph {et~al.}}]{seidel2026component}%
  \BibitemOpen
  \bibfield  {author} {\bibinfo {author} {\bibfnamefont {O.}~\bibnamefont {Seidel}}, \bibinfo {author} {\bibfnamefont {F.}~\bibnamefont {Abouzahr}}, \bibinfo {author} {\bibfnamefont {A.}~\bibnamefont {Chakraborty}}, \bibinfo {author} {\bibfnamefont {S.~A.}\ \bibnamefont {Shanto}}, \bibinfo {author} {\bibfnamefont {S.}~\bibnamefont {Das}}, \bibinfo {author} {\bibfnamefont {D.}~\bibnamefont {Baxter}}, \bibinfo {author} {\bibfnamefont {J.}~\bibnamefont {Asaadi}}, \bibinfo {author} {\bibfnamefont {N.}~\bibnamefont {Pancotti}}, \bibinfo {author} {\bibfnamefont {H.}~\bibnamefont {Yang}}, \bibinfo {author} {\bibfnamefont {B.}~\bibnamefont {Khailany}}, \emph {et~al.},\ }\bibfield  {title} {\bibinfo {title} {Component-level inverse design of transmon qubits using neural networks},\ }\href@noop {} {\bibfield  {journal} {\bibinfo  {journal} {arXiv preprint arXiv:2607.20795}\ } (\bibinfo {year} {2026})}\BibitemShut {NoStop}%
\bibitem [{\citenamefont {Degen}\ \emph {et~al.}(2017)\citenamefont {Degen}, \citenamefont {Reinhard},\ and\ \citenamefont {Cappellaro}}]{degen2017quantum}%
  \BibitemOpen
  \bibfield  {author} {\bibinfo {author} {\bibfnamefont {C.~L.}\ \bibnamefont {Degen}}, \bibinfo {author} {\bibfnamefont {F.}~\bibnamefont {Reinhard}},\ and\ \bibinfo {author} {\bibfnamefont {P.}~\bibnamefont {Cappellaro}},\ }\bibfield  {title} {\bibinfo {title} {Quantum sensing},\ }\href@noop {} {\bibfield  {journal} {\bibinfo  {journal} {Reviews of modern physics}\ }\textbf {\bibinfo {volume} {89}},\ \bibinfo {pages} {035002} (\bibinfo {year} {2017})}\BibitemShut {NoStop}%
\bibitem [{\citenamefont {Cottet}\ \emph {et~al.}(2017)\citenamefont {Cottet}, \citenamefont {Dartiailh}, \citenamefont {Desjardins}, \citenamefont {Cubaynes}, \citenamefont {Contamin}, \citenamefont {Delbecq}, \citenamefont {Viennot}, \citenamefont {Bruhat}, \citenamefont {Dou{\c{c}}ot},\ and\ \citenamefont {Kontos}}]{cottet2017cavity}%
  \BibitemOpen
  \bibfield  {author} {\bibinfo {author} {\bibfnamefont {A.}~\bibnamefont {Cottet}}, \bibinfo {author} {\bibfnamefont {M.~C.}\ \bibnamefont {Dartiailh}}, \bibinfo {author} {\bibfnamefont {M.~M.}\ \bibnamefont {Desjardins}}, \bibinfo {author} {\bibfnamefont {T.}~\bibnamefont {Cubaynes}}, \bibinfo {author} {\bibfnamefont {L.~C.}\ \bibnamefont {Contamin}}, \bibinfo {author} {\bibfnamefont {M.}~\bibnamefont {Delbecq}}, \bibinfo {author} {\bibfnamefont {J.~J.}\ \bibnamefont {Viennot}}, \bibinfo {author} {\bibfnamefont {L.~E.}\ \bibnamefont {Bruhat}}, \bibinfo {author} {\bibfnamefont {B.}~\bibnamefont {Dou{\c{c}}ot}},\ and\ \bibinfo {author} {\bibfnamefont {T.}~\bibnamefont {Kontos}},\ }\bibfield  {title} {\bibinfo {title} {Cavity qed with hybrid nanocircuits: from atomic-like physics to condensed matter phenomena},\ }\href@noop {} {\bibfield  {journal} {\bibinfo  {journal} {Journal of Physics: Condensed Matter}\ }\textbf {\bibinfo {volume} {29}},\ \bibinfo {pages} {433002} (\bibinfo {year} {2017})}\BibitemShut
  {NoStop}%
\bibitem [{\citenamefont {Day}\ \emph {et~al.}(2003)\citenamefont {Day}, \citenamefont {LeDuc}, \citenamefont {Mazin}, \citenamefont {Vayonakis},\ and\ \citenamefont {Zmuidzinas}}]{day2003broadband}%
  \BibitemOpen
  \bibfield  {author} {\bibinfo {author} {\bibfnamefont {P.~K.}\ \bibnamefont {Day}}, \bibinfo {author} {\bibfnamefont {H.~G.}\ \bibnamefont {LeDuc}}, \bibinfo {author} {\bibfnamefont {B.~A.}\ \bibnamefont {Mazin}}, \bibinfo {author} {\bibfnamefont {A.}~\bibnamefont {Vayonakis}},\ and\ \bibinfo {author} {\bibfnamefont {J.}~\bibnamefont {Zmuidzinas}},\ }\bibfield  {title} {\bibinfo {title} {A broadband superconducting detector suitable for use in large arrays},\ }\href@noop {} {\bibfield  {journal} {\bibinfo  {journal} {Nature}\ }\textbf {\bibinfo {volume} {425}},\ \bibinfo {pages} {817} (\bibinfo {year} {2003})}\BibitemShut {NoStop}%
\bibitem [{\citenamefont {Basov}\ \emph {et~al.}(2017)\citenamefont {Basov}, \citenamefont {Averitt},\ and\ \citenamefont {Hsieh}}]{basov2017towards}%
  \BibitemOpen
  \bibfield  {author} {\bibinfo {author} {\bibfnamefont {D.}~\bibnamefont {Basov}}, \bibinfo {author} {\bibfnamefont {R.}~\bibnamefont {Averitt}},\ and\ \bibinfo {author} {\bibfnamefont {D.}~\bibnamefont {Hsieh}},\ }\bibfield  {title} {\bibinfo {title} {Towards properties on demand in quantum materials},\ }\href@noop {} {\bibfield  {journal} {\bibinfo  {journal} {Nature materials}\ }\textbf {\bibinfo {volume} {16}},\ \bibinfo {pages} {1077} (\bibinfo {year} {2017})}\BibitemShut {NoStop}%
\bibitem [{\citenamefont {De~La~Torre}\ \emph {et~al.}(2021)\citenamefont {De~La~Torre}, \citenamefont {Kennes}, \citenamefont {Claassen}, \citenamefont {Gerber}, \citenamefont {McIver},\ and\ \citenamefont {Sentef}}]{de2021colloquium}%
  \BibitemOpen
  \bibfield  {author} {\bibinfo {author} {\bibfnamefont {A.}~\bibnamefont {De~La~Torre}}, \bibinfo {author} {\bibfnamefont {D.~M.}\ \bibnamefont {Kennes}}, \bibinfo {author} {\bibfnamefont {M.}~\bibnamefont {Claassen}}, \bibinfo {author} {\bibfnamefont {S.}~\bibnamefont {Gerber}}, \bibinfo {author} {\bibfnamefont {J.~W.}\ \bibnamefont {McIver}},\ and\ \bibinfo {author} {\bibfnamefont {M.~A.}\ \bibnamefont {Sentef}},\ }\bibfield  {title} {\bibinfo {title} {Colloquium: Nonthermal pathways to ultrafast control in quantum materials},\ }\href@noop {} {\bibfield  {journal} {\bibinfo  {journal} {Reviews of Modern Physics}\ }\textbf {\bibinfo {volume} {93}},\ \bibinfo {pages} {041002} (\bibinfo {year} {2021})}\BibitemShut {NoStop}%
\bibitem [{\citenamefont {Yin}\ \emph {et~al.}(2021)\citenamefont {Yin}, \citenamefont {Pan},\ and\ \citenamefont {Zahid~Hasan}}]{yin2021probing}%
  \BibitemOpen
  \bibfield  {author} {\bibinfo {author} {\bibfnamefont {J.-X.}\ \bibnamefont {Yin}}, \bibinfo {author} {\bibfnamefont {S.~H.}\ \bibnamefont {Pan}},\ and\ \bibinfo {author} {\bibfnamefont {M.}~\bibnamefont {Zahid~Hasan}},\ }\bibfield  {title} {\bibinfo {title} {Probing topological quantum matter with scanning tunnelling microscopy},\ }\href@noop {} {\bibfield  {journal} {\bibinfo  {journal} {Nature Reviews Physics}\ }\textbf {\bibinfo {volume} {3}},\ \bibinfo {pages} {249} (\bibinfo {year} {2021})}\BibitemShut {NoStop}%
\bibitem [{\citenamefont {B{\o}ttcher}\ \emph {et~al.}(2024{\natexlab{a}})\citenamefont {B{\o}ttcher}, \citenamefont {Poniatowski}, \citenamefont {Grankin}, \citenamefont {Wesson}, \citenamefont {Yan}, \citenamefont {Vool}, \citenamefont {Galitski},\ and\ \citenamefont {Yacoby}}]{bottcher2024circuit}%
  \BibitemOpen
  \bibfield  {author} {\bibinfo {author} {\bibfnamefont {C.}~\bibnamefont {B{\o}ttcher}}, \bibinfo {author} {\bibfnamefont {N.}~\bibnamefont {Poniatowski}}, \bibinfo {author} {\bibfnamefont {A.}~\bibnamefont {Grankin}}, \bibinfo {author} {\bibfnamefont {M.}~\bibnamefont {Wesson}}, \bibinfo {author} {\bibfnamefont {Z.}~\bibnamefont {Yan}}, \bibinfo {author} {\bibfnamefont {U.}~\bibnamefont {Vool}}, \bibinfo {author} {\bibfnamefont {V.}~\bibnamefont {Galitski}},\ and\ \bibinfo {author} {\bibfnamefont {A.}~\bibnamefont {Yacoby}},\ }\bibfield  {title} {\bibinfo {title} {Circuit quantum electrodynamics detection of induced two-fold anisotropic pairing in a hybrid superconductor--ferromagnet bilayer},\ }\href@noop {} {\bibfield  {journal} {\bibinfo  {journal} {Nature Physics}\ }\textbf {\bibinfo {volume} {20}},\ \bibinfo {pages} {1609} (\bibinfo {year} {2024}{\natexlab{a}})}\BibitemShut {NoStop}%
\bibitem [{\citenamefont {Tanaka}\ \emph {et~al.}(2025)\citenamefont {Tanaka}, \citenamefont {Wang}, \citenamefont {Dinh}, \citenamefont {Rodan-Legrain}, \citenamefont {Zaman}, \citenamefont {Hays}, \citenamefont {Almanakly}, \citenamefont {Kannan}, \citenamefont {Kim}, \citenamefont {Niedzielski} \emph {et~al.}}]{tanaka2025superfluid}%
  \BibitemOpen
  \bibfield  {author} {\bibinfo {author} {\bibfnamefont {M.}~\bibnamefont {Tanaka}}, \bibinfo {author} {\bibfnamefont {J.~I.-J.}\ \bibnamefont {Wang}}, \bibinfo {author} {\bibfnamefont {T.~H.}\ \bibnamefont {Dinh}}, \bibinfo {author} {\bibfnamefont {D.}~\bibnamefont {Rodan-Legrain}}, \bibinfo {author} {\bibfnamefont {S.}~\bibnamefont {Zaman}}, \bibinfo {author} {\bibfnamefont {M.}~\bibnamefont {Hays}}, \bibinfo {author} {\bibfnamefont {A.}~\bibnamefont {Almanakly}}, \bibinfo {author} {\bibfnamefont {B.}~\bibnamefont {Kannan}}, \bibinfo {author} {\bibfnamefont {D.~K.}\ \bibnamefont {Kim}}, \bibinfo {author} {\bibfnamefont {B.~M.}\ \bibnamefont {Niedzielski}}, \emph {et~al.},\ }\bibfield  {title} {\bibinfo {title} {Superfluid stiffness of magic-angle twisted bilayer graphene},\ }\href@noop {} {\bibfield  {journal} {\bibinfo  {journal} {Nature}\ }\textbf {\bibinfo {volume} {638}},\ \bibinfo {pages} {99} (\bibinfo {year} {2025})}\BibitemShut {NoStop}%
\bibitem [{\citenamefont {Banerjee}\ \emph {et~al.}(2025)\citenamefont {Banerjee}, \citenamefont {Hao}, \citenamefont {Kreidel}, \citenamefont {Ledwith}, \citenamefont {Phinney}, \citenamefont {Park}, \citenamefont {Zimmerman}, \citenamefont {Wesson}, \citenamefont {Watanabe}, \citenamefont {Taniguchi} \emph {et~al.}}]{banerjee2025superfluid}%
  \BibitemOpen
  \bibfield  {author} {\bibinfo {author} {\bibfnamefont {A.}~\bibnamefont {Banerjee}}, \bibinfo {author} {\bibfnamefont {Z.}~\bibnamefont {Hao}}, \bibinfo {author} {\bibfnamefont {M.}~\bibnamefont {Kreidel}}, \bibinfo {author} {\bibfnamefont {P.}~\bibnamefont {Ledwith}}, \bibinfo {author} {\bibfnamefont {I.}~\bibnamefont {Phinney}}, \bibinfo {author} {\bibfnamefont {J.~M.}\ \bibnamefont {Park}}, \bibinfo {author} {\bibfnamefont {A.}~\bibnamefont {Zimmerman}}, \bibinfo {author} {\bibfnamefont {M.~E.}\ \bibnamefont {Wesson}}, \bibinfo {author} {\bibfnamefont {K.}~\bibnamefont {Watanabe}}, \bibinfo {author} {\bibfnamefont {T.}~\bibnamefont {Taniguchi}}, \emph {et~al.},\ }\bibfield  {title} {\bibinfo {title} {Superfluid stiffness of twisted trilayer graphene superconductors},\ }\href@noop {} {\bibfield  {journal} {\bibinfo  {journal} {Nature}\ }\textbf {\bibinfo {volume} {638}},\ \bibinfo {pages} {93} (\bibinfo {year} {2025})}\BibitemShut {NoStop}%
\bibitem [{\citenamefont {Zheng}\ \emph {et~al.}(2026)\citenamefont {Zheng}, \citenamefont {Zaman}, \citenamefont {Zhang}, \citenamefont {Occhialini}, \citenamefont {Xu}, \citenamefont {Wang}, \citenamefont {Li}, \citenamefont {Liu}, \citenamefont {Martins}, \citenamefont {Lim} \emph {et~al.}}]{zheng2026encapsulation}%
  \BibitemOpen
  \bibfield  {author} {\bibinfo {author} {\bibfnamefont {X.}~\bibnamefont {Zheng}}, \bibinfo {author} {\bibfnamefont {S.}~\bibnamefont {Zaman}}, \bibinfo {author} {\bibfnamefont {K.}~\bibnamefont {Zhang}}, \bibinfo {author} {\bibfnamefont {C.~A.}\ \bibnamefont {Occhialini}}, \bibinfo {author} {\bibfnamefont {H.}~\bibnamefont {Xu}}, \bibinfo {author} {\bibfnamefont {Z.}~\bibnamefont {Wang}}, \bibinfo {author} {\bibfnamefont {X.}~\bibnamefont {Li}}, \bibinfo {author} {\bibfnamefont {F.}~\bibnamefont {Liu}}, \bibinfo {author} {\bibfnamefont {L.~G.~P.}\ \bibnamefont {Martins}}, \bibinfo {author} {\bibfnamefont {S.}~\bibnamefont {Lim}}, \emph {et~al.},\ }\bibfield  {title} {\bibinfo {title} {Encapsulation epitaxy of air-stable 2d superconductors for quantum circuits},\ }\href@noop {} {\bibfield  {journal} {\bibinfo  {journal} {Nature}\ ,\ \bibinfo {pages} {1}} (\bibinfo {year} {2026})}\BibitemShut {NoStop}%
\bibitem [{\citenamefont {Probst}\ \emph {et~al.}(2015)\citenamefont {Probst}, \citenamefont {Song}, \citenamefont {Bushev}, \citenamefont {Ustinov},\ and\ \citenamefont {Weides}}]{probst2015efficient}%
  \BibitemOpen
  \bibfield  {author} {\bibinfo {author} {\bibfnamefont {S.}~\bibnamefont {Probst}}, \bibinfo {author} {\bibfnamefont {F.}~\bibnamefont {Song}}, \bibinfo {author} {\bibfnamefont {P.~A.}\ \bibnamefont {Bushev}}, \bibinfo {author} {\bibfnamefont {A.~V.}\ \bibnamefont {Ustinov}},\ and\ \bibinfo {author} {\bibfnamefont {M.}~\bibnamefont {Weides}},\ }\bibfield  {title} {\bibinfo {title} {Efficient and robust analysis of complex scattering data under noise in microwave resonators},\ }\href@noop {} {\bibfield  {journal} {\bibinfo  {journal} {Review of Scientific Instruments}\ }\textbf {\bibinfo {volume} {86}} (\bibinfo {year} {2015})}\BibitemShut {NoStop}%
\bibitem [{\citenamefont {Prozorov}\ and\ \citenamefont {Giannetta}(2006)}]{prozorov2006magnetic}%
  \BibitemOpen
  \bibfield  {author} {\bibinfo {author} {\bibfnamefont {R.}~\bibnamefont {Prozorov}}\ and\ \bibinfo {author} {\bibfnamefont {R.~W.}\ \bibnamefont {Giannetta}},\ }\bibfield  {title} {\bibinfo {title} {Magnetic penetration depth in unconventional superconductors},\ }\href@noop {} {\bibfield  {journal} {\bibinfo  {journal} {Superconductor Science and Technology}\ }\textbf {\bibinfo {volume} {19}},\ \bibinfo {pages} {R41} (\bibinfo {year} {2006})}\BibitemShut {NoStop}%
\bibitem [{\citenamefont {Phan}\ \emph {et~al.}(2022)\citenamefont {Phan}, \citenamefont {Senior}, \citenamefont {Ghazaryan}, \citenamefont {Hatefipour}, \citenamefont {Strickland}, \citenamefont {Shabani}, \citenamefont {Serbyn},\ and\ \citenamefont {Higginbotham}}]{phan2022detecting}%
  \BibitemOpen
  \bibfield  {author} {\bibinfo {author} {\bibfnamefont {D.}~\bibnamefont {Phan}}, \bibinfo {author} {\bibfnamefont {J.}~\bibnamefont {Senior}}, \bibinfo {author} {\bibfnamefont {A.}~\bibnamefont {Ghazaryan}}, \bibinfo {author} {\bibfnamefont {M.}~\bibnamefont {Hatefipour}}, \bibinfo {author} {\bibfnamefont {W.}~\bibnamefont {Strickland}}, \bibinfo {author} {\bibfnamefont {J.}~\bibnamefont {Shabani}}, \bibinfo {author} {\bibfnamefont {M.}~\bibnamefont {Serbyn}},\ and\ \bibinfo {author} {\bibfnamefont {A.~P.}\ \bibnamefont {Higginbotham}},\ }\bibfield  {title} {\bibinfo {title} {Detecting induced p$\pm$ip pairing at the al-inas interface with a quantum microwave circuit},\ }\href@noop {} {\bibfield  {journal} {\bibinfo  {journal} {Physical Review Letters}\ }\textbf {\bibinfo {volume} {128}},\ \bibinfo {pages} {107701} (\bibinfo {year} {2022})}\BibitemShut {NoStop}%
\bibitem [{\citenamefont {Jin}\ \emph {et~al.}(2025)\citenamefont {Jin}, \citenamefont {Serpico}, \citenamefont {Lee}, \citenamefont {Confalone}, \citenamefont {Saggau}, \citenamefont {Lo~Sardo}, \citenamefont {Gu}, \citenamefont {Goodge}, \citenamefont {Lesne}, \citenamefont {Montemurro} \emph {et~al.}}]{jin2025exploring}%
  \BibitemOpen
  \bibfield  {author} {\bibinfo {author} {\bibfnamefont {H.}~\bibnamefont {Jin}}, \bibinfo {author} {\bibfnamefont {G.}~\bibnamefont {Serpico}}, \bibinfo {author} {\bibfnamefont {Y.}~\bibnamefont {Lee}}, \bibinfo {author} {\bibfnamefont {T.}~\bibnamefont {Confalone}}, \bibinfo {author} {\bibfnamefont {C.~N.}\ \bibnamefont {Saggau}}, \bibinfo {author} {\bibfnamefont {F.}~\bibnamefont {Lo~Sardo}}, \bibinfo {author} {\bibfnamefont {G.}~\bibnamefont {Gu}}, \bibinfo {author} {\bibfnamefont {B.~H.}\ \bibnamefont {Goodge}}, \bibinfo {author} {\bibfnamefont {E.}~\bibnamefont {Lesne}}, \bibinfo {author} {\bibfnamefont {D.}~\bibnamefont {Montemurro}}, \emph {et~al.},\ }\bibfield  {title} {\bibinfo {title} {Exploring van der waals cuprate superconductors using a hybrid microwave circuit},\ }\href@noop {} {\bibfield  {journal} {\bibinfo  {journal} {Nano Letters}\ }\textbf {\bibinfo {volume} {25}},\ \bibinfo {pages} {3191} (\bibinfo {year} {2025})}\BibitemShut {NoStop}%
\bibitem [{\citenamefont {Thiemann}\ \emph {et~al.}(2018)\citenamefont {Thiemann}, \citenamefont {Beutel}, \citenamefont {Dressel}, \citenamefont {Lee-Hone}, \citenamefont {Broun}, \citenamefont {Fillis-Tsirakis}, \citenamefont {Boschker}, \citenamefont {Mannhart},\ and\ \citenamefont {Scheffler}}]{thiemann2018single}%
  \BibitemOpen
  \bibfield  {author} {\bibinfo {author} {\bibfnamefont {M.}~\bibnamefont {Thiemann}}, \bibinfo {author} {\bibfnamefont {M.~H.}\ \bibnamefont {Beutel}}, \bibinfo {author} {\bibfnamefont {M.}~\bibnamefont {Dressel}}, \bibinfo {author} {\bibfnamefont {N.~R.}\ \bibnamefont {Lee-Hone}}, \bibinfo {author} {\bibfnamefont {D.~M.}\ \bibnamefont {Broun}}, \bibinfo {author} {\bibfnamefont {E.}~\bibnamefont {Fillis-Tsirakis}}, \bibinfo {author} {\bibfnamefont {H.}~\bibnamefont {Boschker}}, \bibinfo {author} {\bibfnamefont {J.}~\bibnamefont {Mannhart}},\ and\ \bibinfo {author} {\bibfnamefont {M.}~\bibnamefont {Scheffler}},\ }\bibfield  {title} {\bibinfo {title} {Single-gap superconductivity and dome of superfluid density in nb-doped srtio$_3$},\ }\href@noop {} {\bibfield  {journal} {\bibinfo  {journal} {Physical Review Letters}\ }\textbf {\bibinfo {volume} {120}},\ \bibinfo {pages} {237002} (\bibinfo {year} {2018})}\BibitemShut {NoStop}%
\bibitem [{\citenamefont {Kim}\ \emph {et~al.}(2022)\citenamefont {Kim}, \citenamefont {Choi}, \citenamefont {Lewandowski}, \citenamefont {Thomson}, \citenamefont {Zhang}, \citenamefont {Polski}, \citenamefont {Watanabe}, \citenamefont {Taniguchi}, \citenamefont {Alicea},\ and\ \citenamefont {Nadj-Perge}}]{kim2022evidence}%
  \BibitemOpen
  \bibfield  {author} {\bibinfo {author} {\bibfnamefont {H.}~\bibnamefont {Kim}}, \bibinfo {author} {\bibfnamefont {Y.}~\bibnamefont {Choi}}, \bibinfo {author} {\bibfnamefont {C.}~\bibnamefont {Lewandowski}}, \bibinfo {author} {\bibfnamefont {A.}~\bibnamefont {Thomson}}, \bibinfo {author} {\bibfnamefont {Y.}~\bibnamefont {Zhang}}, \bibinfo {author} {\bibfnamefont {R.}~\bibnamefont {Polski}}, \bibinfo {author} {\bibfnamefont {K.}~\bibnamefont {Watanabe}}, \bibinfo {author} {\bibfnamefont {T.}~\bibnamefont {Taniguchi}}, \bibinfo {author} {\bibfnamefont {J.}~\bibnamefont {Alicea}},\ and\ \bibinfo {author} {\bibfnamefont {S.}~\bibnamefont {Nadj-Perge}},\ }\bibfield  {title} {\bibinfo {title} {Evidence for unconventional superconductivity in twisted trilayer graphene},\ }\href@noop {} {\bibfield  {journal} {\bibinfo  {journal} {Nature}\ }\textbf {\bibinfo {volume} {606}},\ \bibinfo {pages} {494} (\bibinfo {year} {2022})}\BibitemShut {NoStop}%
\bibitem [{\citenamefont {Schmidt}\ \emph {et~al.}(2018)\citenamefont {Schmidt}, \citenamefont {Jenkins}, \citenamefont {Watanabe}, \citenamefont {Taniguchi},\ and\ \citenamefont {Steele}}]{schmidt2018ballistic}%
  \BibitemOpen
  \bibfield  {author} {\bibinfo {author} {\bibfnamefont {F.~E.}\ \bibnamefont {Schmidt}}, \bibinfo {author} {\bibfnamefont {M.~D.}\ \bibnamefont {Jenkins}}, \bibinfo {author} {\bibfnamefont {K.}~\bibnamefont {Watanabe}}, \bibinfo {author} {\bibfnamefont {T.}~\bibnamefont {Taniguchi}},\ and\ \bibinfo {author} {\bibfnamefont {G.~A.}\ \bibnamefont {Steele}},\ }\bibfield  {title} {\bibinfo {title} {A ballistic graphene superconducting microwave circuit},\ }\href@noop {} {\bibfield  {journal} {\bibinfo  {journal} {Nature communications}\ }\textbf {\bibinfo {volume} {9}},\ \bibinfo {pages} {4069} (\bibinfo {year} {2018})}\BibitemShut {NoStop}%
\bibitem [{\citenamefont {Zaman}\ \emph {et~al.}(2025)\citenamefont {Zaman}, \citenamefont {Wang}, \citenamefont {Werkmeister}, \citenamefont {Tanaka}, \citenamefont {Dinh}, \citenamefont {Hays}, \citenamefont {Rodan-Legrain}, \citenamefont {Goswami}, \citenamefont {Assouly}, \citenamefont {Demir} \emph {et~al.}}]{zaman2025kinetic}%
  \BibitemOpen
  \bibfield  {author} {\bibinfo {author} {\bibfnamefont {S.}~\bibnamefont {Zaman}}, \bibinfo {author} {\bibfnamefont {J.~I.-J.}\ \bibnamefont {Wang}}, \bibinfo {author} {\bibfnamefont {T.}~\bibnamefont {Werkmeister}}, \bibinfo {author} {\bibfnamefont {M.}~\bibnamefont {Tanaka}}, \bibinfo {author} {\bibfnamefont {T.}~\bibnamefont {Dinh}}, \bibinfo {author} {\bibfnamefont {M.}~\bibnamefont {Hays}}, \bibinfo {author} {\bibfnamefont {D.}~\bibnamefont {Rodan-Legrain}}, \bibinfo {author} {\bibfnamefont {A.}~\bibnamefont {Goswami}}, \bibinfo {author} {\bibfnamefont {R.}~\bibnamefont {Assouly}}, \bibinfo {author} {\bibfnamefont {A.~K.}\ \bibnamefont {Demir}}, \emph {et~al.},\ }\bibfield  {title} {\bibinfo {title} {Kinetic inductance of few-layer nbse$_2$ in the two-dimensional limit},\ }\href@noop {} {\bibfield  {journal} {\bibinfo  {journal} {arXiv preprint arXiv:2511.08466}\ } (\bibinfo {year} {2025})}\BibitemShut {NoStop}%
\bibitem [{\citenamefont {Kreidel}\ \emph {et~al.}(2024)\citenamefont {Kreidel}, \citenamefont {Chu}, \citenamefont {Balgley}, \citenamefont {Antony}, \citenamefont {Verma}, \citenamefont {Ingham}, \citenamefont {Ranzani}, \citenamefont {Queiroz}, \citenamefont {Westervelt}, \citenamefont {Hone} \emph {et~al.}}]{kreidel2024measuring}%
  \BibitemOpen
  \bibfield  {author} {\bibinfo {author} {\bibfnamefont {M.}~\bibnamefont {Kreidel}}, \bibinfo {author} {\bibfnamefont {X.}~\bibnamefont {Chu}}, \bibinfo {author} {\bibfnamefont {J.}~\bibnamefont {Balgley}}, \bibinfo {author} {\bibfnamefont {A.}~\bibnamefont {Antony}}, \bibinfo {author} {\bibfnamefont {N.}~\bibnamefont {Verma}}, \bibinfo {author} {\bibfnamefont {J.}~\bibnamefont {Ingham}}, \bibinfo {author} {\bibfnamefont {L.}~\bibnamefont {Ranzani}}, \bibinfo {author} {\bibfnamefont {R.}~\bibnamefont {Queiroz}}, \bibinfo {author} {\bibfnamefont {R.~M.}\ \bibnamefont {Westervelt}}, \bibinfo {author} {\bibfnamefont {J.}~\bibnamefont {Hone}}, \emph {et~al.},\ }\bibfield  {title} {\bibinfo {title} {Measuring kinetic inductance and superfluid stiffness of two-dimensional superconductors using high-quality transmission-line resonators},\ }\href@noop {} {\bibfield  {journal} {\bibinfo  {journal} {Physical Review Research}\ }\textbf {\bibinfo {volume} {6}},\ \bibinfo {pages} {043245} (\bibinfo {year}
  {2024})}\BibitemShut {NoStop}%
\bibitem [{\citenamefont {Lutchyn}\ \emph {et~al.}(2009)\citenamefont {Lutchyn}, \citenamefont {Nagornykh},\ and\ \citenamefont {Yakovenko}}]{lutchyn2009frequency}%
  \BibitemOpen
  \bibfield  {author} {\bibinfo {author} {\bibfnamefont {R.~M.}\ \bibnamefont {Lutchyn}}, \bibinfo {author} {\bibfnamefont {P.}~\bibnamefont {Nagornykh}},\ and\ \bibinfo {author} {\bibfnamefont {V.~M.}\ \bibnamefont {Yakovenko}},\ }\bibfield  {title} {\bibinfo {title} {Frequency and temperature dependence of the anomalous ac hall conductivity in a chiral px+ ipy superconductor with impurities},\ }\href@noop {} {\bibfield  {journal} {\bibinfo  {journal} {Physical Review B---Condensed Matter and Materials Physics}\ }\textbf {\bibinfo {volume} {80}},\ \bibinfo {pages} {104508} (\bibinfo {year} {2009})}\BibitemShut {NoStop}%
\bibitem [{\citenamefont {Han}\ \emph {et~al.}(2025)\citenamefont {Han}, \citenamefont {Lu}, \citenamefont {Hadjri}, \citenamefont {Shi}, \citenamefont {Wu}, \citenamefont {Xu}, \citenamefont {Yao}, \citenamefont {Cotten}, \citenamefont {Sharifi~Sedeh}, \citenamefont {Weldeyesus} \emph {et~al.}}]{han2025signatures}%
  \BibitemOpen
  \bibfield  {author} {\bibinfo {author} {\bibfnamefont {T.}~\bibnamefont {Han}}, \bibinfo {author} {\bibfnamefont {Z.}~\bibnamefont {Lu}}, \bibinfo {author} {\bibfnamefont {Z.}~\bibnamefont {Hadjri}}, \bibinfo {author} {\bibfnamefont {L.}~\bibnamefont {Shi}}, \bibinfo {author} {\bibfnamefont {Z.}~\bibnamefont {Wu}}, \bibinfo {author} {\bibfnamefont {W.}~\bibnamefont {Xu}}, \bibinfo {author} {\bibfnamefont {Y.}~\bibnamefont {Yao}}, \bibinfo {author} {\bibfnamefont {A.~A.}\ \bibnamefont {Cotten}}, \bibinfo {author} {\bibfnamefont {O.}~\bibnamefont {Sharifi~Sedeh}}, \bibinfo {author} {\bibfnamefont {H.}~\bibnamefont {Weldeyesus}}, \emph {et~al.},\ }\bibfield  {title} {\bibinfo {title} {Signatures of chiral superconductivity in rhombohedral graphene},\ }\href@noop {} {\bibfield  {journal} {\bibinfo  {journal} {Nature}\ }\textbf {\bibinfo {volume} {643}},\ \bibinfo {pages} {654} (\bibinfo {year} {2025})}\BibitemShut {NoStop}%
\bibitem [{\citenamefont {Petrides}\ \emph {et~al.}(2025{\natexlab{a}})\citenamefont {Petrides}, \citenamefont {Curtis}, \citenamefont {Wesson}, \citenamefont {Yacoby},\ and\ \citenamefont {Narang}}]{petrides2025probing}%
  \BibitemOpen
  \bibfield  {author} {\bibinfo {author} {\bibfnamefont {I.}~\bibnamefont {Petrides}}, \bibinfo {author} {\bibfnamefont {J.~B.}\ \bibnamefont {Curtis}}, \bibinfo {author} {\bibfnamefont {M.}~\bibnamefont {Wesson}}, \bibinfo {author} {\bibfnamefont {A.}~\bibnamefont {Yacoby}},\ and\ \bibinfo {author} {\bibfnamefont {P.}~\bibnamefont {Narang}},\ }\bibfield  {title} {\bibinfo {title} {Probing electromagnetic nonreciprocity with quantum geometry of photonic states},\ }\href@noop {} {\bibfield  {journal} {\bibinfo  {journal} {Physical Review Research}\ }\textbf {\bibinfo {volume} {7}},\ \bibinfo {pages} {023006} (\bibinfo {year} {2025}{\natexlab{a}})}\BibitemShut {NoStop}%
\bibitem [{\citenamefont {Petrides}\ \emph {et~al.}(2025{\natexlab{b}})\citenamefont {Petrides}, \citenamefont {Arora},\ and\ \citenamefont {Narang}}]{petrides2025passive}%
  \BibitemOpen
  \bibfield  {author} {\bibinfo {author} {\bibfnamefont {I.}~\bibnamefont {Petrides}}, \bibinfo {author} {\bibfnamefont {A.}~\bibnamefont {Arora}},\ and\ \bibinfo {author} {\bibfnamefont {P.}~\bibnamefont {Narang}},\ }\bibfield  {title} {\bibinfo {title} {Passive detection of schwinger boson dynamics via a qubit},\ }\href@noop {} {\bibfield  {journal} {\bibinfo  {journal} {arXiv preprint arXiv:2510.00108}\ } (\bibinfo {year} {2025}{\natexlab{b}})}\BibitemShut {NoStop}%
\bibitem [{\citenamefont {Dirnegger}\ \emph {et~al.}(2025{\natexlab{b}})\citenamefont {Dirnegger}, \citenamefont {Wesson}, \citenamefont {Arora}, \citenamefont {Petrides}, \citenamefont {Curtis}, \citenamefont {Been}, \citenamefont {Yacoby},\ and\ \citenamefont {Narang}}]{dirnegger2025nonlinear}%
  \BibitemOpen
  \bibfield  {author} {\bibinfo {author} {\bibfnamefont {N.}~\bibnamefont {Dirnegger}}, \bibinfo {author} {\bibfnamefont {M.}~\bibnamefont {Wesson}}, \bibinfo {author} {\bibfnamefont {A.}~\bibnamefont {Arora}}, \bibinfo {author} {\bibfnamefont {I.}~\bibnamefont {Petrides}}, \bibinfo {author} {\bibfnamefont {J.~B.}\ \bibnamefont {Curtis}}, \bibinfo {author} {\bibfnamefont {E.~M.}\ \bibnamefont {Been}}, \bibinfo {author} {\bibfnamefont {A.}~\bibnamefont {Yacoby}},\ and\ \bibinfo {author} {\bibfnamefont {P.}~\bibnamefont {Narang}},\ }\bibfield  {title} {\bibinfo {title} {Nonlinear superconducting ring resonator for sensitive measurement of time reversal symmetry broken order},\ }\href@noop {} {\bibfield  {journal} {\bibinfo  {journal} {arXiv preprint arXiv:2505.21614}\ } (\bibinfo {year} {2025}{\natexlab{b}})}\BibitemShut {NoStop}%
\bibitem [{\citenamefont {Ren}\ \emph {et~al.}(2024)\citenamefont {Ren}, \citenamefont {Copenhaver}, \citenamefont {Rokhinson},\ and\ \citenamefont {V{\"a}yrynen}}]{ren2024microwave}%
  \BibitemOpen
  \bibfield  {author} {\bibinfo {author} {\bibfnamefont {Z.}~\bibnamefont {Ren}}, \bibinfo {author} {\bibfnamefont {J.}~\bibnamefont {Copenhaver}}, \bibinfo {author} {\bibfnamefont {L.}~\bibnamefont {Rokhinson}},\ and\ \bibinfo {author} {\bibfnamefont {J.~I.}\ \bibnamefont {V{\"a}yrynen}},\ }\bibfield  {title} {\bibinfo {title} {Microwave spectroscopy of majorana vortex modes},\ }\href@noop {} {\bibfield  {journal} {\bibinfo  {journal} {Physical Review B}\ }\textbf {\bibinfo {volume} {109}},\ \bibinfo {pages} {L180506} (\bibinfo {year} {2024})}\BibitemShut {NoStop}%
\bibitem [{\citenamefont {Dmytruk}\ and\ \citenamefont {Schir{\`o}}(2024)}]{dmytruk2024hybrid}%
  \BibitemOpen
  \bibfield  {author} {\bibinfo {author} {\bibfnamefont {O.}~\bibnamefont {Dmytruk}}\ and\ \bibinfo {author} {\bibfnamefont {M.}~\bibnamefont {Schir{\`o}}},\ }\bibfield  {title} {\bibinfo {title} {Hybrid light-matter states in topological superconductors coupled to cavity photons},\ }\href@noop {} {\bibfield  {journal} {\bibinfo  {journal} {Physical Review B}\ }\textbf {\bibinfo {volume} {110}},\ \bibinfo {pages} {075416} (\bibinfo {year} {2024})}\BibitemShut {NoStop}%
\bibitem [{\citenamefont {Shulga}\ \emph {et~al.}(2025)\citenamefont {Shulga}, \citenamefont {Nishimura}, \citenamefont {Volkov}, \citenamefont {Hasegawa}, \citenamefont {Hirano}, \citenamefont {Tsuji}, \citenamefont {Iwasaki}, \citenamefont {Hatano}, \citenamefont {Sasaki},\ and\ \citenamefont {Kobayashi}}]{shulga2025observation}%
  \BibitemOpen
  \bibfield  {author} {\bibinfo {author} {\bibfnamefont {K.}~\bibnamefont {Shulga}}, \bibinfo {author} {\bibfnamefont {S.}~\bibnamefont {Nishimura}}, \bibinfo {author} {\bibfnamefont {P.~A.}\ \bibnamefont {Volkov}}, \bibinfo {author} {\bibfnamefont {R.}~\bibnamefont {Hasegawa}}, \bibinfo {author} {\bibfnamefont {M.}~\bibnamefont {Hirano}}, \bibinfo {author} {\bibfnamefont {T.}~\bibnamefont {Tsuji}}, \bibinfo {author} {\bibfnamefont {T.}~\bibnamefont {Iwasaki}}, \bibinfo {author} {\bibfnamefont {M.}~\bibnamefont {Hatano}}, \bibinfo {author} {\bibfnamefont {K.}~\bibnamefont {Sasaki}},\ and\ \bibinfo {author} {\bibfnamefont {K.}~\bibnamefont {Kobayashi}},\ }\bibfield  {title} {\bibinfo {title} {Observation of individual vortex penetration in a coplanar superconducting resonator},\ }\href@noop {} {\bibfield  {journal} {\bibinfo  {journal} {arXiv preprint arXiv:2512.00790}\ } (\bibinfo {year} {2025})}\BibitemShut {NoStop}%
\bibitem [{\citenamefont {Wang}\ \emph {et~al.}(2023)\citenamefont {Wang}, \citenamefont {Balembois}, \citenamefont {Ran{\v{c}}i{\'c}}, \citenamefont {Billaud}, \citenamefont {Le~Dantec}, \citenamefont {Ferrier}, \citenamefont {Goldner}, \citenamefont {Bertaina}, \citenamefont {Chaneli{\`e}re}, \citenamefont {Est{\`e}ve} \emph {et~al.}}]{wang2023single}%
  \BibitemOpen
  \bibfield  {author} {\bibinfo {author} {\bibfnamefont {Z.}~\bibnamefont {Wang}}, \bibinfo {author} {\bibfnamefont {L.}~\bibnamefont {Balembois}}, \bibinfo {author} {\bibfnamefont {M.}~\bibnamefont {Ran{\v{c}}i{\'c}}}, \bibinfo {author} {\bibfnamefont {E.}~\bibnamefont {Billaud}}, \bibinfo {author} {\bibfnamefont {M.}~\bibnamefont {Le~Dantec}}, \bibinfo {author} {\bibfnamefont {A.}~\bibnamefont {Ferrier}}, \bibinfo {author} {\bibfnamefont {P.}~\bibnamefont {Goldner}}, \bibinfo {author} {\bibfnamefont {S.}~\bibnamefont {Bertaina}}, \bibinfo {author} {\bibfnamefont {T.}~\bibnamefont {Chaneli{\`e}re}}, \bibinfo {author} {\bibfnamefont {D.}~\bibnamefont {Est{\`e}ve}}, \emph {et~al.},\ }\bibfield  {title} {\bibinfo {title} {Single-electron spin resonance detection by microwave photon counting},\ }\href@noop {} {\bibfield  {journal} {\bibinfo  {journal} {Nature}\ }\textbf {\bibinfo {volume} {619}},\ \bibinfo {pages} {276} (\bibinfo {year} {2023})}\BibitemShut {NoStop}%
\bibitem [{\citenamefont {Budakian}\ \emph {et~al.}(2024)\citenamefont {Budakian}, \citenamefont {Finkler}, \citenamefont {Eichler}, \citenamefont {Poggio}, \citenamefont {Degen}, \citenamefont {Tabatabaei}, \citenamefont {Lee}, \citenamefont {Hammel}, \citenamefont {Eugene}, \citenamefont {Taminiau} \emph {et~al.}}]{budakian2024roadmap}%
  \BibitemOpen
  \bibfield  {author} {\bibinfo {author} {\bibfnamefont {R.}~\bibnamefont {Budakian}}, \bibinfo {author} {\bibfnamefont {A.}~\bibnamefont {Finkler}}, \bibinfo {author} {\bibfnamefont {A.}~\bibnamefont {Eichler}}, \bibinfo {author} {\bibfnamefont {M.}~\bibnamefont {Poggio}}, \bibinfo {author} {\bibfnamefont {C.~L.}\ \bibnamefont {Degen}}, \bibinfo {author} {\bibfnamefont {S.}~\bibnamefont {Tabatabaei}}, \bibinfo {author} {\bibfnamefont {I.}~\bibnamefont {Lee}}, \bibinfo {author} {\bibfnamefont {P.~C.}\ \bibnamefont {Hammel}}, \bibinfo {author} {\bibfnamefont {S.~P.}\ \bibnamefont {Eugene}}, \bibinfo {author} {\bibfnamefont {T.~H.}\ \bibnamefont {Taminiau}}, \emph {et~al.},\ }\bibfield  {title} {\bibinfo {title} {Roadmap on nanoscale magnetic resonance imaging},\ }\href@noop {} {\bibfield  {journal} {\bibinfo  {journal} {Nanotechnology}\ }\textbf {\bibinfo {volume} {35}},\ \bibinfo {pages} {412001} (\bibinfo {year} {2024})}\BibitemShut {NoStop}%
\bibitem [{\citenamefont {Travesedo}\ \emph {et~al.}(2025)\citenamefont {Travesedo}, \citenamefont {O'Sullivan}, \citenamefont {Pallegoix}, \citenamefont {Huang}, \citenamefont {Hogan}, \citenamefont {Goldner}, \citenamefont {Chaneliere}, \citenamefont {Bertaina}, \citenamefont {Est{\`e}ve}, \citenamefont {Abgrall} \emph {et~al.}}]{travesedo2025all}%
  \BibitemOpen
  \bibfield  {author} {\bibinfo {author} {\bibfnamefont {J.}~\bibnamefont {Travesedo}}, \bibinfo {author} {\bibfnamefont {J.}~\bibnamefont {O'Sullivan}}, \bibinfo {author} {\bibfnamefont {L.}~\bibnamefont {Pallegoix}}, \bibinfo {author} {\bibfnamefont {Z.~W.}\ \bibnamefont {Huang}}, \bibinfo {author} {\bibfnamefont {P.}~\bibnamefont {Hogan}}, \bibinfo {author} {\bibfnamefont {P.}~\bibnamefont {Goldner}}, \bibinfo {author} {\bibfnamefont {T.}~\bibnamefont {Chaneliere}}, \bibinfo {author} {\bibfnamefont {S.}~\bibnamefont {Bertaina}}, \bibinfo {author} {\bibfnamefont {D.}~\bibnamefont {Est{\`e}ve}}, \bibinfo {author} {\bibfnamefont {P.}~\bibnamefont {Abgrall}}, \emph {et~al.},\ }\bibfield  {title} {\bibinfo {title} {All-microwave spectroscopy and polarization of individual nuclear spins in a solid},\ }\href@noop {} {\bibfield  {journal} {\bibinfo  {journal} {Science Advances}\ }\textbf {\bibinfo {volume} {11}},\ \bibinfo {pages} {eadu0581} (\bibinfo {year} {2025})}\BibitemShut {NoStop}%
\bibitem [{\citenamefont {Gorgi~Zadeh}\ \emph {et~al.}(2025)\citenamefont {Gorgi~Zadeh}, \citenamefont {Ghirri}, \citenamefont {Pagano}, \citenamefont {Tocci}, \citenamefont {Gatti},\ and\ \citenamefont {Cassinese}}]{gorgi2025high}%
  \BibitemOpen
  \bibfield  {author} {\bibinfo {author} {\bibfnamefont {S.}~\bibnamefont {Gorgi~Zadeh}}, \bibinfo {author} {\bibfnamefont {A.}~\bibnamefont {Ghirri}}, \bibinfo {author} {\bibfnamefont {S.}~\bibnamefont {Pagano}}, \bibinfo {author} {\bibfnamefont {S.}~\bibnamefont {Tocci}}, \bibinfo {author} {\bibfnamefont {C.}~\bibnamefont {Gatti}},\ and\ \bibinfo {author} {\bibfnamefont {A.}~\bibnamefont {Cassinese}},\ }\bibfield  {title} {\bibinfo {title} {High-q cavity coupled to a high permittivity dielectric resonator for sensing applications},\ }\href@noop {} {\bibfield  {journal} {\bibinfo  {journal} {Annalen der Physik}\ }\textbf {\bibinfo {volume} {537}},\ \bibinfo {pages} {e00241} (\bibinfo {year} {2025})}\BibitemShut {NoStop}%
\bibitem [{\citenamefont {Wang}\ \emph {et~al.}(2022)\citenamefont {Wang}, \citenamefont {Yamoah}, \citenamefont {Li}, \citenamefont {Karamlou}, \citenamefont {Dinh}, \citenamefont {Kannan}, \citenamefont {Braum{\"u}ller}, \citenamefont {Kim}, \citenamefont {Melville}, \citenamefont {Muschinske} \emph {et~al.}}]{wang2022hexagonal}%
  \BibitemOpen
  \bibfield  {author} {\bibinfo {author} {\bibfnamefont {J.~I.}\ \bibnamefont {Wang}}, \bibinfo {author} {\bibfnamefont {M.~A.}\ \bibnamefont {Yamoah}}, \bibinfo {author} {\bibfnamefont {Q.}~\bibnamefont {Li}}, \bibinfo {author} {\bibfnamefont {A.~H.}\ \bibnamefont {Karamlou}}, \bibinfo {author} {\bibfnamefont {T.}~\bibnamefont {Dinh}}, \bibinfo {author} {\bibfnamefont {B.}~\bibnamefont {Kannan}}, \bibinfo {author} {\bibfnamefont {J.}~\bibnamefont {Braum{\"u}ller}}, \bibinfo {author} {\bibfnamefont {D.}~\bibnamefont {Kim}}, \bibinfo {author} {\bibfnamefont {A.~J.}\ \bibnamefont {Melville}}, \bibinfo {author} {\bibfnamefont {S.~E.}\ \bibnamefont {Muschinske}}, \emph {et~al.},\ }\bibfield  {title} {\bibinfo {title} {Hexagonal boron nitride as a low-loss dielectric for superconducting quantum circuits and qubits},\ }\href@noop {} {\bibfield  {journal} {\bibinfo  {journal} {Nature materials}\ }\textbf {\bibinfo {volume} {21}},\ \bibinfo {pages} {398} (\bibinfo {year} {2022})}\BibitemShut {NoStop}%
\bibitem [{\citenamefont {Antony}\ \emph {et~al.}(2021)\citenamefont {Antony}, \citenamefont {Gustafsson}, \citenamefont {Ribeill}, \citenamefont {Ware}, \citenamefont {Rajendran}, \citenamefont {Govia}, \citenamefont {Ohki}, \citenamefont {Taniguchi}, \citenamefont {Watanabe}, \citenamefont {Hone} \emph {et~al.}}]{antony2021miniaturizing}%
  \BibitemOpen
  \bibfield  {author} {\bibinfo {author} {\bibfnamefont {A.}~\bibnamefont {Antony}}, \bibinfo {author} {\bibfnamefont {M.~V.}\ \bibnamefont {Gustafsson}}, \bibinfo {author} {\bibfnamefont {G.~J.}\ \bibnamefont {Ribeill}}, \bibinfo {author} {\bibfnamefont {M.}~\bibnamefont {Ware}}, \bibinfo {author} {\bibfnamefont {A.}~\bibnamefont {Rajendran}}, \bibinfo {author} {\bibfnamefont {L.~C.}\ \bibnamefont {Govia}}, \bibinfo {author} {\bibfnamefont {T.~A.}\ \bibnamefont {Ohki}}, \bibinfo {author} {\bibfnamefont {T.}~\bibnamefont {Taniguchi}}, \bibinfo {author} {\bibfnamefont {K.}~\bibnamefont {Watanabe}}, \bibinfo {author} {\bibfnamefont {J.}~\bibnamefont {Hone}}, \emph {et~al.},\ }\bibfield  {title} {\bibinfo {title} {Miniaturizing transmon qubits using van der waals materials},\ }\href@noop {} {\bibfield  {journal} {\bibinfo  {journal} {Nano letters}\ }\textbf {\bibinfo {volume} {21}},\ \bibinfo {pages} {10122} (\bibinfo {year} {2021})}\BibitemShut {NoStop}%
\bibitem [{\citenamefont {Maji}\ \emph {et~al.}(2024)\citenamefont {Maji}, \citenamefont {Sarkar}, \citenamefont {Mandal}, \citenamefont {Hingankar}, \citenamefont {Mukherjee}, \citenamefont {Samal}, \citenamefont {Bhattacharjee}, \citenamefont {Patankar}, \citenamefont {Watanabe}, \citenamefont {Taniguchi} \emph {et~al.}}]{maji2024superconducting}%
  \BibitemOpen
  \bibfield  {author} {\bibinfo {author} {\bibfnamefont {K.}~\bibnamefont {Maji}}, \bibinfo {author} {\bibfnamefont {J.}~\bibnamefont {Sarkar}}, \bibinfo {author} {\bibfnamefont {S.}~\bibnamefont {Mandal}}, \bibinfo {author} {\bibfnamefont {M.}~\bibnamefont {Hingankar}}, \bibinfo {author} {\bibfnamefont {A.}~\bibnamefont {Mukherjee}}, \bibinfo {author} {\bibfnamefont {S.}~\bibnamefont {Samal}}, \bibinfo {author} {\bibfnamefont {A.}~\bibnamefont {Bhattacharjee}}, \bibinfo {author} {\bibfnamefont {M.~P.}\ \bibnamefont {Patankar}}, \bibinfo {author} {\bibfnamefont {K.}~\bibnamefont {Watanabe}}, \bibinfo {author} {\bibfnamefont {T.}~\bibnamefont {Taniguchi}}, \emph {et~al.},\ }\bibfield  {title} {\bibinfo {title} {Superconducting cavity-based sensing of band gaps in 2d materials},\ }\href@noop {} {\bibfield  {journal} {\bibinfo  {journal} {Nano Letters}\ }\textbf {\bibinfo {volume} {24}},\ \bibinfo {pages} {4369} (\bibinfo {year} {2024})}\BibitemShut {NoStop}%
\bibitem [{\citenamefont {Cai}\ \emph {et~al.}(2023)\citenamefont {Cai}, \citenamefont {Anderson}, \citenamefont {Wang}, \citenamefont {Zhang}, \citenamefont {Liu}, \citenamefont {Holtzmann}, \citenamefont {Zhang}, \citenamefont {Fan}, \citenamefont {Taniguchi}, \citenamefont {Watanabe} \emph {et~al.}}]{cai2023signatures}%
  \BibitemOpen
  \bibfield  {author} {\bibinfo {author} {\bibfnamefont {J.}~\bibnamefont {Cai}}, \bibinfo {author} {\bibfnamefont {E.}~\bibnamefont {Anderson}}, \bibinfo {author} {\bibfnamefont {C.}~\bibnamefont {Wang}}, \bibinfo {author} {\bibfnamefont {X.}~\bibnamefont {Zhang}}, \bibinfo {author} {\bibfnamefont {X.}~\bibnamefont {Liu}}, \bibinfo {author} {\bibfnamefont {W.}~\bibnamefont {Holtzmann}}, \bibinfo {author} {\bibfnamefont {Y.}~\bibnamefont {Zhang}}, \bibinfo {author} {\bibfnamefont {F.}~\bibnamefont {Fan}}, \bibinfo {author} {\bibfnamefont {T.}~\bibnamefont {Taniguchi}}, \bibinfo {author} {\bibfnamefont {K.}~\bibnamefont {Watanabe}}, \emph {et~al.},\ }\bibfield  {title} {\bibinfo {title} {Signatures of fractional quantum anomalous hall states in twisted mote2},\ }\href@noop {} {\bibfield  {journal} {\bibinfo  {journal} {Nature}\ }\textbf {\bibinfo {volume} {622}},\ \bibinfo {pages} {63} (\bibinfo {year} {2023})}\BibitemShut {NoStop}%
\bibitem [{\citenamefont {Lu}\ \emph {et~al.}(2024)\citenamefont {Lu}, \citenamefont {Han}, \citenamefont {Yao}, \citenamefont {Reddy}, \citenamefont {Yang}, \citenamefont {Seo}, \citenamefont {Watanabe}, \citenamefont {Taniguchi}, \citenamefont {Fu},\ and\ \citenamefont {Ju}}]{lu2024fractional}%
  \BibitemOpen
  \bibfield  {author} {\bibinfo {author} {\bibfnamefont {Z.}~\bibnamefont {Lu}}, \bibinfo {author} {\bibfnamefont {T.}~\bibnamefont {Han}}, \bibinfo {author} {\bibfnamefont {Y.}~\bibnamefont {Yao}}, \bibinfo {author} {\bibfnamefont {A.~P.}\ \bibnamefont {Reddy}}, \bibinfo {author} {\bibfnamefont {J.}~\bibnamefont {Yang}}, \bibinfo {author} {\bibfnamefont {J.}~\bibnamefont {Seo}}, \bibinfo {author} {\bibfnamefont {K.}~\bibnamefont {Watanabe}}, \bibinfo {author} {\bibfnamefont {T.}~\bibnamefont {Taniguchi}}, \bibinfo {author} {\bibfnamefont {L.}~\bibnamefont {Fu}},\ and\ \bibinfo {author} {\bibfnamefont {L.}~\bibnamefont {Ju}},\ }\bibfield  {title} {\bibinfo {title} {Fractional quantum anomalous hall effect in multilayer graphene},\ }\href@noop {} {\bibfield  {journal} {\bibinfo  {journal} {Nature}\ }\textbf {\bibinfo {volume} {626}},\ \bibinfo {pages} {759} (\bibinfo {year} {2024})}\BibitemShut {NoStop}%
\bibitem [{\citenamefont {Lee}\ and\ \citenamefont {Rice}(1979)}]{lee1979electric}%
  \BibitemOpen
  \bibfield  {author} {\bibinfo {author} {\bibfnamefont {P.}~\bibnamefont {Lee}}\ and\ \bibinfo {author} {\bibfnamefont {T.}~\bibnamefont {Rice}},\ }\bibfield  {title} {\bibinfo {title} {Electric field depinning of charge density waves},\ }\href@noop {} {\bibfield  {journal} {\bibinfo  {journal} {Physical Review B}\ }\textbf {\bibinfo {volume} {19}},\ \bibinfo {pages} {3970} (\bibinfo {year} {1979})}\BibitemShut {NoStop}%
\bibitem [{\citenamefont {Littlewood}(1987)}]{littlewood1987screened}%
  \BibitemOpen
  \bibfield  {author} {\bibinfo {author} {\bibfnamefont {P.}~\bibnamefont {Littlewood}},\ }\bibfield  {title} {\bibinfo {title} {Screened dielectric response of sliding charge-density waves},\ }\href@noop {} {\bibfield  {journal} {\bibinfo  {journal} {Physical Review B}\ }\textbf {\bibinfo {volume} {36}},\ \bibinfo {pages} {3108} (\bibinfo {year} {1987})}\BibitemShut {NoStop}%
\bibitem [{\citenamefont {Mazza}\ \emph {et~al.}(2024)\citenamefont {Mazza}, \citenamefont {Biswas}, \citenamefont {Yan}, \citenamefont {Prokofiev}, \citenamefont {Steffens}, \citenamefont {Si}, \citenamefont {Assaad},\ and\ \citenamefont {Paschen}}]{mazza2024quantum}%
  \BibitemOpen
  \bibfield  {author} {\bibinfo {author} {\bibfnamefont {F.}~\bibnamefont {Mazza}}, \bibinfo {author} {\bibfnamefont {S.}~\bibnamefont {Biswas}}, \bibinfo {author} {\bibfnamefont {X.}~\bibnamefont {Yan}}, \bibinfo {author} {\bibfnamefont {A.}~\bibnamefont {Prokofiev}}, \bibinfo {author} {\bibfnamefont {P.}~\bibnamefont {Steffens}}, \bibinfo {author} {\bibfnamefont {Q.}~\bibnamefont {Si}}, \bibinfo {author} {\bibfnamefont {F.~F.}\ \bibnamefont {Assaad}},\ and\ \bibinfo {author} {\bibfnamefont {S.}~\bibnamefont {Paschen}},\ }\bibfield  {title} {\bibinfo {title} {Quantum fisher information in a strange metal},\ }\href@noop {} {\bibfield  {journal} {\bibinfo  {journal} {arXiv preprint arXiv:2403.12779}\ } (\bibinfo {year} {2024})}\BibitemShut {NoStop}%
\bibitem [{\citenamefont {Fang}\ \emph {et~al.}(2025)\citenamefont {Fang}, \citenamefont {Mahankali}, \citenamefont {Wang}, \citenamefont {Chen}, \citenamefont {Hu}, \citenamefont {Paschen},\ and\ \citenamefont {Si}}]{fang2025amplified}%
  \BibitemOpen
  \bibfield  {author} {\bibinfo {author} {\bibfnamefont {Y.}~\bibnamefont {Fang}}, \bibinfo {author} {\bibfnamefont {M.}~\bibnamefont {Mahankali}}, \bibinfo {author} {\bibfnamefont {Y.}~\bibnamefont {Wang}}, \bibinfo {author} {\bibfnamefont {L.}~\bibnamefont {Chen}}, \bibinfo {author} {\bibfnamefont {H.}~\bibnamefont {Hu}}, \bibinfo {author} {\bibfnamefont {S.}~\bibnamefont {Paschen}},\ and\ \bibinfo {author} {\bibfnamefont {Q.}~\bibnamefont {Si}},\ }\bibfield  {title} {\bibinfo {title} {Amplified multipartite entanglement witnessed in a quantum critical metal},\ }\href@noop {} {\bibfield  {journal} {\bibinfo  {journal} {Nature Communications}\ }\textbf {\bibinfo {volume} {16}},\ \bibinfo {pages} {2498} (\bibinfo {year} {2025})}\BibitemShut {NoStop}%
\bibitem [{\citenamefont {T{\"o}rm{\"a}}\ \emph {et~al.}(2022)\citenamefont {T{\"o}rm{\"a}}, \citenamefont {Peotta},\ and\ \citenamefont {Bernevig}}]{torma2022superconductivity}%
  \BibitemOpen
  \bibfield  {author} {\bibinfo {author} {\bibfnamefont {P.}~\bibnamefont {T{\"o}rm{\"a}}}, \bibinfo {author} {\bibfnamefont {S.}~\bibnamefont {Peotta}},\ and\ \bibinfo {author} {\bibfnamefont {B.~A.}\ \bibnamefont {Bernevig}},\ }\bibfield  {title} {\bibinfo {title} {Superconductivity, superfluidity and quantum geometry in twisted multilayer systems},\ }\href@noop {} {\bibfield  {journal} {\bibinfo  {journal} {Nature Reviews Physics}\ }\textbf {\bibinfo {volume} {4}},\ \bibinfo {pages} {528} (\bibinfo {year} {2022})}\BibitemShut {NoStop}%
\bibitem [{\citenamefont {Regnault}\ \emph {et~al.}(2022)\citenamefont {Regnault}, \citenamefont {Xu}, \citenamefont {Li}, \citenamefont {Ma}, \citenamefont {Jovanovic}, \citenamefont {Yazdani}, \citenamefont {Parkin}, \citenamefont {Felser}, \citenamefont {Schoop}, \citenamefont {Ong} \emph {et~al.}}]{regnault2022catalogue}%
  \BibitemOpen
  \bibfield  {author} {\bibinfo {author} {\bibfnamefont {N.}~\bibnamefont {Regnault}}, \bibinfo {author} {\bibfnamefont {Y.}~\bibnamefont {Xu}}, \bibinfo {author} {\bibfnamefont {M.-R.}\ \bibnamefont {Li}}, \bibinfo {author} {\bibfnamefont {D.-S.}\ \bibnamefont {Ma}}, \bibinfo {author} {\bibfnamefont {M.}~\bibnamefont {Jovanovic}}, \bibinfo {author} {\bibfnamefont {A.}~\bibnamefont {Yazdani}}, \bibinfo {author} {\bibfnamefont {S.~S.}\ \bibnamefont {Parkin}}, \bibinfo {author} {\bibfnamefont {C.}~\bibnamefont {Felser}}, \bibinfo {author} {\bibfnamefont {L.~M.}\ \bibnamefont {Schoop}}, \bibinfo {author} {\bibfnamefont {N.~P.}\ \bibnamefont {Ong}}, \emph {et~al.},\ }\bibfield  {title} {\bibinfo {title} {Catalogue of flat-band stoichiometric materials},\ }\href@noop {} {\bibfield  {journal} {\bibinfo  {journal} {Nature}\ }\textbf {\bibinfo {volume} {603}},\ \bibinfo {pages} {824} (\bibinfo {year} {2022})}\BibitemShut {NoStop}%
\bibitem [{\citenamefont {Checkelsky}\ \emph {et~al.}(2024)\citenamefont {Checkelsky}, \citenamefont {Bernevig}, \citenamefont {Coleman}, \citenamefont {Si},\ and\ \citenamefont {Paschen}}]{checkelsky2024flat}%
  \BibitemOpen
  \bibfield  {author} {\bibinfo {author} {\bibfnamefont {J.~G.}\ \bibnamefont {Checkelsky}}, \bibinfo {author} {\bibfnamefont {B.~A.}\ \bibnamefont {Bernevig}}, \bibinfo {author} {\bibfnamefont {P.}~\bibnamefont {Coleman}}, \bibinfo {author} {\bibfnamefont {Q.}~\bibnamefont {Si}},\ and\ \bibinfo {author} {\bibfnamefont {S.}~\bibnamefont {Paschen}},\ }\bibfield  {title} {\bibinfo {title} {Flat bands, strange metals and the kondo effect},\ }\href@noop {} {\bibfield  {journal} {\bibinfo  {journal} {Nature Reviews Materials}\ }\textbf {\bibinfo {volume} {9}},\ \bibinfo {pages} {509} (\bibinfo {year} {2024})}\BibitemShut {NoStop}%
\bibitem [{\citenamefont {Peotta}\ and\ \citenamefont {T{\"o}rm{\"a}}(2015)}]{peotta2015superfluidity}%
  \BibitemOpen
  \bibfield  {author} {\bibinfo {author} {\bibfnamefont {S.}~\bibnamefont {Peotta}}\ and\ \bibinfo {author} {\bibfnamefont {P.}~\bibnamefont {T{\"o}rm{\"a}}},\ }\bibfield  {title} {\bibinfo {title} {Superfluidity in topologically nontrivial flat bands},\ }\href@noop {} {\bibfield  {journal} {\bibinfo  {journal} {Nature communications}\ }\textbf {\bibinfo {volume} {6}},\ \bibinfo {pages} {8944} (\bibinfo {year} {2015})}\BibitemShut {NoStop}%
\bibitem [{\citenamefont {Yu}\ \emph {et~al.}(2025)\citenamefont {Yu}, \citenamefont {Bernevig}, \citenamefont {Queiroz}, \citenamefont {Rossi}, \citenamefont {T{\"o}rm{\"a}},\ and\ \citenamefont {Yang}}]{yu2025quantum}%
  \BibitemOpen
  \bibfield  {author} {\bibinfo {author} {\bibfnamefont {J.}~\bibnamefont {Yu}}, \bibinfo {author} {\bibfnamefont {B.~A.}\ \bibnamefont {Bernevig}}, \bibinfo {author} {\bibfnamefont {R.}~\bibnamefont {Queiroz}}, \bibinfo {author} {\bibfnamefont {E.}~\bibnamefont {Rossi}}, \bibinfo {author} {\bibfnamefont {P.}~\bibnamefont {T{\"o}rm{\"a}}},\ and\ \bibinfo {author} {\bibfnamefont {B.-J.}\ \bibnamefont {Yang}},\ }\bibfield  {title} {\bibinfo {title} {Quantum geometry in quantum materials},\ }\href@noop {} {\bibfield  {journal} {\bibinfo  {journal} {npj Quantum Materials}\ }\textbf {\bibinfo {volume} {10}},\ \bibinfo {pages} {101} (\bibinfo {year} {2025})}\BibitemShut {NoStop}%
\bibitem [{\citenamefont {Arora}\ \emph {et~al.}(2025)\citenamefont {Arora}, \citenamefont {Curtis},\ and\ \citenamefont {Narang}}]{arora2025quantum}%
  \BibitemOpen
  \bibfield  {author} {\bibinfo {author} {\bibfnamefont {A.}~\bibnamefont {Arora}}, \bibinfo {author} {\bibfnamefont {J.~B.}\ \bibnamefont {Curtis}},\ and\ \bibinfo {author} {\bibfnamefont {P.}~\bibnamefont {Narang}},\ }\bibfield  {title} {\bibinfo {title} {Quantum geometry induced microwave enhancement of superconducting order in flat bands},\ }\href@noop {} {\bibfield  {journal} {\bibinfo  {journal} {Communications Physics}\ }\textbf {\bibinfo {volume} {8}},\ \bibinfo {pages} {327} (\bibinfo {year} {2025})}\BibitemShut {NoStop}%
\bibitem [{\citenamefont {Verma}\ \emph {et~al.}(2025)\citenamefont {Verma}, \citenamefont {Moll}, \citenamefont {Holder},\ and\ \citenamefont {Queiroz}}]{verma2025quantum}%
  \BibitemOpen
  \bibfield  {author} {\bibinfo {author} {\bibfnamefont {N.}~\bibnamefont {Verma}}, \bibinfo {author} {\bibfnamefont {P.~J.}\ \bibnamefont {Moll}}, \bibinfo {author} {\bibfnamefont {T.}~\bibnamefont {Holder}},\ and\ \bibinfo {author} {\bibfnamefont {R.}~\bibnamefont {Queiroz}},\ }\bibfield  {title} {\bibinfo {title} {Quantum geometry: Revisiting electronic scales in quantum matter},\ }\href@noop {} {\bibfield  {journal} {\bibinfo  {journal} {arXiv preprint arXiv:2504.07173}\ } (\bibinfo {year} {2025})}\BibitemShut {NoStop}%
\bibitem [{\citenamefont {Kim}\ \emph {et~al.}(2025)\citenamefont {Kim}, \citenamefont {Chung}, \citenamefont {Qian}, \citenamefont {Park}, \citenamefont {Jozwiak}, \citenamefont {Rotenberg}, \citenamefont {Bostwick}, \citenamefont {Kim},\ and\ \citenamefont {Yang}}]{kim2025direct}%
  \BibitemOpen
  \bibfield  {author} {\bibinfo {author} {\bibfnamefont {S.}~\bibnamefont {Kim}}, \bibinfo {author} {\bibfnamefont {Y.}~\bibnamefont {Chung}}, \bibinfo {author} {\bibfnamefont {Y.}~\bibnamefont {Qian}}, \bibinfo {author} {\bibfnamefont {S.}~\bibnamefont {Park}}, \bibinfo {author} {\bibfnamefont {C.}~\bibnamefont {Jozwiak}}, \bibinfo {author} {\bibfnamefont {E.}~\bibnamefont {Rotenberg}}, \bibinfo {author} {\bibfnamefont {A.}~\bibnamefont {Bostwick}}, \bibinfo {author} {\bibfnamefont {K.~S.}\ \bibnamefont {Kim}},\ and\ \bibinfo {author} {\bibfnamefont {B.-J.}\ \bibnamefont {Yang}},\ }\bibfield  {title} {\bibinfo {title} {Direct measurement of the quantum metric tensor in solids},\ }\href@noop {} {\bibfield  {journal} {\bibinfo  {journal} {Science}\ }\textbf {\bibinfo {volume} {388}},\ \bibinfo {pages} {1050} (\bibinfo {year} {2025})}\BibitemShut {NoStop}%
\bibitem [{\citenamefont {Cai}\ \emph {et~al.}(2025)\citenamefont {Cai}, \citenamefont {Pan}, \citenamefont {Wang}, \citenamefont {Rasmita}, \citenamefont {Yang}, \citenamefont {Zhao}, \citenamefont {Wang}, \citenamefont {Duan}, \citenamefont {He}, \citenamefont {Watanabe} \emph {et~al.}}]{cai2025optical}%
  \BibitemOpen
  \bibfield  {author} {\bibinfo {author} {\bibfnamefont {X.}~\bibnamefont {Cai}}, \bibinfo {author} {\bibfnamefont {H.}~\bibnamefont {Pan}}, \bibinfo {author} {\bibfnamefont {Y.}~\bibnamefont {Wang}}, \bibinfo {author} {\bibfnamefont {A.}~\bibnamefont {Rasmita}}, \bibinfo {author} {\bibfnamefont {S.}~\bibnamefont {Yang}}, \bibinfo {author} {\bibfnamefont {Y.}~\bibnamefont {Zhao}}, \bibinfo {author} {\bibfnamefont {W.}~\bibnamefont {Wang}}, \bibinfo {author} {\bibfnamefont {R.}~\bibnamefont {Duan}}, \bibinfo {author} {\bibfnamefont {R.}~\bibnamefont {He}}, \bibinfo {author} {\bibfnamefont {K.}~\bibnamefont {Watanabe}}, \emph {et~al.},\ }\bibfield  {title} {\bibinfo {title} {Optical switching of moir\'e chern ferromagnet},\ }\href@noop {} {\bibfield  {journal} {\bibinfo  {journal} {arXiv preprint arXiv:2508.19602}\ } (\bibinfo {year} {2025})}\BibitemShut {NoStop}%
\bibitem [{\citenamefont {Huber}\ \emph {et~al.}(2025)\citenamefont {Huber}, \citenamefont {Kuhlbrodt}, \citenamefont {Anderson}, \citenamefont {Li}, \citenamefont {Watanabe}, \citenamefont {Taniguchi}, \citenamefont {Kroner}, \citenamefont {Xu}, \citenamefont {Imamoglu},\ and\ \citenamefont {Smolenski}}]{huber2025optical}%
  \BibitemOpen
  \bibfield  {author} {\bibinfo {author} {\bibfnamefont {O.}~\bibnamefont {Huber}}, \bibinfo {author} {\bibfnamefont {K.}~\bibnamefont {Kuhlbrodt}}, \bibinfo {author} {\bibfnamefont {E.}~\bibnamefont {Anderson}}, \bibinfo {author} {\bibfnamefont {W.}~\bibnamefont {Li}}, \bibinfo {author} {\bibfnamefont {K.}~\bibnamefont {Watanabe}}, \bibinfo {author} {\bibfnamefont {T.}~\bibnamefont {Taniguchi}}, \bibinfo {author} {\bibfnamefont {M.}~\bibnamefont {Kroner}}, \bibinfo {author} {\bibfnamefont {X.}~\bibnamefont {Xu}}, \bibinfo {author} {\bibfnamefont {A.}~\bibnamefont {Imamoglu}},\ and\ \bibinfo {author} {\bibfnamefont {T.}~\bibnamefont {Smolenski}},\ }\bibfield  {title} {\bibinfo {title} {Optical control over topological chern number in moir\'e materials},\ }\href@noop {} {\bibfield  {journal} {\bibinfo  {journal} {arXiv preprint arXiv:2508.19063}\ } (\bibinfo {year} {2025})}\BibitemShut {NoStop}%
\bibitem [{\citenamefont {Holtzmann}\ \emph {et~al.}(2025)\citenamefont {Holtzmann}, \citenamefont {Li}, \citenamefont {Anderson}, \citenamefont {Cai}, \citenamefont {Park}, \citenamefont {Hu}, \citenamefont {Taniguchi}, \citenamefont {Watanabe}, \citenamefont {Chu}, \citenamefont {Xiao} \emph {et~al.}}]{holtzmann2025optical}%
  \BibitemOpen
  \bibfield  {author} {\bibinfo {author} {\bibfnamefont {W.}~\bibnamefont {Holtzmann}}, \bibinfo {author} {\bibfnamefont {W.}~\bibnamefont {Li}}, \bibinfo {author} {\bibfnamefont {E.}~\bibnamefont {Anderson}}, \bibinfo {author} {\bibfnamefont {J.}~\bibnamefont {Cai}}, \bibinfo {author} {\bibfnamefont {H.}~\bibnamefont {Park}}, \bibinfo {author} {\bibfnamefont {C.}~\bibnamefont {Hu}}, \bibinfo {author} {\bibfnamefont {T.}~\bibnamefont {Taniguchi}}, \bibinfo {author} {\bibfnamefont {K.}~\bibnamefont {Watanabe}}, \bibinfo {author} {\bibfnamefont {J.-H.}\ \bibnamefont {Chu}}, \bibinfo {author} {\bibfnamefont {D.}~\bibnamefont {Xiao}}, \emph {et~al.},\ }\bibfield  {title} {\bibinfo {title} {Optical control of integer and fractional chern insulators},\ }\href@noop {} {\bibfield  {journal} {\bibinfo  {journal} {arXiv preprint arXiv:2508.18639}\ } (\bibinfo {year} {2025})}\BibitemShut {NoStop}%
\bibitem [{\citenamefont {B{\o}ttcher}\ \emph {et~al.}(2024{\natexlab{b}})\citenamefont {B{\o}ttcher}, \citenamefont {Nichele}, \citenamefont {Shabani}, \citenamefont {Palmstr{\o}m},\ and\ \citenamefont {Marcus}}]{bottcher2024berezinskii}%
  \BibitemOpen
  \bibfield  {author} {\bibinfo {author} {\bibfnamefont {C.}~\bibnamefont {B{\o}ttcher}}, \bibinfo {author} {\bibfnamefont {F.}~\bibnamefont {Nichele}}, \bibinfo {author} {\bibfnamefont {J.}~\bibnamefont {Shabani}}, \bibinfo {author} {\bibfnamefont {C.}~\bibnamefont {Palmstr{\o}m}},\ and\ \bibinfo {author} {\bibfnamefont {C.}~\bibnamefont {Marcus}},\ }\bibfield  {title} {\bibinfo {title} {Berezinskii-kosterlitz-thouless transition and anomalous metallic phase in a hybrid josephson junction array},\ }\href@noop {} {\bibfield  {journal} {\bibinfo  {journal} {Physical Review B}\ }\textbf {\bibinfo {volume} {110}},\ \bibinfo {pages} {L180502} (\bibinfo {year} {2024}{\natexlab{b}})}\BibitemShut {NoStop}%
\bibitem [{\citenamefont {Sasmal}\ \emph {et~al.}(2025)\citenamefont {Sasmal}, \citenamefont {Efthymiou-Tsironi}, \citenamefont {Nagda}, \citenamefont {Fugl}, \citenamefont {Olsen}, \citenamefont {Krizek}, \citenamefont {Marcus},\ and\ \citenamefont {Vaitiek{\.e}nas}}]{sasmal2025voltage}%
  \BibitemOpen
  \bibfield  {author} {\bibinfo {author} {\bibfnamefont {S.}~\bibnamefont {Sasmal}}, \bibinfo {author} {\bibfnamefont {M.}~\bibnamefont {Efthymiou-Tsironi}}, \bibinfo {author} {\bibfnamefont {G.}~\bibnamefont {Nagda}}, \bibinfo {author} {\bibfnamefont {E.}~\bibnamefont {Fugl}}, \bibinfo {author} {\bibfnamefont {L.}~\bibnamefont {Olsen}}, \bibinfo {author} {\bibfnamefont {F.}~\bibnamefont {Krizek}}, \bibinfo {author} {\bibfnamefont {C.}~\bibnamefont {Marcus}},\ and\ \bibinfo {author} {\bibfnamefont {S.}~\bibnamefont {Vaitiek{\.e}nas}},\ }\bibfield  {title} {\bibinfo {title} {Voltage-tuned anomalous-metal to metal transition in hybrid josephson junction arrays},\ }\href@noop {} {\bibfield  {journal} {\bibinfo  {journal} {Physical Review Letters}\ }\textbf {\bibinfo {volume} {135}},\ \bibinfo {pages} {156301} (\bibinfo {year} {2025})}\BibitemShut {NoStop}%
\bibitem [{\citenamefont {Wang}\ \emph {et~al.}(2026)\citenamefont {Wang}, \citenamefont {Drachmann}, \citenamefont {Thomas}, \citenamefont {Manfra}, \citenamefont {Trivedi}, \citenamefont {Marcus}, \citenamefont {Vaitiek{\.e}nas},\ and\ \citenamefont {Kalisky}}]{wang2026long}%
  \BibitemOpen
  \bibfield  {author} {\bibinfo {author} {\bibfnamefont {X.}~\bibnamefont {Wang}}, \bibinfo {author} {\bibfnamefont {A.~C.}\ \bibnamefont {Drachmann}}, \bibinfo {author} {\bibfnamefont {C.}~\bibnamefont {Thomas}}, \bibinfo {author} {\bibfnamefont {M.~J.}\ \bibnamefont {Manfra}}, \bibinfo {author} {\bibfnamefont {N.}~\bibnamefont {Trivedi}}, \bibinfo {author} {\bibfnamefont {C.~M.}\ \bibnamefont {Marcus}}, \bibinfo {author} {\bibfnamefont {S.}~\bibnamefont {Vaitiek{\.e}nas}},\ and\ \bibinfo {author} {\bibfnamefont {B.}~\bibnamefont {Kalisky}},\ }\bibfield  {title} {\bibinfo {title} {Long-range phase coherence and phase patterns in hybrid josephson junction arrays},\ }\href@noop {} {\bibfield  {journal} {\bibinfo  {journal} {arXiv preprint arXiv:2602.02255}\ } (\bibinfo {year} {2026})}\BibitemShut {NoStop}%
\bibitem [{\citenamefont {Sarker}\ \emph {et~al.}(2026)\citenamefont {Sarker}, \citenamefont {Khriyenko}, \citenamefont {Mikkonen},\ and\ \citenamefont {Haghparast}}]{sarker2026machine}%
  \BibitemOpen
  \bibfield  {author} {\bibinfo {author} {\bibfnamefont {S.}~\bibnamefont {Sarker}}, \bibinfo {author} {\bibfnamefont {O.}~\bibnamefont {Khriyenko}}, \bibinfo {author} {\bibfnamefont {T.}~\bibnamefont {Mikkonen}},\ and\ \bibinfo {author} {\bibfnamefont {M.}~\bibnamefont {Haghparast}},\ }\bibfield  {title} {\bibinfo {title} {A machine learning approach to detecting crosstalk-induced fidelity degradation in quantum computing systems},\ }\href@noop {} {\bibfield  {journal} {\bibinfo  {journal} {Physica Scripta}\ } (\bibinfo {year} {2026})}\BibitemShut {NoStop}%
\bibitem [{\citenamefont {Gupta}\ \emph {et~al.}(2025)\citenamefont {Gupta}, \citenamefont {Joshi}, \citenamefont {Kandpal}, \citenamefont {Mandayam}, \citenamefont {Gheeraert},\ and\ \citenamefont {Dhomkar}}]{gupta2025expedited}%
  \BibitemOpen
  \bibfield  {author} {\bibinfo {author} {\bibfnamefont {B.}~\bibnamefont {Gupta}}, \bibinfo {author} {\bibfnamefont {V.}~\bibnamefont {Joshi}}, \bibinfo {author} {\bibfnamefont {U.}~\bibnamefont {Kandpal}}, \bibinfo {author} {\bibfnamefont {P.}~\bibnamefont {Mandayam}}, \bibinfo {author} {\bibfnamefont {N.}~\bibnamefont {Gheeraert}},\ and\ \bibinfo {author} {\bibfnamefont {S.}~\bibnamefont {Dhomkar}},\ }\bibfield  {title} {\bibinfo {title} {Expedited noise spectroscopy of transmon qubits},\ }\href@noop {} {\bibfield  {journal} {\bibinfo  {journal} {Advanced Quantum Technologies}\ }\textbf {\bibinfo {volume} {8}},\ \bibinfo {pages} {e00109} (\bibinfo {year} {2025})}\BibitemShut {NoStop}%
\bibitem [{\citenamefont {Koolstra}\ \emph {et~al.}(2022)\citenamefont {Koolstra}, \citenamefont {Stevenson}, \citenamefont {Barzili}, \citenamefont {Burns}, \citenamefont {Siva}, \citenamefont {Greenfield}, \citenamefont {Livingston}, \citenamefont {Hashim}, \citenamefont {Naik}, \citenamefont {Kreikebaum} \emph {et~al.}}]{koolstra2022monitoring}%
  \BibitemOpen
  \bibfield  {author} {\bibinfo {author} {\bibfnamefont {G.}~\bibnamefont {Koolstra}}, \bibinfo {author} {\bibfnamefont {N.}~\bibnamefont {Stevenson}}, \bibinfo {author} {\bibfnamefont {S.}~\bibnamefont {Barzili}}, \bibinfo {author} {\bibfnamefont {L.}~\bibnamefont {Burns}}, \bibinfo {author} {\bibfnamefont {K.}~\bibnamefont {Siva}}, \bibinfo {author} {\bibfnamefont {S.}~\bibnamefont {Greenfield}}, \bibinfo {author} {\bibfnamefont {W.}~\bibnamefont {Livingston}}, \bibinfo {author} {\bibfnamefont {A.}~\bibnamefont {Hashim}}, \bibinfo {author} {\bibfnamefont {R.}~\bibnamefont {Naik}}, \bibinfo {author} {\bibfnamefont {J.}~\bibnamefont {Kreikebaum}}, \emph {et~al.},\ }\bibfield  {title} {\bibinfo {title} {Monitoring fast superconducting qubit dynamics using a neural network},\ }\href@noop {} {\bibfield  {journal} {\bibinfo  {journal} {Physical Review X}\ }\textbf {\bibinfo {volume} {12}},\ \bibinfo {pages} {031017} (\bibinfo {year} {2022})}\BibitemShut {NoStop}%
\bibitem [{\citenamefont {Barzanjeh}\ \emph {et~al.}(2025)\citenamefont {Barzanjeh}, \citenamefont {Xuereb}, \citenamefont {Al{\`u}}, \citenamefont {Mann}, \citenamefont {Nefedkin}, \citenamefont {Peano},\ and\ \citenamefont {Rabl}}]{barzanjeh2025nonreciprocity}%
  \BibitemOpen
  \bibfield  {author} {\bibinfo {author} {\bibfnamefont {S.}~\bibnamefont {Barzanjeh}}, \bibinfo {author} {\bibfnamefont {A.}~\bibnamefont {Xuereb}}, \bibinfo {author} {\bibfnamefont {A.}~\bibnamefont {Al{\`u}}}, \bibinfo {author} {\bibfnamefont {S.~A.}\ \bibnamefont {Mann}}, \bibinfo {author} {\bibfnamefont {N.}~\bibnamefont {Nefedkin}}, \bibinfo {author} {\bibfnamefont {V.}~\bibnamefont {Peano}},\ and\ \bibinfo {author} {\bibfnamefont {P.}~\bibnamefont {Rabl}},\ }\bibfield  {title} {\bibinfo {title} {Nonreciprocity in quantum technology},\ }\href@noop {} {\bibfield  {journal} {\bibinfo  {journal} {arXiv preprint arXiv:2508.03945}\ } (\bibinfo {year} {2025})}\BibitemShut {NoStop}%
\bibitem [{\citenamefont {Hovhannisyan}\ \emph {et~al.}(2025)\citenamefont {Hovhannisyan}, \citenamefont {Lotfian}, \citenamefont {Golod},\ and\ \citenamefont {Krasnov}}]{hovhannisyan2025demonstration}%
  \BibitemOpen
  \bibfield  {author} {\bibinfo {author} {\bibfnamefont {R.~A.}\ \bibnamefont {Hovhannisyan}}, \bibinfo {author} {\bibfnamefont {A.}~\bibnamefont {Lotfian}}, \bibinfo {author} {\bibfnamefont {T.}~\bibnamefont {Golod}},\ and\ \bibinfo {author} {\bibfnamefont {V.~M.}\ \bibnamefont {Krasnov}},\ }\bibfield  {title} {\bibinfo {title} {Demonstration of an optical microwave rectification by a superconducting diode with near 100\% efficiency},\ }\href@noop {} {\bibfield  {journal} {\bibinfo  {journal} {arXiv preprint arXiv:2508.21696}\ } (\bibinfo {year} {2025})}\BibitemShut {NoStop}%
\bibitem [{\citenamefont {Genois}\ \emph {et~al.}(2021)\citenamefont {Genois}, \citenamefont {Gross}, \citenamefont {Di~Paolo}, \citenamefont {Stevenson}, \citenamefont {Koolstra}, \citenamefont {Hashim}, \citenamefont {Siddiqi},\ and\ \citenamefont {Blais}}]{genois2021quantum}%
  \BibitemOpen
  \bibfield  {author} {\bibinfo {author} {\bibfnamefont {{\'E}.}~\bibnamefont {Genois}}, \bibinfo {author} {\bibfnamefont {J.~A.}\ \bibnamefont {Gross}}, \bibinfo {author} {\bibfnamefont {A.}~\bibnamefont {Di~Paolo}}, \bibinfo {author} {\bibfnamefont {N.~J.}\ \bibnamefont {Stevenson}}, \bibinfo {author} {\bibfnamefont {G.}~\bibnamefont {Koolstra}}, \bibinfo {author} {\bibfnamefont {A.}~\bibnamefont {Hashim}}, \bibinfo {author} {\bibfnamefont {I.}~\bibnamefont {Siddiqi}},\ and\ \bibinfo {author} {\bibfnamefont {A.}~\bibnamefont {Blais}},\ }\bibfield  {title} {\bibinfo {title} {Quantum-tailored machine-learning characterization of a superconducting qubit},\ }\href@noop {} {\bibfield  {journal} {\bibinfo  {journal} {PRX Quantum}\ }\textbf {\bibinfo {volume} {2}},\ \bibinfo {pages} {040355} (\bibinfo {year} {2021})}\BibitemShut {NoStop}%
\bibitem [{\citenamefont {Shah}\ \emph {et~al.}(2026)\citenamefont {Shah}, \citenamefont {Patti}, \citenamefont {Berner}, \citenamefont {Tolooshams}, \citenamefont {Kossaifi},\ and\ \citenamefont {Anandkumar}}]{shah2026fourier}%
  \BibitemOpen
  \bibfield  {author} {\bibinfo {author} {\bibfnamefont {F.}~\bibnamefont {Shah}}, \bibinfo {author} {\bibfnamefont {T.~L.}\ \bibnamefont {Patti}}, \bibinfo {author} {\bibfnamefont {J.}~\bibnamefont {Berner}}, \bibinfo {author} {\bibfnamefont {B.}~\bibnamefont {Tolooshams}}, \bibinfo {author} {\bibfnamefont {J.}~\bibnamefont {Kossaifi}},\ and\ \bibinfo {author} {\bibfnamefont {A.}~\bibnamefont {Anandkumar}},\ }\bibfield  {title} {\bibinfo {title} {Fourier neural operators for learning dynamics in quantum spin systems},\ }\href@noop {} {\bibfield  {journal} {\bibinfo  {journal} {Communications Physics}\ } (\bibinfo {year} {2026})}\BibitemShut {NoStop}%
\bibitem [{\citenamefont {Waugh}\ \emph {et~al.}(2026)\citenamefont {Waugh}, \citenamefont {Wang}, \citenamefont {Andrei}, \citenamefont {Islam}, \citenamefont {Patti}, \citenamefont {Demler},\ and\ \citenamefont {Anandkumar}}]{waugh2026toward}%
  \BibitemOpen
  \bibfield  {author} {\bibinfo {author} {\bibfnamefont {M.}~\bibnamefont {Waugh}}, \bibinfo {author} {\bibfnamefont {C.}~\bibnamefont {Wang}}, \bibinfo {author} {\bibfnamefont {R.}~\bibnamefont {Andrei}}, \bibinfo {author} {\bibfnamefont {N.}~\bibnamefont {Islam}}, \bibinfo {author} {\bibfnamefont {T.~L.}\ \bibnamefont {Patti}}, \bibinfo {author} {\bibfnamefont {E.}~\bibnamefont {Demler}},\ and\ \bibinfo {author} {\bibfnamefont {A.}~\bibnamefont {Anandkumar}},\ }\bibfield  {title} {\bibinfo {title} {Toward the thermodynamic limit: Neural operators for non-equilibrium dynamics of mott insulators},\ }\href@noop {} {\bibfield  {journal} {\bibinfo  {journal} {arXiv preprint arXiv:2602.19484}\ } (\bibinfo {year} {2026})}\BibitemShut {NoStop}%
\bibitem [{\citenamefont {Pipi}\ \emph {et~al.}(2025)\citenamefont {Pipi}, \citenamefont {Nivedha}, \citenamefont {Duruisseaux}, \citenamefont {Marmarelis}, \citenamefont {Patti}, \citenamefont {Khailany}, \citenamefont {Narang},\ and\ \citenamefont {Anandkumar}}]{pipi2025inverse}%
  \BibitemOpen
  \bibfield  {author} {\bibinfo {author} {\bibfnamefont {A.}~\bibnamefont {Pipi}}, \bibinfo {author} {\bibfnamefont {G.}~\bibnamefont {Nivedha}}, \bibinfo {author} {\bibfnamefont {V.}~\bibnamefont {Duruisseaux}}, \bibinfo {author} {\bibfnamefont {M.~G.}\ \bibnamefont {Marmarelis}}, \bibinfo {author} {\bibfnamefont {T.~L.}\ \bibnamefont {Patti}}, \bibinfo {author} {\bibfnamefont {B.}~\bibnamefont {Khailany}}, \bibinfo {author} {\bibfnamefont {P.}~\bibnamefont {Narang}},\ and\ \bibinfo {author} {\bibfnamefont {A.}~\bibnamefont {Anandkumar}},\ }\bibfield  {title} {\bibinfo {title} {Inverse design with fourier neural operators for quantum system control},\ }in\ \href@noop {} {\emph {\bibinfo {booktitle} {Machine Learning and the Physical Sciences Workshop at NeurIPS 2025}}}\ (\bibinfo {year} {2025})\BibitemShut {NoStop}%
\bibitem [{\citenamefont {Mitridate}\ \emph {et~al.}(2021)\citenamefont {Mitridate}, \citenamefont {Trickle}, \citenamefont {Zhang},\ and\ \citenamefont {Zurek}}]{mitridate_dark_2021}%
  \BibitemOpen
  \bibfield  {author} {\bibinfo {author} {\bibfnamefont {A.}~\bibnamefont {Mitridate}}, \bibinfo {author} {\bibfnamefont {T.}~\bibnamefont {Trickle}}, \bibinfo {author} {\bibfnamefont {Z.}~\bibnamefont {Zhang}},\ and\ \bibinfo {author} {\bibfnamefont {K.~M.}\ \bibnamefont {Zurek}},\ }\bibfield  {title} {\bibinfo {title} {Dark matter absorption via electronic excitations},\ }\href {https://doi.org/10.1007/JHEP09(2021)123} {\bibfield  {journal} {\bibinfo  {journal} {Journal of High Energy Physics}\ }\textbf {\bibinfo {volume} {2021}},\ \bibinfo {pages} {123} (\bibinfo {year} {2021})}\BibitemShut {NoStop}%
\bibitem [{\citenamefont {O'Hare}(2020)}]{AxionLimits}%
  \BibitemOpen
  \bibfield  {author} {\bibinfo {author} {\bibfnamefont {C.}~\bibnamefont {O'Hare}},\ }\href {https://doi.org/10.5281/zenodo.3932430} {\bibinfo {title} {cajohare/axionlimits: Axionlimits}},\ \bibinfo {howpublished} {\url{https://cajohare.github.io/AxionLimits/}} (\bibinfo {year} {2020})\BibitemShut {NoStop}%
\bibitem [{\citenamefont {Peccei}\ and\ \citenamefont {Quinn}(1977)}]{peccei_mathrmcp_1977}%
  \BibitemOpen
  \bibfield  {author} {\bibinfo {author} {\bibfnamefont {R.~D.}\ \bibnamefont {Peccei}}\ and\ \bibinfo {author} {\bibfnamefont {H.~R.}\ \bibnamefont {Quinn}},\ }\bibfield  {title} {\bibinfo {title} {\${\textbackslash}mathrm\{{CP}\}\$ {Conservation} in the {Presence} of {Pseudoparticles}},\ }\href {https://doi.org/10.1103/PhysRevLett.38.1440} {\bibfield  {journal} {\bibinfo  {journal} {Physical Review Letters}\ }\textbf {\bibinfo {volume} {38}},\ \bibinfo {pages} {1440} (\bibinfo {year} {1977})},\ \bibinfo {note} {publisher: American Physical Society}\BibitemShut {NoStop}%
\bibitem [{\citenamefont {Preskill}\ \emph {et~al.}(1983)\citenamefont {Preskill}, \citenamefont {Wise},\ and\ \citenamefont {Wilczek}}]{preskill_cosmology_1983}%
  \BibitemOpen
  \bibfield  {author} {\bibinfo {author} {\bibfnamefont {J.}~\bibnamefont {Preskill}}, \bibinfo {author} {\bibfnamefont {M.~B.}\ \bibnamefont {Wise}},\ and\ \bibinfo {author} {\bibfnamefont {F.}~\bibnamefont {Wilczek}},\ }\bibfield  {title} {\bibinfo {title} {Cosmology of the invisible axion},\ }\href {https://doi.org/10.1016/0370-2693(83)90637-8} {\bibfield  {journal} {\bibinfo  {journal} {Physics Letters B}\ }\textbf {\bibinfo {volume} {120}},\ \bibinfo {pages} {127} (\bibinfo {year} {1983})}\BibitemShut {NoStop}%
\bibitem [{\citenamefont {Abbott}\ and\ \citenamefont {Sikivie}(1983)}]{abbott_cosmological_1983}%
  \BibitemOpen
  \bibfield  {author} {\bibinfo {author} {\bibfnamefont {L.~F.}\ \bibnamefont {Abbott}}\ and\ \bibinfo {author} {\bibfnamefont {P.}~\bibnamefont {Sikivie}},\ }\bibfield  {title} {\bibinfo {title} {A cosmological bound on the invisible axion},\ }\href {https://doi.org/10.1016/0370-2693(83)90638-X} {\bibfield  {journal} {\bibinfo  {journal} {Physics Letters B}\ }\textbf {\bibinfo {volume} {120}},\ \bibinfo {pages} {133} (\bibinfo {year} {1983})}\BibitemShut {NoStop}%
\bibitem [{\citenamefont {Dine}\ and\ \citenamefont {Fischler}(1983)}]{dine_not-so-harmless_1983}%
  \BibitemOpen
  \bibfield  {author} {\bibinfo {author} {\bibfnamefont {M.}~\bibnamefont {Dine}}\ and\ \bibinfo {author} {\bibfnamefont {W.}~\bibnamefont {Fischler}},\ }\bibfield  {title} {\bibinfo {title} {The not-so-harmless axion},\ }\href {https://doi.org/10.1016/0370-2693(83)90639-1} {\bibfield  {journal} {\bibinfo  {journal} {Physics Letters B}\ }\textbf {\bibinfo {volume} {120}},\ \bibinfo {pages} {137} (\bibinfo {year} {1983})}\BibitemShut {NoStop}%
\bibitem [{\citenamefont {Sikivie}(2008)}]{sikivie_axion_2008}%
  \BibitemOpen
  \bibfield  {author} {\bibinfo {author} {\bibfnamefont {P.}~\bibnamefont {Sikivie}},\ }\bibfield  {title} {\bibinfo {title} {Axion {Cosmology}},\ }in\ \href {https://doi.org/10.1007/978-3-540-73518-2_2} {\emph {\bibinfo {booktitle} {Axions: {Theory}, {Cosmology}, and {Experimental} {Searches}}}},\ \bibinfo {editor} {edited by\ \bibinfo {editor} {\bibfnamefont {M.}~\bibnamefont {Kuster}}, \bibinfo {editor} {\bibfnamefont {G.}~\bibnamefont {Raffelt}},\ and\ \bibinfo {editor} {\bibfnamefont {B.}~\bibnamefont {Beltr\'an}}}\ (\bibinfo  {publisher} {Springer},\ \bibinfo {address} {Berlin, Heidelberg},\ \bibinfo {year} {2008})\ pp.\ \bibinfo {pages} {19--50}\BibitemShut {NoStop}%
\bibitem [{\citenamefont {Caves}(1982)}]{caves_quantum_1982}%
  \BibitemOpen
  \bibfield  {author} {\bibinfo {author} {\bibfnamefont {C.~M.}\ \bibnamefont {Caves}},\ }\bibfield  {title} {\bibinfo {title} {Quantum limits on noise in linear amplifiers},\ }\href {https://doi.org/10.1103/PhysRevD.26.1817} {\bibfield  {journal} {\bibinfo  {journal} {Physical Review D}\ }\textbf {\bibinfo {volume} {26}},\ \bibinfo {pages} {1817} (\bibinfo {year} {1982})},\ \bibinfo {note} {publisher: American Physical Society}\BibitemShut {NoStop}%
\bibitem [{\citenamefont {Lamoreaux}\ \emph {et~al.}(2013)\citenamefont {Lamoreaux}, \citenamefont {van Bibber}, \citenamefont {Lehnert},\ and\ \citenamefont {Carosi}}]{lamoreaux_analysis_2013}%
  \BibitemOpen
  \bibfield  {author} {\bibinfo {author} {\bibfnamefont {S.~K.}\ \bibnamefont {Lamoreaux}}, \bibinfo {author} {\bibfnamefont {K.~A.}\ \bibnamefont {van Bibber}}, \bibinfo {author} {\bibfnamefont {K.~W.}\ \bibnamefont {Lehnert}},\ and\ \bibinfo {author} {\bibfnamefont {G.}~\bibnamefont {Carosi}},\ }\bibfield  {title} {\bibinfo {title} {Analysis of single-photon and linear amplifier detectors for microwave cavity dark matter axion searches},\ }\href {https://doi.org/10.1103/PhysRevD.88.035020} {\bibfield  {journal} {\bibinfo  {journal} {Physical Review D}\ }\textbf {\bibinfo {volume} {88}},\ \bibinfo {pages} {035020} (\bibinfo {year} {2013})},\ \bibinfo {note} {publisher: American Physical Society}\BibitemShut {NoStop}%
\bibitem [{\citenamefont {Baudis}\ \emph {et~al.}(2025)\citenamefont {Baudis}, \citenamefont {Bismark}, \citenamefont {Brugger}, \citenamefont {Capelli}, \citenamefont {Charaev}, \citenamefont {Garc\'{\i}a}, \citenamefont {Hadas}, \citenamefont {Hochberg}, \citenamefont {Hohmann}, \citenamefont {Kavner}, \citenamefont {Koos}, \citenamefont {Kuzmin}, \citenamefont {Lehmann}, \citenamefont {N\"ageli}, \citenamefont {Neupert}, \citenamefont {Penning}, \citenamefont {Garc\'{\i}a},\ and\ \citenamefont {Schilling}}]{Baudis_QROCODILE_2025}%
  \BibitemOpen
  \bibfield  {author} {\bibinfo {author} {\bibfnamefont {L.}~\bibnamefont {Baudis}}, \bibinfo {author} {\bibfnamefont {A.}~\bibnamefont {Bismark}}, \bibinfo {author} {\bibfnamefont {N.}~\bibnamefont {Brugger}}, \bibinfo {author} {\bibfnamefont {C.}~\bibnamefont {Capelli}}, \bibinfo {author} {\bibfnamefont {I.}~\bibnamefont {Charaev}}, \bibinfo {author} {\bibfnamefont {J.~C.}\ \bibnamefont {Garc\'{\i}a}}, \bibinfo {author} {\bibfnamefont {G.~D.}\ \bibnamefont {Hadas}}, \bibinfo {author} {\bibfnamefont {Y.}~\bibnamefont {Hochberg}}, \bibinfo {author} {\bibfnamefont {J.~K.}\ \bibnamefont {Hohmann}}, \bibinfo {author} {\bibfnamefont {A.}~\bibnamefont {Kavner}}, \bibinfo {author} {\bibfnamefont {C.}~\bibnamefont {Koos}}, \bibinfo {author} {\bibfnamefont {A.}~\bibnamefont {Kuzmin}}, \bibinfo {author} {\bibfnamefont {B.~V.}\ \bibnamefont {Lehmann}}, \bibinfo {author} {\bibfnamefont {S.}~\bibnamefont {N\"ageli}}, \bibinfo {author} {\bibfnamefont {T.}~\bibnamefont {Neupert}}, \bibinfo {author} {\bibfnamefont
  {B.}~\bibnamefont {Penning}}, \bibinfo {author} {\bibfnamefont {D.~R.}\ \bibnamefont {Garc\'{\i}a}},\ and\ \bibinfo {author} {\bibfnamefont {A.}~\bibnamefont {Schilling}},\ }\bibfield  {title} {\bibinfo {title} {First sub-mev dark matter search with the qrocodile experiment using superconducting nanowire single-photon detectors},\ }\href {https://doi.org/10.1103/4hb6-f6jl} {\bibfield  {journal} {\bibinfo  {journal} {Phys. Rev. Lett.}\ }\textbf {\bibinfo {volume} {135}},\ \bibinfo {pages} {081002} (\bibinfo {year} {2025})}\BibitemShut {NoStop}%
\bibitem [{\citenamefont {Kinion}\ and\ \citenamefont {Clarke}(2011)}]{ADMX-MSA}%
  \BibitemOpen
  \bibfield  {author} {\bibinfo {author} {\bibfnamefont {D.}~\bibnamefont {Kinion}}\ and\ \bibinfo {author} {\bibfnamefont {J.}~\bibnamefont {Clarke}},\ }\bibfield  {title} {\bibinfo {title} {Superconducting quantum interference device as a near-quantum-limited amplifier for the axion dark-matter experiment},\ }\href {https://doi.org/10.1063/1.3583380} {\bibfield  {journal} {\bibinfo  {journal} {Applied Physics Letters}\ }\textbf {\bibinfo {volume} {98}},\ \bibinfo {pages} {202503} (\bibinfo {year} {2011})},\ \Eprint {https://arxiv.org/abs/https://pubs.aip.org/aip/apl/article-pdf/doi/10.1063/1.3583380/14449319/202503\_1\_online.pdf} {https://pubs.aip.org/aip/apl/article-pdf/doi/10.1063/1.3583380/14449319/202503\_1\_online.pdf} \BibitemShut {NoStop}%
\bibitem [{\citenamefont {Zheng}\ \emph {et~al.}(2016)\citenamefont {Zheng}, \citenamefont {Silveri}, \citenamefont {Brierley}, \citenamefont {Girvin},\ and\ \citenamefont {Lehnert}}]{zheng_accelerating_2016}%
  \BibitemOpen
  \bibfield  {author} {\bibinfo {author} {\bibfnamefont {H.}~\bibnamefont {Zheng}}, \bibinfo {author} {\bibfnamefont {M.}~\bibnamefont {Silveri}}, \bibinfo {author} {\bibfnamefont {R.~T.}\ \bibnamefont {Brierley}}, \bibinfo {author} {\bibfnamefont {S.~M.}\ \bibnamefont {Girvin}},\ and\ \bibinfo {author} {\bibfnamefont {K.~W.}\ \bibnamefont {Lehnert}},\ }\href {https://doi.org/10.48550/arXiv.1607.02529} {\bibinfo {title} {Accelerating dark-matter axion searches with quantum measurement technology}} (\bibinfo {year} {2016}),\ \bibinfo {note} {arXiv:1607.02529 [hep-ph]}\BibitemShut {NoStop}%
\bibitem [{\citenamefont {Dixit}\ \emph {et~al.}(2018)\citenamefont {Dixit}, \citenamefont {Chou},\ and\ \citenamefont {Schuster}}]{dixit_detecting_2018}%
  \BibitemOpen
  \bibfield  {author} {\bibinfo {author} {\bibfnamefont {A.}~\bibnamefont {Dixit}}, \bibinfo {author} {\bibfnamefont {A.}~\bibnamefont {Chou}},\ and\ \bibinfo {author} {\bibfnamefont {D.}~\bibnamefont {Schuster}},\ }\bibfield  {title} {\bibinfo {title} {Detecting {Axion} {Dark} {Matter} with {Superconducting} {Qubits}},\ }in\ \href {https://doi.org/10.1007/978-3-319-92726-8_11} {\emph {\bibinfo {booktitle} {Springer {Proceedings} in {Physics}}}}\ (\bibinfo  {publisher} {Springer International Publishing},\ \bibinfo {address} {Cham},\ \bibinfo {year} {2018})\ pp.\ \bibinfo {pages} {97--103},\ \bibinfo {note} {iSSN: 0930-8989, 1867-4941}\BibitemShut {NoStop}%
\bibitem [{\citenamefont {Braggio}\ \emph {et~al.}(2025)\citenamefont {Braggio}, \citenamefont {Balembois}, \citenamefont {Di~Vora}, \citenamefont {Wang}, \citenamefont {Travesedo}, \citenamefont {Pallegoix}, \citenamefont {Carugno}, \citenamefont {Ortolan}, \citenamefont {Ruoso}, \citenamefont {Gambardella}, \citenamefont {D'Agostino}, \citenamefont {Bertet},\ and\ \citenamefont {Flurin}}]{braggio_quantum-enhanced_2025}%
  \BibitemOpen
  \bibfield  {author} {\bibinfo {author} {\bibfnamefont {C.}~\bibnamefont {Braggio}}, \bibinfo {author} {\bibfnamefont {L.}~\bibnamefont {Balembois}}, \bibinfo {author} {\bibfnamefont {R.}~\bibnamefont {Di~Vora}}, \bibinfo {author} {\bibfnamefont {Z.}~\bibnamefont {Wang}}, \bibinfo {author} {\bibfnamefont {J.}~\bibnamefont {Travesedo}}, \bibinfo {author} {\bibfnamefont {L.}~\bibnamefont {Pallegoix}}, \bibinfo {author} {\bibfnamefont {G.}~\bibnamefont {Carugno}}, \bibinfo {author} {\bibfnamefont {A.}~\bibnamefont {Ortolan}}, \bibinfo {author} {\bibfnamefont {G.}~\bibnamefont {Ruoso}}, \bibinfo {author} {\bibfnamefont {U.}~\bibnamefont {Gambardella}}, \bibinfo {author} {\bibfnamefont {D.}~\bibnamefont {D'Agostino}}, \bibinfo {author} {\bibfnamefont {P.}~\bibnamefont {Bertet}},\ and\ \bibinfo {author} {\bibfnamefont {E.}~\bibnamefont {Flurin}},\ }\bibfield  {title} {\bibinfo {title} {Quantum-{Enhanced} {Sensing} of {Axion} {Dark} {Matter} with a {Transmon}-{Based} {Single} {Microwave} {Photon} {Counter}},\
  }\href {https://doi.org/10.1103/PhysRevX.15.021031} {\bibfield  {journal} {\bibinfo  {journal} {Physical Review X}\ }\textbf {\bibinfo {volume} {15}},\ \bibinfo {pages} {021031} (\bibinfo {year} {2025})},\ \bibinfo {note} {publisher: American Physical Society}\BibitemShut {NoStop}%
\bibitem [{\citenamefont {Fabbrichesi}\ \emph {et~al.}(2020)\citenamefont {Fabbrichesi}, \citenamefont {Gabrielli},\ and\ \citenamefont {Lanfranchi}}]{Fabbrichesi:2020wbt}%
  \BibitemOpen
  \bibfield  {author} {\bibinfo {author} {\bibfnamefont {M.}~\bibnamefont {Fabbrichesi}}, \bibinfo {author} {\bibfnamefont {E.}~\bibnamefont {Gabrielli}},\ and\ \bibinfo {author} {\bibfnamefont {G.}~\bibnamefont {Lanfranchi}},\ }\href {https://doi.org/10.1007/978-3-030-62519-1} {\emph {\bibinfo {title} {{The Dark Photon}}}}\ (\bibinfo  {publisher} {Springer},\ \bibinfo {year} {2020})\ \Eprint {https://arxiv.org/abs/2005.01515} {arXiv:2005.01515 [hep-ph]} \BibitemShut {NoStop}%
\bibitem [{\citenamefont {Dixit}\ \emph {et~al.}(2021)\citenamefont {Dixit}, \citenamefont {Chakram}, \citenamefont {He}, \citenamefont {Agrawal}, \citenamefont {Naik}, \citenamefont {Schuster},\ and\ \citenamefont {Chou}}]{dixit_searching_2021}%
  \BibitemOpen
  \bibfield  {author} {\bibinfo {author} {\bibfnamefont {A.~V.}\ \bibnamefont {Dixit}}, \bibinfo {author} {\bibfnamefont {S.}~\bibnamefont {Chakram}}, \bibinfo {author} {\bibfnamefont {K.}~\bibnamefont {He}}, \bibinfo {author} {\bibfnamefont {A.}~\bibnamefont {Agrawal}}, \bibinfo {author} {\bibfnamefont {R.~K.}\ \bibnamefont {Naik}}, \bibinfo {author} {\bibfnamefont {D.~I.}\ \bibnamefont {Schuster}},\ and\ \bibinfo {author} {\bibfnamefont {A.}~\bibnamefont {Chou}},\ }\bibfield  {title} {\bibinfo {title} {Searching for {Dark} {Matter} with a {Superconducting} {Qubit}},\ }\href {https://doi.org/10.1103/PhysRevLett.126.141302} {\bibfield  {journal} {\bibinfo  {journal} {Physical Review Letters}\ }\textbf {\bibinfo {volume} {126}},\ \bibinfo {pages} {141302} (\bibinfo {year} {2021})},\ \bibinfo {note} {publisher: American Physical Society}\BibitemShut {NoStop}%
\bibitem [{\citenamefont {Shi}\ and\ \citenamefont {Zhuang}(2023)}]{shi_ultimate_2023}%
  \BibitemOpen
  \bibfield  {author} {\bibinfo {author} {\bibfnamefont {H.}~\bibnamefont {Shi}}\ and\ \bibinfo {author} {\bibfnamefont {Q.}~\bibnamefont {Zhuang}},\ }\bibfield  {title} {\bibinfo {title} {Ultimate precision limit of noise sensing and dark matter search},\ }\href {https://doi.org/10.1038/s41534-023-00693-w} {\bibfield  {journal} {\bibinfo  {journal} {npj Quantum Information}\ }\textbf {\bibinfo {volume} {9}},\ \bibinfo {pages} {27} (\bibinfo {year} {2023})},\ \bibinfo {note} {publisher: Nature Publishing Group}\BibitemShut {NoStop}%
\bibitem [{\citenamefont {Shi}\ \emph {et~al.}(2025)\citenamefont {Shi}, \citenamefont {Brady}, \citenamefont {G\'orecki}, \citenamefont {Maccone}, \citenamefont {Di~Candia},\ and\ \citenamefont {Zhuang}}]{shi_quantum-enhanced_2025}%
  \BibitemOpen
  \bibfield  {author} {\bibinfo {author} {\bibfnamefont {H.}~\bibnamefont {Shi}}, \bibinfo {author} {\bibfnamefont {A.~J.}\ \bibnamefont {Brady}}, \bibinfo {author} {\bibfnamefont {W.}~\bibnamefont {G\'orecki}}, \bibinfo {author} {\bibfnamefont {L.}~\bibnamefont {Maccone}}, \bibinfo {author} {\bibfnamefont {R.}~\bibnamefont {Di~Candia}},\ and\ \bibinfo {author} {\bibfnamefont {Q.}~\bibnamefont {Zhuang}},\ }\bibfield  {title} {\bibinfo {title} {Quantum-enhanced dark matter detection with in-cavity control: mitigating the {Rayleigh} curse},\ }\href {https://doi.org/10.1038/s41534-025-01004-1} {\bibfield  {journal} {\bibinfo  {journal} {npj Quantum Information}\ }\textbf {\bibinfo {volume} {11}},\ \bibinfo {pages} {48} (\bibinfo {year} {2025})},\ \bibinfo {note} {publisher: Nature Publishing Group}\BibitemShut {NoStop}%
\bibitem [{\citenamefont {Gardner}\ \emph {et~al.}(2025)\citenamefont {Gardner}, \citenamefont {Gefen}, \citenamefont {Haine}, \citenamefont {Hope}, \citenamefont {Preskill}, \citenamefont {Chen},\ and\ \citenamefont {McCuller}}]{gardner_stochastic_2025}%
  \BibitemOpen
  \bibfield  {author} {\bibinfo {author} {\bibfnamefont {J.~W.}\ \bibnamefont {Gardner}}, \bibinfo {author} {\bibfnamefont {T.}~\bibnamefont {Gefen}}, \bibinfo {author} {\bibfnamefont {S.~A.}\ \bibnamefont {Haine}}, \bibinfo {author} {\bibfnamefont {J.~J.}\ \bibnamefont {Hope}}, \bibinfo {author} {\bibfnamefont {J.}~\bibnamefont {Preskill}}, \bibinfo {author} {\bibfnamefont {Y.}~\bibnamefont {Chen}},\ and\ \bibinfo {author} {\bibfnamefont {L.}~\bibnamefont {McCuller}},\ }\bibfield  {title} {\bibinfo {title} {Stochastic waveform estimation at the fundamental quantum limit},\ }\href {https://doi.org/10.1103/h91r-4ws9} {\bibfield  {journal} {\bibinfo  {journal} {PRX Quantum}\ }\textbf {\bibinfo {volume} {6}},\ \bibinfo {pages} {030311} (\bibinfo {year} {2025})}\BibitemShut {NoStop}%
\bibitem [{\citenamefont {Labarca}\ \emph {et~al.}(2026)\citenamefont {Labarca}, \citenamefont {Turcotte}, \citenamefont {Blais},\ and\ \citenamefont {Royer}}]{qmss-lc5x}%
  \BibitemOpen
  \bibfield  {author} {\bibinfo {author} {\bibfnamefont {L.}~\bibnamefont {Labarca}}, \bibinfo {author} {\bibfnamefont {S.}~\bibnamefont {Turcotte}}, \bibinfo {author} {\bibfnamefont {A.}~\bibnamefont {Blais}},\ and\ \bibinfo {author} {\bibfnamefont {B.}~\bibnamefont {Royer}},\ }\bibfield  {title} {\bibinfo {title} {Quantum sensing of displacements with stabilized gottesman-kitaev-preskill states},\ }\href {https://doi.org/10.1103/qmss-lc5x} {\bibfield  {journal} {\bibinfo  {journal} {PRX Quantum}\ }\textbf {\bibinfo {volume} {7}},\ \bibinfo {pages} {020301} (\bibinfo {year} {2026})}\BibitemShut {NoStop}%
\bibitem [{\citenamefont {Agrawal}\ \emph {et~al.}(2024)\citenamefont {Agrawal}, \citenamefont {Dixit}, \citenamefont {Roy}, \citenamefont {Chakram}, \citenamefont {He}, \citenamefont {Naik}, \citenamefont {Schuster},\ and\ \citenamefont {Chou}}]{Agrawal_2024}%
  \BibitemOpen
  \bibfield  {author} {\bibinfo {author} {\bibfnamefont {A.}~\bibnamefont {Agrawal}}, \bibinfo {author} {\bibfnamefont {A.~V.}\ \bibnamefont {Dixit}}, \bibinfo {author} {\bibfnamefont {T.}~\bibnamefont {Roy}}, \bibinfo {author} {\bibfnamefont {S.}~\bibnamefont {Chakram}}, \bibinfo {author} {\bibfnamefont {K.}~\bibnamefont {He}}, \bibinfo {author} {\bibfnamefont {R.~K.}\ \bibnamefont {Naik}}, \bibinfo {author} {\bibfnamefont {D.~I.}\ \bibnamefont {Schuster}},\ and\ \bibinfo {author} {\bibfnamefont {A.}~\bibnamefont {Chou}},\ }\bibfield  {title} {\bibinfo {title} {Stimulated emission of signal photons from dark matter waves},\ }\bibfield  {journal} {\bibinfo  {journal} {Physical Review Letters}\ }\textbf {\bibinfo {volume} {132}},\ \href {https://doi.org/10.1103/physrevlett.132.140801} {10.1103/physrevlett.132.140801} (\bibinfo {year} {2024})\BibitemShut {NoStop}%
\bibitem [{\citenamefont {Hua}\ \emph {et~al.}(2026)\citenamefont {Hua}, \citenamefont {Ma}, \citenamefont {Zhou}, \citenamefont {Xu}, \citenamefont {Chen}, \citenamefont {Cai}, \citenamefont {Chen}, \citenamefont {Xiao}, \citenamefont {Huang}, \citenamefont {Wang}, \citenamefont {Li}, \citenamefont {Wang}, \citenamefont {Li}, \citenamefont {Zou},\ and\ \citenamefont {Sun}}]{hua2026quantumconfocalmicroscopyfock}%
  \BibitemOpen
  \bibfield  {author} {\bibinfo {author} {\bibfnamefont {Z.}~\bibnamefont {Hua}}, \bibinfo {author} {\bibfnamefont {C.}~\bibnamefont {Ma}}, \bibinfo {author} {\bibfnamefont {Y.}~\bibnamefont {Zhou}}, \bibinfo {author} {\bibfnamefont {Y.}~\bibnamefont {Xu}}, \bibinfo {author} {\bibfnamefont {Z.-J.}\ \bibnamefont {Chen}}, \bibinfo {author} {\bibfnamefont {W.}~\bibnamefont {Cai}}, \bibinfo {author} {\bibfnamefont {J.}~\bibnamefont {Chen}}, \bibinfo {author} {\bibfnamefont {L.}~\bibnamefont {Xiao}}, \bibinfo {author} {\bibfnamefont {H.}~\bibnamefont {Huang}}, \bibinfo {author} {\bibfnamefont {W.}~\bibnamefont {Wang}}, \bibinfo {author} {\bibfnamefont {H.}~\bibnamefont {Li}}, \bibinfo {author} {\bibfnamefont {H.}~\bibnamefont {Wang}}, \bibinfo {author} {\bibfnamefont {M.}~\bibnamefont {Li}}, \bibinfo {author} {\bibfnamefont {C.-L.}\ \bibnamefont {Zou}},\ and\ \bibinfo {author} {\bibfnamefont {L.}~\bibnamefont {Sun}},\ }\href {https://arxiv.org/abs/2602.23254} {\bibinfo {title} {Quantum confocal microscopy in fock
  space with a 19 db metrological gain}} (\bibinfo {year} {2026}),\ \bibinfo {note} {arXiv:2602.23254 [quant-ph]},\ \Eprint {https://arxiv.org/abs/2602.23254} {arXiv:2602.23254 [quant-ph]} \BibitemShut {NoStop}%
\bibitem [{\citenamefont {Campagne-Ibarcq}\ \emph {et~al.}(2020)\citenamefont {Campagne-Ibarcq}, \citenamefont {Eickbusch}, \citenamefont {Touzard}, \citenamefont {Zalys-Geller}, \citenamefont {Frattini}, \citenamefont {Sivak}, \citenamefont {Reinhold}, \citenamefont {Puri}, \citenamefont {Shankar}, \citenamefont {Schoelkopf}, \citenamefont {Frunzio}, \citenamefont {Mirrahimi},\ and\ \citenamefont {Devoret}}]{campagne-ibarcq_quantum_2020}%
  \BibitemOpen
  \bibfield  {author} {\bibinfo {author} {\bibfnamefont {P.}~\bibnamefont {Campagne-Ibarcq}}, \bibinfo {author} {\bibfnamefont {A.}~\bibnamefont {Eickbusch}}, \bibinfo {author} {\bibfnamefont {S.}~\bibnamefont {Touzard}}, \bibinfo {author} {\bibfnamefont {E.}~\bibnamefont {Zalys-Geller}}, \bibinfo {author} {\bibfnamefont {N.~E.}\ \bibnamefont {Frattini}}, \bibinfo {author} {\bibfnamefont {V.~V.}\ \bibnamefont {Sivak}}, \bibinfo {author} {\bibfnamefont {P.}~\bibnamefont {Reinhold}}, \bibinfo {author} {\bibfnamefont {S.}~\bibnamefont {Puri}}, \bibinfo {author} {\bibfnamefont {S.}~\bibnamefont {Shankar}}, \bibinfo {author} {\bibfnamefont {R.~J.}\ \bibnamefont {Schoelkopf}}, \bibinfo {author} {\bibfnamefont {L.}~\bibnamefont {Frunzio}}, \bibinfo {author} {\bibfnamefont {M.}~\bibnamefont {Mirrahimi}},\ and\ \bibinfo {author} {\bibfnamefont {M.~H.}\ \bibnamefont {Devoret}},\ }\bibfield  {title} {\bibinfo {title} {Quantum error correction of a qubit encoded in grid states of an oscillator},\ }\href
  {https://doi.org/10.1038/s41586-020-2603-3} {\bibfield  {journal} {\bibinfo  {journal} {Nature}\ }\textbf {\bibinfo {volume} {584}},\ \bibinfo {pages} {368} (\bibinfo {year} {2020})}\BibitemShut {NoStop}%
\bibitem [{\citenamefont {Milul}\ \emph {et~al.}(2023)\citenamefont {Milul}, \citenamefont {Guttel}, \citenamefont {Goldblatt}, \citenamefont {Hazanov}, \citenamefont {Joshi}, \citenamefont {Chausovsky}, \citenamefont {Kahn}, \citenamefont {\c{C}ifty\"urek}, \citenamefont {Lafont},\ and\ \citenamefont {Rosenblum}}]{milul_superconducting_2023}%
  \BibitemOpen
  \bibfield  {author} {\bibinfo {author} {\bibfnamefont {O.}~\bibnamefont {Milul}}, \bibinfo {author} {\bibfnamefont {B.}~\bibnamefont {Guttel}}, \bibinfo {author} {\bibfnamefont {U.}~\bibnamefont {Goldblatt}}, \bibinfo {author} {\bibfnamefont {S.}~\bibnamefont {Hazanov}}, \bibinfo {author} {\bibfnamefont {L.~M.}\ \bibnamefont {Joshi}}, \bibinfo {author} {\bibfnamefont {D.}~\bibnamefont {Chausovsky}}, \bibinfo {author} {\bibfnamefont {N.}~\bibnamefont {Kahn}}, \bibinfo {author} {\bibfnamefont {E.}~\bibnamefont {\c{C}ifty\"urek}}, \bibinfo {author} {\bibfnamefont {F.}~\bibnamefont {Lafont}},\ and\ \bibinfo {author} {\bibfnamefont {S.}~\bibnamefont {Rosenblum}},\ }\bibfield  {title} {\bibinfo {title} {Superconducting {Cavity} {Qubit} with {Tens} of {Milliseconds} {Single}-{Photon} {Coherence} {Time}},\ }\href {https://doi.org/10.1103/PRXQuantum.4.030336} {\bibfield  {journal} {\bibinfo  {journal} {PRX Quantum}\ }\textbf {\bibinfo {volume} {4}},\ \bibinfo {pages} {030336} (\bibinfo {year} {2023})},\ \bibinfo
  {note} {publisher: American Physical Society}\BibitemShut {NoStop}%
\bibitem [{\citenamefont {Balembois}\ \emph {et~al.}(2024)\citenamefont {Balembois}, \citenamefont {Travesedo}, \citenamefont {Pallegoix}, \citenamefont {May}, \citenamefont {Billaud}, \citenamefont {Villiers}, \citenamefont {Est\`eve}, \citenamefont {Vion}, \citenamefont {Bertet},\ and\ \citenamefont {Flurin}}]{balembois_cyclically_2024}%
  \BibitemOpen
  \bibfield  {author} {\bibinfo {author} {\bibfnamefont {L.}~\bibnamefont {Balembois}}, \bibinfo {author} {\bibfnamefont {J.}~\bibnamefont {Travesedo}}, \bibinfo {author} {\bibfnamefont {L.}~\bibnamefont {Pallegoix}}, \bibinfo {author} {\bibfnamefont {A.}~\bibnamefont {May}}, \bibinfo {author} {\bibfnamefont {E.}~\bibnamefont {Billaud}}, \bibinfo {author} {\bibfnamefont {M.}~\bibnamefont {Villiers}}, \bibinfo {author} {\bibfnamefont {D.}~\bibnamefont {Est\`eve}}, \bibinfo {author} {\bibfnamefont {D.}~\bibnamefont {Vion}}, \bibinfo {author} {\bibfnamefont {P.}~\bibnamefont {Bertet}},\ and\ \bibinfo {author} {\bibfnamefont {E.}~\bibnamefont {Flurin}},\ }\bibfield  {title} {\bibinfo {title} {Cyclically operated microwave single-photon counter with sensitivity of $10^{-22}\,\mathrm{W}/\sqrt{\mathrm{hz}}$},\ }\href {https://doi.org/10.1103/PhysRevApplied.21.014043} {\bibfield  {journal} {\bibinfo  {journal} {Physical Review Applied}\ }\textbf {\bibinfo {volume} {21}},\ \bibinfo {pages} {014043} (\bibinfo {year}
  {2024})},\ \bibinfo {note} {publisher: American Physical Society}\BibitemShut {NoStop}%
\bibitem [{\citenamefont {Pallegoix}\ \emph {et~al.}(2025)\citenamefont {Pallegoix}, \citenamefont {Travesedo}, \citenamefont {May}, \citenamefont {Balembois}, \citenamefont {Vion}, \citenamefont {Bertet},\ and\ \citenamefont {Flurin}}]{pallegoix_enhancing_2025}%
  \BibitemOpen
  \bibfield  {author} {\bibinfo {author} {\bibfnamefont {L.}~\bibnamefont {Pallegoix}}, \bibinfo {author} {\bibfnamefont {J.}~\bibnamefont {Travesedo}}, \bibinfo {author} {\bibfnamefont {A.~S.}\ \bibnamefont {May}}, \bibinfo {author} {\bibfnamefont {L.}~\bibnamefont {Balembois}}, \bibinfo {author} {\bibfnamefont {D.}~\bibnamefont {Vion}}, \bibinfo {author} {\bibfnamefont {P.}~\bibnamefont {Bertet}},\ and\ \bibinfo {author} {\bibfnamefont {E.}~\bibnamefont {Flurin}},\ }\bibfield  {title} {\bibinfo {title} {Enhancing the sensitivity of single-microwave-photon detection with bandwidth tunability},\ }\bibfield  {journal} {\bibinfo  {journal} {Phys. Rev. Appl.}\ }\href {https://doi.org/10.1103/dbqj-qld5} {10.1103/dbqj-qld5} (\bibinfo {year} {2025})\BibitemShut {NoStop}%
\bibitem [{\citenamefont {May}\ \emph {et~al.}(2025)\citenamefont {May}, \citenamefont {Sutevski}, \citenamefont {Solard}, \citenamefont {Cardoso}, \citenamefont {Carde}, \citenamefont {Pallegoix}, \citenamefont {Lescanne}, \citenamefont {Vion}, \citenamefont {Bertet},\ and\ \citenamefont {Flurin}}]{may2025noisemitigationsinglemicrowave}%
  \BibitemOpen
  \bibfield  {author} {\bibinfo {author} {\bibfnamefont {A.~S.}\ \bibnamefont {May}}, \bibinfo {author} {\bibfnamefont {L.}~\bibnamefont {Sutevski}}, \bibinfo {author} {\bibfnamefont {J.}~\bibnamefont {Solard}}, \bibinfo {author} {\bibfnamefont {G.}~\bibnamefont {Cardoso}}, \bibinfo {author} {\bibfnamefont {L.}~\bibnamefont {Carde}}, \bibinfo {author} {\bibfnamefont {L.}~\bibnamefont {Pallegoix}}, \bibinfo {author} {\bibfnamefont {R.}~\bibnamefont {Lescanne}}, \bibinfo {author} {\bibfnamefont {D.}~\bibnamefont {Vion}}, \bibinfo {author} {\bibfnamefont {P.}~\bibnamefont {Bertet}},\ and\ \bibinfo {author} {\bibfnamefont {E.}~\bibnamefont {Flurin}},\ }\href {https://arxiv.org/abs/2502.14804} {\bibinfo {title} {Noise mitigation in single microwave photon counting by cascaded quantum measurements}} (\bibinfo {year} {2025}),\ \bibinfo {note} {arXiv:2502.14804 [quant-ph]},\ \Eprint {https://arxiv.org/abs/2502.14804} {arXiv:2502.14804 [quant-ph]} \BibitemShut {NoStop}%
\bibitem [{\citenamefont {Kuo}\ \emph {et~al.}(2025)\citenamefont {Kuo}, \citenamefont {Bartram}, \citenamefont {Chou}, \citenamefont {Dyson}, \citenamefont {Kurinsky}, \citenamefont {Rybka}, \citenamefont {Ruppert}, \citenamefont {Wen}, \citenamefont {Withers}, \citenamefont {Yi},\ and\ \citenamefont {Zhang}}]{kuo_maximizing_2025}%
  \BibitemOpen
  \bibfield  {author} {\bibinfo {author} {\bibfnamefont {C.-L.}\ \bibnamefont {Kuo}}, \bibinfo {author} {\bibfnamefont {C.~L.}\ \bibnamefont {Bartram}}, \bibinfo {author} {\bibfnamefont {A.~S.}\ \bibnamefont {Chou}}, \bibinfo {author} {\bibfnamefont {T.~A.}\ \bibnamefont {Dyson}}, \bibinfo {author} {\bibfnamefont {N.~A.}\ \bibnamefont {Kurinsky}}, \bibinfo {author} {\bibfnamefont {G.}~\bibnamefont {Rybka}}, \bibinfo {author} {\bibfnamefont {S.}~\bibnamefont {Ruppert}}, \bibinfo {author} {\bibfnamefont {O.}~\bibnamefont {Wen}}, \bibinfo {author} {\bibfnamefont {M.~O.}\ \bibnamefont {Withers}}, \bibinfo {author} {\bibfnamefont {A.~K.}\ \bibnamefont {Yi}},\ and\ \bibinfo {author} {\bibfnamefont {C.}~\bibnamefont {Zhang}},\ }\bibfield  {title} {\bibinfo {title} {Maximizing quantum enhancement in axion dark matter experiments},\ }\href {https://doi.org/10.1103/pgtg-3lhd} {\bibfield  {journal} {\bibinfo  {journal} {Physical Review D}\ }\textbf {\bibinfo {volume} {111}},\ \bibinfo {pages} {123018} (\bibinfo {year}
  {2025})},\ \bibinfo {note} {publisher: American Physical Society}\BibitemShut {NoStop}%
\bibitem [{\citenamefont {Andersen}\ \emph {et~al.}(2016)\citenamefont {Andersen}, \citenamefont {Gehring}, \citenamefont {Marquardt},\ and\ \citenamefont {Leuchs}}]{andersen_30_2016}%
  \BibitemOpen
  \bibfield  {author} {\bibinfo {author} {\bibfnamefont {U.~L.}\ \bibnamefont {Andersen}}, \bibinfo {author} {\bibfnamefont {T.}~\bibnamefont {Gehring}}, \bibinfo {author} {\bibfnamefont {C.}~\bibnamefont {Marquardt}},\ and\ \bibinfo {author} {\bibfnamefont {G.}~\bibnamefont {Leuchs}},\ }\bibfield  {title} {\bibinfo {title} {30 years of squeezed light generation},\ }\href {https://doi.org/10.1088/0031-8949/91/5/053001} {\bibfield  {journal} {\bibinfo  {journal} {Physica Scripta}\ }\textbf {\bibinfo {volume} {91}},\ \bibinfo {pages} {053001} (\bibinfo {year} {2016})},\ \bibinfo {note} {publisher: IOP Publishing}\BibitemShut {NoStop}%
\bibitem [{\citenamefont {{HAYSTAC Collaboration}}\ \emph {et~al.}(2023)\citenamefont {{HAYSTAC Collaboration}}, \citenamefont {Jewell}, \citenamefont {Leder}, \citenamefont {Backes}, \citenamefont {Bai}, \citenamefont {van Bibber}, \citenamefont {Brubaker}, \citenamefont {Cahn}, \citenamefont {Droster}, \citenamefont {Esmat}, \citenamefont {Ghosh}, \citenamefont {Graham}, \citenamefont {Hilton}, \citenamefont {Jackson}, \citenamefont {Laffan}, \citenamefont {Lamoreaux}, \citenamefont {Lehnert}, \citenamefont {Lewis}, \citenamefont {Malnou}, \citenamefont {Maruyama}, \citenamefont {Palken}, \citenamefont {Rapidis}, \citenamefont {Ruddy}, \citenamefont {Simanovskaia}, \citenamefont {Singh}, \citenamefont {Speller}, \citenamefont {Vale}, \citenamefont {Wang},\ and\ \citenamefont {Zhu}}]{haystac_collaboration_new_2023}%
  \BibitemOpen
  \bibfield  {author} {\bibinfo {author} {\bibnamefont {{HAYSTAC Collaboration}}}, \bibinfo {author} {\bibfnamefont {M.~J.}\ \bibnamefont {Jewell}}, \bibinfo {author} {\bibfnamefont {A.~F.}\ \bibnamefont {Leder}}, \bibinfo {author} {\bibfnamefont {K.~M.}\ \bibnamefont {Backes}}, \bibinfo {author} {\bibfnamefont {X.}~\bibnamefont {Bai}}, \bibinfo {author} {\bibfnamefont {K.}~\bibnamefont {van Bibber}}, \bibinfo {author} {\bibfnamefont {B.~M.}\ \bibnamefont {Brubaker}}, \bibinfo {author} {\bibfnamefont {S.~B.}\ \bibnamefont {Cahn}}, \bibinfo {author} {\bibfnamefont {A.}~\bibnamefont {Droster}}, \bibinfo {author} {\bibfnamefont {M.~H.}\ \bibnamefont {Esmat}}, \bibinfo {author} {\bibfnamefont {S.}~\bibnamefont {Ghosh}}, \bibinfo {author} {\bibfnamefont {E.}~\bibnamefont {Graham}}, \bibinfo {author} {\bibfnamefont {G.~C.}\ \bibnamefont {Hilton}}, \bibinfo {author} {\bibfnamefont {H.}~\bibnamefont {Jackson}}, \bibinfo {author} {\bibfnamefont {C.}~\bibnamefont {Laffan}}, \bibinfo {author} {\bibfnamefont {S.~K.}\
  \bibnamefont {Lamoreaux}}, \bibinfo {author} {\bibfnamefont {K.~W.}\ \bibnamefont {Lehnert}}, \bibinfo {author} {\bibfnamefont {S.~M.}\ \bibnamefont {Lewis}}, \bibinfo {author} {\bibfnamefont {M.}~\bibnamefont {Malnou}}, \bibinfo {author} {\bibfnamefont {R.~H.}\ \bibnamefont {Maruyama}}, \bibinfo {author} {\bibfnamefont {D.~A.}\ \bibnamefont {Palken}}, \bibinfo {author} {\bibfnamefont {N.~M.}\ \bibnamefont {Rapidis}}, \bibinfo {author} {\bibfnamefont {E.~P.}\ \bibnamefont {Ruddy}}, \bibinfo {author} {\bibfnamefont {M.}~\bibnamefont {Simanovskaia}}, \bibinfo {author} {\bibfnamefont {S.}~\bibnamefont {Singh}}, \bibinfo {author} {\bibfnamefont {D.~H.}\ \bibnamefont {Speller}}, \bibinfo {author} {\bibfnamefont {L.~R.}\ \bibnamefont {Vale}}, \bibinfo {author} {\bibfnamefont {H.}~\bibnamefont {Wang}},\ and\ \bibinfo {author} {\bibfnamefont {Y.}~\bibnamefont {Zhu}},\ }\bibfield  {title} {\bibinfo {title} {New results from {HAYSTAC}'s phase {II} operation with a squeezed state receiver},\ }\href
  {https://doi.org/10.1103/PhysRevD.107.072007} {\bibfield  {journal} {\bibinfo  {journal} {Physical Review D}\ }\textbf {\bibinfo {volume} {107}},\ \bibinfo {pages} {072007} (\bibinfo {year} {2023})},\ \bibinfo {note} {publisher: American Physical Society}\BibitemShut {NoStop}%
\bibitem [{\citenamefont {Sushkov}(2023)}]{sushkov_quantum_2023}%
  \BibitemOpen
  \bibfield  {author} {\bibinfo {author} {\bibfnamefont {A.~O.}\ \bibnamefont {Sushkov}},\ }\bibfield  {title} {\bibinfo {title} {Quantum {Science} and the {Search} for {Axion} {Dark} {Matter}},\ }\href {https://doi.org/10.1103/PRXQuantum.4.020101} {\bibfield  {journal} {\bibinfo  {journal} {PRX Quantum}\ }\textbf {\bibinfo {volume} {4}},\ \bibinfo {pages} {020101} (\bibinfo {year} {2023})},\ \bibinfo {note} {publisher: American Physical Society}\BibitemShut {NoStop}%
\bibitem [{\citenamefont {Shokair}\ \emph {et~al.}(2014)\citenamefont {Shokair}, \citenamefont {Root}, \citenamefont {Van~Bibber}, \citenamefont {Brubaker}, \citenamefont {Gurevich}, \citenamefont {Cahn}, \citenamefont {Lamoreaux}, \citenamefont {Anil}, \citenamefont {Lehnert}, \citenamefont {Mitchell}, \citenamefont {Reed},\ and\ \citenamefont {Carosi}}]{Shokair_2014}%
  \BibitemOpen
  \bibfield  {author} {\bibinfo {author} {\bibfnamefont {T.~M.}\ \bibnamefont {Shokair}}, \bibinfo {author} {\bibfnamefont {J.}~\bibnamefont {Root}}, \bibinfo {author} {\bibfnamefont {K.~A.}\ \bibnamefont {Van~Bibber}}, \bibinfo {author} {\bibfnamefont {B.}~\bibnamefont {Brubaker}}, \bibinfo {author} {\bibfnamefont {Y.~V.}\ \bibnamefont {Gurevich}}, \bibinfo {author} {\bibfnamefont {S.~B.}\ \bibnamefont {Cahn}}, \bibinfo {author} {\bibfnamefont {S.~K.}\ \bibnamefont {Lamoreaux}}, \bibinfo {author} {\bibfnamefont {M.~A.}\ \bibnamefont {Anil}}, \bibinfo {author} {\bibfnamefont {K.~W.}\ \bibnamefont {Lehnert}}, \bibinfo {author} {\bibfnamefont {B.~K.}\ \bibnamefont {Mitchell}}, \bibinfo {author} {\bibfnamefont {A.}~\bibnamefont {Reed}},\ and\ \bibinfo {author} {\bibfnamefont {G.}~\bibnamefont {Carosi}},\ }\bibfield  {title} {\bibinfo {title} {Future directions in the microwave cavity search for dark matter axions},\ }\href {https://doi.org/10.1142/s0217751x14430040} {\bibfield  {journal} {\bibinfo  {journal}
  {International Journal of Modern Physics A}\ }\textbf {\bibinfo {volume} {29}},\ \bibinfo {pages} {1443004} (\bibinfo {year} {2014})}\BibitemShut {NoStop}%
\bibitem [{\citenamefont {Backes}\ \emph {et~al.}(2021)\citenamefont {Backes}, \citenamefont {Palken}, \citenamefont {Kenany}, \citenamefont {Brubaker}, \citenamefont {Cahn}, \citenamefont {Droster}, \citenamefont {Hilton}, \citenamefont {Ghosh}, \citenamefont {Jackson}, \citenamefont {Lamoreaux}, \citenamefont {Leder}, \citenamefont {Lehnert}, \citenamefont {Lewis}, \citenamefont {Malnou}, \citenamefont {Maruyama}, \citenamefont {Rapidis}, \citenamefont {Simanovskaia}, \citenamefont {Singh}, \citenamefont {Speller}, \citenamefont {Urdinaran}, \citenamefont {Vale}, \citenamefont {van Assendelft}, \citenamefont {van Bibber},\ and\ \citenamefont {Wang}}]{backes_quantum_2021}%
  \BibitemOpen
  \bibfield  {author} {\bibinfo {author} {\bibfnamefont {K.~M.}\ \bibnamefont {Backes}}, \bibinfo {author} {\bibfnamefont {D.~A.}\ \bibnamefont {Palken}}, \bibinfo {author} {\bibfnamefont {S.~A.}\ \bibnamefont {Kenany}}, \bibinfo {author} {\bibfnamefont {B.~M.}\ \bibnamefont {Brubaker}}, \bibinfo {author} {\bibfnamefont {S.~B.}\ \bibnamefont {Cahn}}, \bibinfo {author} {\bibfnamefont {A.}~\bibnamefont {Droster}}, \bibinfo {author} {\bibfnamefont {G.~C.}\ \bibnamefont {Hilton}}, \bibinfo {author} {\bibfnamefont {S.}~\bibnamefont {Ghosh}}, \bibinfo {author} {\bibfnamefont {H.}~\bibnamefont {Jackson}}, \bibinfo {author} {\bibfnamefont {S.~K.}\ \bibnamefont {Lamoreaux}}, \bibinfo {author} {\bibfnamefont {A.~F.}\ \bibnamefont {Leder}}, \bibinfo {author} {\bibfnamefont {K.~W.}\ \bibnamefont {Lehnert}}, \bibinfo {author} {\bibfnamefont {S.~M.}\ \bibnamefont {Lewis}}, \bibinfo {author} {\bibfnamefont {M.}~\bibnamefont {Malnou}}, \bibinfo {author} {\bibfnamefont {R.~H.}\ \bibnamefont {Maruyama}}, \bibinfo {author}
  {\bibfnamefont {N.~M.}\ \bibnamefont {Rapidis}}, \bibinfo {author} {\bibfnamefont {M.}~\bibnamefont {Simanovskaia}}, \bibinfo {author} {\bibfnamefont {S.}~\bibnamefont {Singh}}, \bibinfo {author} {\bibfnamefont {D.~H.}\ \bibnamefont {Speller}}, \bibinfo {author} {\bibfnamefont {I.}~\bibnamefont {Urdinaran}}, \bibinfo {author} {\bibfnamefont {L.~R.}\ \bibnamefont {Vale}}, \bibinfo {author} {\bibfnamefont {E.~C.}\ \bibnamefont {van Assendelft}}, \bibinfo {author} {\bibfnamefont {K.}~\bibnamefont {van Bibber}},\ and\ \bibinfo {author} {\bibfnamefont {H.}~\bibnamefont {Wang}},\ }\bibfield  {title} {\bibinfo {title} {A quantum enhanced search for dark matter axions},\ }\href {https://doi.org/10.1038/s41586-021-03226-7} {\bibfield  {journal} {\bibinfo  {journal} {Nature}\ }\textbf {\bibinfo {volume} {590}},\ \bibinfo {pages} {238} (\bibinfo {year} {2021})},\ \bibinfo {note} {publisher: Nature Publishing Group}\BibitemShut {NoStop}%
\bibitem [{\citenamefont {Silva-Feaver}\ \emph {et~al.}(2017)\citenamefont {Silva-Feaver}, \citenamefont {Chaudhuri}, \citenamefont {Cho}, \citenamefont {Dawson}, \citenamefont {Graham}, \citenamefont {Irwin}, \citenamefont {Kuenstner}, \citenamefont {Li}, \citenamefont {Mardon}, \citenamefont {Moseley}, \citenamefont {Mule}, \citenamefont {Phipps}, \citenamefont {Rajendran}, \citenamefont {Steffen},\ and\ \citenamefont {Young}}]{silva-feaver_design_2017}%
  \BibitemOpen
  \bibfield  {author} {\bibinfo {author} {\bibfnamefont {M.}~\bibnamefont {Silva-Feaver}}, \bibinfo {author} {\bibfnamefont {S.}~\bibnamefont {Chaudhuri}}, \bibinfo {author} {\bibfnamefont {H.-M.}\ \bibnamefont {Cho}}, \bibinfo {author} {\bibfnamefont {C.}~\bibnamefont {Dawson}}, \bibinfo {author} {\bibfnamefont {P.}~\bibnamefont {Graham}}, \bibinfo {author} {\bibfnamefont {K.}~\bibnamefont {Irwin}}, \bibinfo {author} {\bibfnamefont {S.}~\bibnamefont {Kuenstner}}, \bibinfo {author} {\bibfnamefont {D.}~\bibnamefont {Li}}, \bibinfo {author} {\bibfnamefont {J.}~\bibnamefont {Mardon}}, \bibinfo {author} {\bibfnamefont {H.}~\bibnamefont {Moseley}}, \bibinfo {author} {\bibfnamefont {R.}~\bibnamefont {Mule}}, \bibinfo {author} {\bibfnamefont {A.}~\bibnamefont {Phipps}}, \bibinfo {author} {\bibfnamefont {S.}~\bibnamefont {Rajendran}}, \bibinfo {author} {\bibfnamefont {Z.}~\bibnamefont {Steffen}},\ and\ \bibinfo {author} {\bibfnamefont {B.}~\bibnamefont {Young}},\ }\bibfield  {title} {\bibinfo {title} {Design
  {Overview} of {DM} {Radio} {Pathfinder} {Experiment}},\ }\href {https://doi.org/10.1109/TASC.2016.2631425} {\bibfield  {journal} {\bibinfo  {journal} {IEEE Transactions on Applied Superconductivity}\ }\textbf {\bibinfo {volume} {27}},\ \bibinfo {pages} {1} (\bibinfo {year} {2017})}\BibitemShut {NoStop}%
\bibitem [{\citenamefont {Ouellet}\ \emph {et~al.}(2019)\citenamefont {Ouellet}, \citenamefont {Salemi}, \citenamefont {Foster}, \citenamefont {Henning}, \citenamefont {Bogorad}, \citenamefont {Conrad}, \citenamefont {Formaggio}, \citenamefont {Kahn}, \citenamefont {Minervini}, \citenamefont {Radovinsky}, \citenamefont {Rodd}, \citenamefont {Safdi}, \citenamefont {Thaler}, \citenamefont {Winklehner},\ and\ \citenamefont {Winslow}}]{ouellet_first_2019}%
  \BibitemOpen
  \bibfield  {author} {\bibinfo {author} {\bibfnamefont {J.~L.}\ \bibnamefont {Ouellet}}, \bibinfo {author} {\bibfnamefont {C.~P.}\ \bibnamefont {Salemi}}, \bibinfo {author} {\bibfnamefont {J.~W.}\ \bibnamefont {Foster}}, \bibinfo {author} {\bibfnamefont {R.}~\bibnamefont {Henning}}, \bibinfo {author} {\bibfnamefont {Z.}~\bibnamefont {Bogorad}}, \bibinfo {author} {\bibfnamefont {J.~M.}\ \bibnamefont {Conrad}}, \bibinfo {author} {\bibfnamefont {J.~A.}\ \bibnamefont {Formaggio}}, \bibinfo {author} {\bibfnamefont {Y.}~\bibnamefont {Kahn}}, \bibinfo {author} {\bibfnamefont {J.}~\bibnamefont {Minervini}}, \bibinfo {author} {\bibfnamefont {A.}~\bibnamefont {Radovinsky}}, \bibinfo {author} {\bibfnamefont {N.~L.}\ \bibnamefont {Rodd}}, \bibinfo {author} {\bibfnamefont {B.~R.}\ \bibnamefont {Safdi}}, \bibinfo {author} {\bibfnamefont {J.}~\bibnamefont {Thaler}}, \bibinfo {author} {\bibfnamefont {D.}~\bibnamefont {Winklehner}},\ and\ \bibinfo {author} {\bibfnamefont {L.}~\bibnamefont {Winslow}},\ }\bibfield  {title}
  {\bibinfo {title} {First {Results} from {ABRACADABRA}-10 cm: {A} {Search} for {Sub}-{\textmu}ev {Axion} {Dark} {Matter}},\ }\href {https://doi.org/10.1103/PhysRevLett.122.121802} {\bibfield  {journal} {\bibinfo  {journal} {Physical Review Letters}\ }\textbf {\bibinfo {volume} {122}},\ \bibinfo {pages} {121802} (\bibinfo {year} {2019})},\ \bibinfo {note} {publisher: American Physical Society}\BibitemShut {NoStop}%
\bibitem [{\citenamefont {Salemi}\ \emph {et~al.}(2021)\citenamefont {Salemi}, \citenamefont {Foster}, \citenamefont {Ouellet}, \citenamefont {Gavin}, \citenamefont {Pappas}, \citenamefont {Cheng}, \citenamefont {Richardson}, \citenamefont {Henning}, \citenamefont {Kahn}, \citenamefont {Nguyen}, \citenamefont {Rodd}, \citenamefont {Safdi},\ and\ \citenamefont {Winslow}}]{salemi_search_2021}%
  \BibitemOpen
  \bibfield  {author} {\bibinfo {author} {\bibfnamefont {C.~P.}\ \bibnamefont {Salemi}}, \bibinfo {author} {\bibfnamefont {J.~W.}\ \bibnamefont {Foster}}, \bibinfo {author} {\bibfnamefont {J.~L.}\ \bibnamefont {Ouellet}}, \bibinfo {author} {\bibfnamefont {A.}~\bibnamefont {Gavin}}, \bibinfo {author} {\bibfnamefont {K.~M.~W.}\ \bibnamefont {Pappas}}, \bibinfo {author} {\bibfnamefont {S.}~\bibnamefont {Cheng}}, \bibinfo {author} {\bibfnamefont {K.~A.}\ \bibnamefont {Richardson}}, \bibinfo {author} {\bibfnamefont {R.}~\bibnamefont {Henning}}, \bibinfo {author} {\bibfnamefont {Y.}~\bibnamefont {Kahn}}, \bibinfo {author} {\bibfnamefont {R.}~\bibnamefont {Nguyen}}, \bibinfo {author} {\bibfnamefont {N.~L.}\ \bibnamefont {Rodd}}, \bibinfo {author} {\bibfnamefont {B.~R.}\ \bibnamefont {Safdi}},\ and\ \bibinfo {author} {\bibfnamefont {L.}~\bibnamefont {Winslow}},\ }\bibfield  {title} {\bibinfo {title} {Search for {Low}-{Mass} {Axion} {Dark} {Matter} with {ABRACADABRA}-10 cm},\ }\href
  {https://doi.org/10.1103/PhysRevLett.127.081801} {\bibfield  {journal} {\bibinfo  {journal} {Physical Review Letters}\ }\textbf {\bibinfo {volume} {127}},\ \bibinfo {pages} {081801} (\bibinfo {year} {2021})}\BibitemShut {NoStop}%
\bibitem [{\citenamefont {Chen}\ \emph {et~al.}(2024{\natexlab{a}})\citenamefont {Chen}, \citenamefont {Fukuda}, \citenamefont {Inada}, \citenamefont {Moroi}, \citenamefont {Nitta},\ and\ \citenamefont {Sichanugrist}}]{chen_quantum_2024}%
  \BibitemOpen
  \bibfield  {author} {\bibinfo {author} {\bibfnamefont {S.}~\bibnamefont {Chen}}, \bibinfo {author} {\bibfnamefont {H.}~\bibnamefont {Fukuda}}, \bibinfo {author} {\bibfnamefont {T.}~\bibnamefont {Inada}}, \bibinfo {author} {\bibfnamefont {T.}~\bibnamefont {Moroi}}, \bibinfo {author} {\bibfnamefont {T.}~\bibnamefont {Nitta}},\ and\ \bibinfo {author} {\bibfnamefont {T.}~\bibnamefont {Sichanugrist}},\ }\bibfield  {title} {\bibinfo {title} {Quantum {Enhancement} in {Dark} {Matter} {Detection} with {Quantum} {Computation}},\ }\href {https://doi.org/10.1103/PhysRevLett.133.021801} {\bibfield  {journal} {\bibinfo  {journal} {Physical Review Letters}\ }\textbf {\bibinfo {volume} {133}},\ \bibinfo {pages} {021801} (\bibinfo {year} {2024}{\natexlab{a}})},\ \bibinfo {note} {publisher: American Physical Society}\BibitemShut {NoStop}%
\bibitem [{\citenamefont {Chen}\ \emph {et~al.}(2024{\natexlab{b}})\citenamefont {Chen}, \citenamefont {Fukuda}, \citenamefont {Inada}, \citenamefont {Moroi}, \citenamefont {Nitta},\ and\ \citenamefont {Sichanugrist}}]{chen_search_2024}%
  \BibitemOpen
  \bibfield  {author} {\bibinfo {author} {\bibfnamefont {S.}~\bibnamefont {Chen}}, \bibinfo {author} {\bibfnamefont {H.}~\bibnamefont {Fukuda}}, \bibinfo {author} {\bibfnamefont {T.}~\bibnamefont {Inada}}, \bibinfo {author} {\bibfnamefont {T.}~\bibnamefont {Moroi}}, \bibinfo {author} {\bibfnamefont {T.}~\bibnamefont {Nitta}},\ and\ \bibinfo {author} {\bibfnamefont {T.}~\bibnamefont {Sichanugrist}},\ }\href {https://doi.org/10.48550/arXiv.2407.19755} {\bibinfo {title} {Search for {QCD} axion dark matter with transmon qubits and quantum circuit}} (\bibinfo {year} {2024}{\natexlab{b}}),\ \bibinfo {note} {arXiv:2407.19755 [hep-ph]}\BibitemShut {NoStop}%
\bibitem [{\citenamefont {Bodas}\ \emph {et~al.}(2025)\citenamefont {Bodas}, \citenamefont {Ghosh},\ and\ \citenamefont {Harnik}}]{bodas2025speedupwavelikedarkmatter}%
  \BibitemOpen
  \bibfield  {author} {\bibinfo {author} {\bibfnamefont {A.}~\bibnamefont {Bodas}}, \bibinfo {author} {\bibfnamefont {S.}~\bibnamefont {Ghosh}},\ and\ \bibinfo {author} {\bibfnamefont {R.}~\bibnamefont {Harnik}},\ }\href {https://arxiv.org/abs/2510.11795} {\bibinfo {title} {On the speed-up of wave-like dark matter searches with entangled qubits}} (\bibinfo {year} {2025}),\ \bibinfo {note} {arXiv:2510.11795 [hep-ph]},\ \Eprint {https://arxiv.org/abs/2510.11795} {arXiv:2510.11795 [hep-ph]} \BibitemShut {NoStop}%
\bibitem [{\citenamefont {Chen}\ \emph {et~al.}(2025)\citenamefont {Chen}, \citenamefont {Fukuda}, \citenamefont {Iiyama}, \citenamefont {Mino}, \citenamefont {Moroi}, \citenamefont {Nakahara}, \citenamefont {Nitta},\ and\ \citenamefont {Sichanugrist}}]{chen_background_2025}%
  \BibitemOpen
  \bibfield  {author} {\bibinfo {author} {\bibfnamefont {S.}~\bibnamefont {Chen}}, \bibinfo {author} {\bibfnamefont {H.}~\bibnamefont {Fukuda}}, \bibinfo {author} {\bibfnamefont {Y.}~\bibnamefont {Iiyama}}, \bibinfo {author} {\bibfnamefont {Y.}~\bibnamefont {Mino}}, \bibinfo {author} {\bibfnamefont {T.}~\bibnamefont {Moroi}}, \bibinfo {author} {\bibfnamefont {M.}~\bibnamefont {Nakahara}}, \bibinfo {author} {\bibfnamefont {T.}~\bibnamefont {Nitta}},\ and\ \bibinfo {author} {\bibfnamefont {T.}~\bibnamefont {Sichanugrist}},\ }\href {https://doi.org/10.48550/arXiv.2510.01816} {\bibinfo {title} {Background {Suppression} in {Quantum} {Sensing} of {Dark} {Matter} via \${W}\$ {State} {Projection}}} (\bibinfo {year} {2025}),\ \bibinfo {note} {arXiv:2510.01816 [hep-ph]}\BibitemShut {NoStop}%
\bibitem [{\citenamefont {Fukuda}\ \emph {et~al.}(2025)\citenamefont {Fukuda}, \citenamefont {Moroi},\ and\ \citenamefont {Sichanugrist}}]{fukuda_quantum_2025}%
  \BibitemOpen
  \bibfield  {author} {\bibinfo {author} {\bibfnamefont {H.}~\bibnamefont {Fukuda}}, \bibinfo {author} {\bibfnamefont {T.}~\bibnamefont {Moroi}},\ and\ \bibinfo {author} {\bibfnamefont {T.}~\bibnamefont {Sichanugrist}},\ }\href {https://doi.org/10.48550/arXiv.2511.03253} {\bibinfo {title} {Quantum {Error} {Correction}-like {Noise} {Mitigation} for {Wave}-like {Dark} {Matter} {Searches} with {Quantum} {Sensors}}} (\bibinfo {year} {2025}),\ \bibinfo {note} {arXiv:2511.03253 [hep-ph]}\BibitemShut {NoStop}%
\bibitem [{\citenamefont {Yan}\ and\ \citenamefont {Feng}(2025)}]{Yan:2025etq}%
  \BibitemOpen
  \bibfield  {author} {\bibinfo {author} {\bibfnamefont {R.-Y.}\ \bibnamefont {Yan}}\ and\ \bibinfo {author} {\bibfnamefont {Z.-B.}\ \bibnamefont {Feng}},\ }\bibfield  {title} {\bibinfo {title} {{Efficient Entanglement Generation of Two Superconducting Qubits in a Circuit QED}},\ }\href {https://doi.org/10.1002/andp.202500128} {\bibfield  {journal} {\bibinfo  {journal} {Annalen Phys.}\ }\textbf {\bibinfo {volume} {537}},\ \bibinfo {pages} {e2500128} (\bibinfo {year} {2025})}\BibitemShut {NoStop}%
\bibitem [{\citenamefont {Rossatto}\ and\ \citenamefont {Villas-Boas}(2013)}]{PhysRevA.88.042324}%
  \BibitemOpen
  \bibfield  {author} {\bibinfo {author} {\bibfnamefont {D.~Z.}\ \bibnamefont {Rossatto}}\ and\ \bibinfo {author} {\bibfnamefont {C.~J.}\ \bibnamefont {Villas-Boas}},\ }\bibfield  {title} {\bibinfo {title} {Method for preparing two-atom entangled states in circuit qed and probing it via quantum nondemolition measurements},\ }\href {https://doi.org/10.1103/PhysRevA.88.042324} {\bibfield  {journal} {\bibinfo  {journal} {Phys. Rev. A}\ }\textbf {\bibinfo {volume} {88}},\ \bibinfo {pages} {042324} (\bibinfo {year} {2013})}\BibitemShut {NoStop}%
\bibitem [{\citenamefont {Wallraff}\ \emph {et~al.}(2004)\citenamefont {Wallraff} \emph {et~al.}}]{Wallraff_2004}%
  \BibitemOpen
  \bibfield  {author} {\bibinfo {author} {\bibfnamefont {A.}~\bibnamefont {Wallraff}} \emph {et~al.},\ }\bibfield  {title} {\bibinfo {title} {Strong coupling of a single photon to a superconducting qubit using circuit quantum electrodynamics},\ }\href@noop {} {\bibfield  {journal} {\bibinfo  {journal} {Nature}\ }\textbf {\bibinfo {volume} {431}},\ \bibinfo {pages} {162} (\bibinfo {year} {2004})}\BibitemShut {NoStop}%
\bibitem [{\citenamefont {Devoret}\ and\ \citenamefont {Schoelkopf}(2013)}]{Devoret_2013}%
  \BibitemOpen
  \bibfield  {author} {\bibinfo {author} {\bibfnamefont {M.~H.}\ \bibnamefont {Devoret}}\ and\ \bibinfo {author} {\bibfnamefont {R.~J.}\ \bibnamefont {Schoelkopf}},\ }\bibfield  {title} {\bibinfo {title} {Superconducting circuits for quantum information: An outlook},\ }\href@noop {} {\bibfield  {journal} {\bibinfo  {journal} {Science}\ }\textbf {\bibinfo {volume} {339}},\ \bibinfo {pages} {1169} (\bibinfo {year} {2013})}\BibitemShut {NoStop}%
\bibitem [{\citenamefont {Barends}\ \emph {et~al.}(2014)\citenamefont {Barends} \emph {et~al.}}]{Barends_2014}%
  \BibitemOpen
  \bibfield  {author} {\bibinfo {author} {\bibfnamefont {R.}~\bibnamefont {Barends}} \emph {et~al.},\ }\bibfield  {title} {\bibinfo {title} {Superconducting quantum circuits at the surface code threshold},\ }\href@noop {} {\bibfield  {journal} {\bibinfo  {journal} {Nature}\ }\textbf {\bibinfo {volume} {508}},\ \bibinfo {pages} {500} (\bibinfo {year} {2014})}\BibitemShut {NoStop}%
\bibitem [{\citenamefont {Chow}\ \emph {et~al.}(2013)\citenamefont {Chow} \emph {et~al.}}]{Chow_2013}%
  \BibitemOpen
  \bibfield  {author} {\bibinfo {author} {\bibfnamefont {J.~M.}\ \bibnamefont {Chow}} \emph {et~al.},\ }\bibfield  {title} {\bibinfo {title} {Microwave-activated conditional-phase gate for superconducting qubits},\ }\href@noop {} {\bibfield  {journal} {\bibinfo  {journal} {Physical Review Letters}\ }\textbf {\bibinfo {volume} {111}},\ \bibinfo {pages} {090502} (\bibinfo {year} {2013})}\BibitemShut {NoStop}%
\bibitem [{\citenamefont {Reagor}\ \emph {et~al.}(2016)\citenamefont {Reagor} \emph {et~al.}}]{Reagor_2016}%
  \BibitemOpen
  \bibfield  {author} {\bibinfo {author} {\bibfnamefont {M.}~\bibnamefont {Reagor}} \emph {et~al.},\ }\bibfield  {title} {\bibinfo {title} {Quantum memory with millisecond coherence in circuit qed},\ }\href@noop {} {\bibfield  {journal} {\bibinfo  {journal} {Physical Review B}\ }\textbf {\bibinfo {volume} {94}},\ \bibinfo {pages} {014506} (\bibinfo {year} {2016})}\BibitemShut {NoStop}%
\bibitem [{\citenamefont {Krantz}\ \emph {et~al.}(2019)\citenamefont {Krantz} \emph {et~al.}}]{Krantz_2019}%
  \BibitemOpen
  \bibfield  {author} {\bibinfo {author} {\bibfnamefont {P.}~\bibnamefont {Krantz}} \emph {et~al.},\ }\bibfield  {title} {\bibinfo {title} {A quantum engineer's guide to superconducting qubits},\ }\href@noop {} {\bibfield  {journal} {\bibinfo  {journal} {Applied Physics Reviews}\ }\textbf {\bibinfo {volume} {6}},\ \bibinfo {pages} {021318} (\bibinfo {year} {2019})}\BibitemShut {NoStop}%
\bibitem [{\citenamefont {Araki}\ and\ \citenamefont {Lieb}(1970)}]{Araki:1970}%
  \BibitemOpen
  \bibfield  {author} {\bibinfo {author} {\bibfnamefont {H.}~\bibnamefont {Araki}}\ and\ \bibinfo {author} {\bibfnamefont {E.}~\bibnamefont {Lieb}},\ }\bibfield  {title} {\bibinfo {title} {Entropy inequalities},\ }\href {https://doi.org/10.1007/BF01646092} {\bibfield  {journal} {\bibinfo  {journal} {Communications In Mathematical Physics}\ }\textbf {\bibinfo {volume} {18}},\ \bibinfo {pages} {160} (\bibinfo {year} {1970})}\BibitemShut {NoStop}%
\bibitem [{\citenamefont {Lieb}\ and\ \citenamefont {Ruskai}(1973)}]{Lieb:1973}%
  \BibitemOpen
  \bibfield  {author} {\bibinfo {author} {\bibfnamefont {E.}~\bibnamefont {Lieb}}\ and\ \bibinfo {author} {\bibfnamefont {M.}~\bibnamefont {Ruskai}},\ }\bibfield  {title} {\bibinfo {title} {Proof of the strong subadditivity of quantum-mechanical entropy},\ }\href {https://doi.org/10.1063/1.1666274} {\bibfield  {journal} {\bibinfo  {journal} {Journal of Mathematical Physics}\ }\textbf {\bibinfo {volume} {14}},\ \bibinfo {pages} {1938} (\bibinfo {year} {1973})}\BibitemShut {NoStop}%
\bibitem [{\citenamefont {Hayden}\ \emph {et~al.}(2013)\citenamefont {Hayden}, \citenamefont {Headrick},\ and\ \citenamefont {Maloney}}]{Hayden:2013}%
  \BibitemOpen
  \bibfield  {author} {\bibinfo {author} {\bibfnamefont {P.}~\bibnamefont {Hayden}}, \bibinfo {author} {\bibfnamefont {M.}~\bibnamefont {Headrick}},\ and\ \bibinfo {author} {\bibfnamefont {A.}~\bibnamefont {Maloney}},\ }\bibfield  {title} {\bibinfo {title} {Holographic mutual information is monogamous},\ }\href {https://doi.org/10.1103/PhysRevD.87.046003} {\bibfield  {journal} {\bibinfo  {journal} {Phys. Rev. D}\ }\textbf {\bibinfo {volume} {87}},\ \bibinfo {pages} {046003} (\bibinfo {year} {2013})},\ \Eprint {https://arxiv.org/abs/1107.2940} {arXiv:1107.2940 [hep-th]} \BibitemShut {NoStop}%
\bibitem [{\citenamefont {Ingleton}(1971)}]{Ingleton:1971}%
  \BibitemOpen
  \bibfield  {author} {\bibinfo {author} {\bibfnamefont {A.~W.}\ \bibnamefont {Ingleton}},\ }\bibfield  {title} {\bibinfo {title} {Representation of matroids},\ }in\ \href@noop {} {\emph {\bibinfo {booktitle} {Combinatorial Mathematics and its Applications, Proceedings of the Conference at the Mathematical Institute, Oxford, 1969}}}\ (\bibinfo {year} {1971})\ pp.\ \bibinfo {pages} {149--167},\ \bibinfo {note} {zbl: 0222.05025}\BibitemShut {NoStop}%
\bibitem [{\citenamefont {Fong}\ \emph {et~al.}(2008)\citenamefont {Fong}, \citenamefont {Shadbakht},\ and\ \citenamefont {Hassibi}}]{Fong2008Ingleton}%
  \BibitemOpen
  \bibfield  {author} {\bibinfo {author} {\bibfnamefont {D.}~\bibnamefont {Fong}}, \bibinfo {author} {\bibfnamefont {S.}~\bibnamefont {Shadbakht}},\ and\ \bibinfo {author} {\bibfnamefont {B.}~\bibnamefont {Hassibi}},\ }\bibfield  {title} {\bibinfo {title} {On the entropy region and the ingleton inequality},\ }in\ \href {https://scholar.lib.vt.edu/MTNS/Papers/164.pdf} {\emph {\bibinfo {booktitle} {Proceedings of the 19th International Symposium on Mathematical Theory of Networks and Systems (MTNS 2008)}}}\ (\bibinfo {year} {2008})\BibitemShut {NoStop}%
\bibitem [{\citenamefont {Dougherty}\ \emph {et~al.}(2009)\citenamefont {Dougherty}, \citenamefont {Freiling},\ and\ \citenamefont {Zeger}}]{Dougherty2009}%
  \BibitemOpen
  \bibfield  {author} {\bibinfo {author} {\bibfnamefont {R.}~\bibnamefont {Dougherty}}, \bibinfo {author} {\bibfnamefont {C.}~\bibnamefont {Freiling}},\ and\ \bibinfo {author} {\bibfnamefont {K.}~\bibnamefont {Zeger}},\ }\href@noop {} {\bibinfo {title} {Linear rank inequalities on five or more variables}} (\bibinfo {year} {2009}),\ \bibinfo {note} {arXiv:0910.0284 [cs.IT]},\ \Eprint {https://arxiv.org/abs/0910.0284} {arXiv:0910.0284 [cs.IT]} \BibitemShut {NoStop}%
\bibitem [{\citenamefont {Zhang}\ and\ \citenamefont {Yeung}(1998)}]{ZhangYeung1998}%
  \BibitemOpen
  \bibfield  {author} {\bibinfo {author} {\bibfnamefont {Z.}~\bibnamefont {Zhang}}\ and\ \bibinfo {author} {\bibfnamefont {R.}~\bibnamefont {Yeung}},\ }\bibfield  {title} {\bibinfo {title} {On characterization of entropy function via information inequalities},\ }\href {https://doi.org/10.1109/18.681320} {\bibfield  {journal} {\bibinfo  {journal} {IEEE Transactions on Information Theory}\ }\textbf {\bibinfo {volume} {44}},\ \bibinfo {pages} {1440} (\bibinfo {year} {1998})}\BibitemShut {NoStop}%
\bibitem [{\citenamefont {Mat{\'u}s}(2007)}]{Matus2007}%
  \BibitemOpen
  \bibfield  {author} {\bibinfo {author} {\bibfnamefont {F.}~\bibnamefont {Mat{\'u}s}},\ }\bibfield  {title} {\bibinfo {title} {Infinitely many information inequalities},\ }\href {https://api.semanticscholar.org/CorpusID:14959910} {\bibfield  {journal} {\bibinfo  {journal} {2007 IEEE International Symposium on Information Theory}\ ,\ \bibinfo {pages} {41}} (\bibinfo {year} {2007})}\BibitemShut {NoStop}%
\bibitem [{\citenamefont {Fuentes}\ \emph {et~al.}(2025)\citenamefont {Fuentes}, \citenamefont {Keeler}, \citenamefont {Munizzi},\ and\ \citenamefont {Pollack}}]{Fuentes:2025kwg}%
  \BibitemOpen
  \bibfield  {author} {\bibinfo {author} {\bibfnamefont {J.}~\bibnamefont {Fuentes}}, \bibinfo {author} {\bibfnamefont {C.}~\bibnamefont {Keeler}}, \bibinfo {author} {\bibfnamefont {W.}~\bibnamefont {Munizzi}},\ and\ \bibinfo {author} {\bibfnamefont {J.}~\bibnamefont {Pollack}},\ }\href@noop {} {\bibinfo {title} {{Monogamy of Mutual Information in Graph States}}} (\bibinfo {year} {2025}),\ \bibinfo {note} {arXiv:2511.19585 [quant-ph]},\ \Eprint {https://arxiv.org/abs/2511.19585} {arXiv:2511.19585 [quant-ph]} \BibitemShut {NoStop}%
\bibitem [{\citenamefont {Khumalo}\ \emph {et~al.}(2025)\citenamefont {Khumalo}, \citenamefont {Mehta}, \citenamefont {Munizzi},\ and\ \citenamefont {Narang}}]{Khumalo:2025xfv}%
  \BibitemOpen
  \bibfield  {author} {\bibinfo {author} {\bibfnamefont {N.}~\bibnamefont {Khumalo}}, \bibinfo {author} {\bibfnamefont {A.}~\bibnamefont {Mehta}}, \bibinfo {author} {\bibfnamefont {W.}~\bibnamefont {Munizzi}},\ and\ \bibinfo {author} {\bibfnamefont {P.}~\bibnamefont {Narang}},\ }\bibfield  {title} {\bibinfo {title} {{Navigating the Quantum Resource Landscape of Entropy Vector Space Using Machine Learning and Optimization}},\ }\href@noop {} {\bibfield  {journal} {\bibinfo  {journal} {arXiv preprint arXiv:2511.16724}\ } (\bibinfo {year} {2025})}\BibitemShut {NoStop}%
\bibitem [{\citenamefont {Linden}\ \emph {et~al.}(2013)\citenamefont {Linden}, \citenamefont {Mat{\'u}{\v{s}}}, \citenamefont {Ruskai},\ and\ \citenamefont {Winter}}]{Linden:2013kal}%
  \BibitemOpen
  \bibfield  {author} {\bibinfo {author} {\bibfnamefont {N.}~\bibnamefont {Linden}}, \bibinfo {author} {\bibfnamefont {F.}~\bibnamefont {Mat{\'u}{\v{s}}}}, \bibinfo {author} {\bibfnamefont {M.~B.}\ \bibnamefont {Ruskai}},\ and\ \bibinfo {author} {\bibfnamefont {A.}~\bibnamefont {Winter}},\ }\bibfield  {title} {\bibinfo {title} {{The Quantum Entropy Cone of Stabiliser States}},\ }\href {https://doi.org/10.4230/LIPIcs.TQC.2013.270} {\bibfield  {journal} {\bibinfo  {journal} {Leibniz Int. Proc. Inf.}\ }\textbf {\bibinfo {volume} {22}},\ \bibinfo {pages} {270} (\bibinfo {year} {2013})},\ \Eprint {https://arxiv.org/abs/1302.5453} {arXiv:1302.5453 [quant-ph]} \BibitemShut {NoStop}%
\bibitem [{\citenamefont {Bao}\ \emph {et~al.}(2015)\citenamefont {Bao}, \citenamefont {Nezami}, \citenamefont {Ooguri}, \citenamefont {Stoica}, \citenamefont {Sully},\ and\ \citenamefont {Walter}}]{Bao:2015bfa}%
  \BibitemOpen
  \bibfield  {author} {\bibinfo {author} {\bibfnamefont {N.}~\bibnamefont {Bao}}, \bibinfo {author} {\bibfnamefont {S.}~\bibnamefont {Nezami}}, \bibinfo {author} {\bibfnamefont {H.}~\bibnamefont {Ooguri}}, \bibinfo {author} {\bibfnamefont {B.}~\bibnamefont {Stoica}}, \bibinfo {author} {\bibfnamefont {J.}~\bibnamefont {Sully}},\ and\ \bibinfo {author} {\bibfnamefont {M.}~\bibnamefont {Walter}},\ }\bibfield  {title} {\bibinfo {title} {{The Holographic Entropy Cone}},\ }\href {https://doi.org/10.1007/JHEP09(2015)130} {\bibfield  {journal} {\bibinfo  {journal} {JHEP}\ }\textbf {\bibinfo {volume} {09}},\ \bibinfo {pages} {130}},\ \Eprint {https://arxiv.org/abs/1505.07839} {arXiv:1505.07839 [hep-th]} \BibitemShut {NoStop}%
\bibitem [{\citenamefont {Bao}\ \emph {et~al.}(2020)\citenamefont {Bao}, \citenamefont {Cheng}, \citenamefont {Hern{\'a}ndez-Cuenca},\ and\ \citenamefont {Su}}]{Bao:2020mqq}%
  \BibitemOpen
  \bibfield  {author} {\bibinfo {author} {\bibfnamefont {N.}~\bibnamefont {Bao}}, \bibinfo {author} {\bibfnamefont {N.}~\bibnamefont {Cheng}}, \bibinfo {author} {\bibfnamefont {S.}~\bibnamefont {Hern{\'a}ndez-Cuenca}},\ and\ \bibinfo {author} {\bibfnamefont {V.~P.}\ \bibnamefont {Su}},\ }\href@noop {} {\bibinfo {title} {{A Gap Between the Hypergraph and Stabilizer Entropy Cones}}} (\bibinfo {year} {2020}),\ \bibinfo {note} {arXiv:2006.16292 [quant-ph]},\ \Eprint {https://arxiv.org/abs/2006.16292} {arXiv:2006.16292 [quant-ph]} \BibitemShut {NoStop}%
\bibitem [{\citenamefont {Hern\'andez-Cuenca}\ \emph {et~al.}(2024)\citenamefont {Hern\'andez-Cuenca}, \citenamefont {Hubeny},\ and\ \citenamefont {Jia}}]{Hern_ndez_Cuenca_2024}%
  \BibitemOpen
  \bibfield  {author} {\bibinfo {author} {\bibfnamefont {S.}~\bibnamefont {Hern\'andez-Cuenca}}, \bibinfo {author} {\bibfnamefont {V.~E.}\ \bibnamefont {Hubeny}},\ and\ \bibinfo {author} {\bibfnamefont {H.~F.}\ \bibnamefont {Jia}},\ }\bibfield  {title} {\bibinfo {title} {Holographic entropy inequalities and multipartite entanglement},\ }\bibfield  {journal} {\bibinfo  {journal} {Journal of High Energy Physics}\ }\textbf {\bibinfo {volume} {2024}},\ \href {https://doi.org/10.1007/jhep08(2024)238} {10.1007/jhep08(2024)238} (\bibinfo {year} {2024})\BibitemShut {NoStop}%
\bibitem [{\citenamefont {Czech}\ and\ \citenamefont {Wang}(2023)}]{Czech:2022fzb}%
  \BibitemOpen
  \bibfield  {author} {\bibinfo {author} {\bibfnamefont {B.}~\bibnamefont {Czech}}\ and\ \bibinfo {author} {\bibfnamefont {Y.}~\bibnamefont {Wang}},\ }\bibfield  {title} {\bibinfo {title} {{A holographic inequality for N = 7 regions}},\ }\href {https://doi.org/10.1007/JHEP01(2023)101} {\bibfield  {journal} {\bibinfo  {journal} {JHEP}\ }\textbf {\bibinfo {volume} {01}},\ \bibinfo {pages} {101}},\ \Eprint {https://arxiv.org/abs/2209.10547} {arXiv:2209.10547 [hep-th]} \BibitemShut {NoStop}%
\bibitem [{\citenamefont {He}\ \emph {et~al.}(2020)\citenamefont {He}, \citenamefont {Hubeny},\ and\ \citenamefont {Rangamani}}]{He:2020xuo}%
  \BibitemOpen
  \bibfield  {author} {\bibinfo {author} {\bibfnamefont {T.}~\bibnamefont {He}}, \bibinfo {author} {\bibfnamefont {V.~E.}\ \bibnamefont {Hubeny}},\ and\ \bibinfo {author} {\bibfnamefont {M.}~\bibnamefont {Rangamani}},\ }\bibfield  {title} {\bibinfo {title} {{Superbalance of Holographic Entropy Inequalities}},\ }\href {https://doi.org/10.1007/JHEP07(2020)245} {\bibfield  {journal} {\bibinfo  {journal} {JHEP}\ }\textbf {\bibinfo {volume} {07}},\ \bibinfo {pages} {245}},\ \Eprint {https://arxiv.org/abs/2002.04558} {arXiv:2002.04558 [hep-th]} \BibitemShut {NoStop}%
\bibitem [{\citenamefont {Keeler}\ \emph {et~al.}(2022)\citenamefont {Keeler}, \citenamefont {Munizzi},\ and\ \citenamefont {Pollack}}]{Keeler:2022ajf}%
  \BibitemOpen
  \bibfield  {author} {\bibinfo {author} {\bibfnamefont {C.}~\bibnamefont {Keeler}}, \bibinfo {author} {\bibfnamefont {W.}~\bibnamefont {Munizzi}},\ and\ \bibinfo {author} {\bibfnamefont {J.}~\bibnamefont {Pollack}},\ }\bibfield  {title} {\bibinfo {title} {{Entropic lens on stabilizer states}},\ }\href {https://doi.org/10.1103/PhysRevA.106.062418} {\bibfield  {journal} {\bibinfo  {journal} {Phys. Rev. A}\ }\textbf {\bibinfo {volume} {106}},\ \bibinfo {pages} {062418} (\bibinfo {year} {2022})},\ \Eprint {https://arxiv.org/abs/2204.07593} {arXiv:2204.07593 [quant-ph]} \BibitemShut {NoStop}%
\bibitem [{\citenamefont {Keeler}\ \emph {et~al.}(2024)\citenamefont {Keeler}, \citenamefont {Munizzi},\ and\ \citenamefont {Pollack}}]{Keeler:2023xcx}%
  \BibitemOpen
  \bibfield  {author} {\bibinfo {author} {\bibfnamefont {C.}~\bibnamefont {Keeler}}, \bibinfo {author} {\bibfnamefont {W.}~\bibnamefont {Munizzi}},\ and\ \bibinfo {author} {\bibfnamefont {J.}~\bibnamefont {Pollack}},\ }\bibfield  {title} {\bibinfo {title} {{Clifford orbits from cayley graph quotients}},\ }\href {https://doi.org/10.26421/qic24.1-2-1} {\bibfield  {journal} {\bibinfo  {journal} {Quant. Inf. Comput.}\ }\textbf {\bibinfo {volume} {24}},\ \bibinfo {pages} {1} (\bibinfo {year} {2024})},\ \Eprint {https://arxiv.org/abs/2306.01043} {arXiv:2306.01043 [quant-ph]} \BibitemShut {NoStop}%
\bibitem [{\citenamefont {Keeler}\ \emph {et~al.}(2025)\citenamefont {Keeler}, \citenamefont {Munizzi},\ and\ \citenamefont {Pollack}}]{Keeler:2023shl}%
  \BibitemOpen
  \bibfield  {author} {\bibinfo {author} {\bibfnamefont {C.}~\bibnamefont {Keeler}}, \bibinfo {author} {\bibfnamefont {W.}~\bibnamefont {Munizzi}},\ and\ \bibinfo {author} {\bibfnamefont {J.}~\bibnamefont {Pollack}},\ }\bibfield  {title} {\bibinfo {title} {{Bounding entanglement entropy with Clifford double cosets}},\ }\href {https://doi.org/10.1103/hfhg-2z68} {\bibfield  {journal} {\bibinfo  {journal} {Phys. Rev. A}\ }\textbf {\bibinfo {volume} {112}},\ \bibinfo {pages} {022413} (\bibinfo {year} {2025})},\ \Eprint {https://arxiv.org/abs/2310.19874} {arXiv:2310.19874 [quant-ph]} \BibitemShut {NoStop}%
\bibitem [{\citenamefont {Munizzi}\ and\ \citenamefont {Schnitzer}(2024)}]{Munizzi:2023ihc}%
  \BibitemOpen
  \bibfield  {author} {\bibinfo {author} {\bibfnamefont {W.}~\bibnamefont {Munizzi}}\ and\ \bibinfo {author} {\bibfnamefont {H.~J.}\ \bibnamefont {Schnitzer}},\ }\bibfield  {title} {\bibinfo {title} {{Entropy cones and entanglement evolution for Dicke states}},\ }\href {https://doi.org/10.1103/PhysRevA.109.012405} {\bibfield  {journal} {\bibinfo  {journal} {Phys. Rev. A}\ }\textbf {\bibinfo {volume} {109}},\ \bibinfo {pages} {012405} (\bibinfo {year} {2024})},\ \Eprint {https://arxiv.org/abs/2306.13146} {arXiv:2306.13146 [quant-ph]} \BibitemShut {NoStop}%
\bibitem [{\citenamefont {Walter}\ \emph {et~al.}(2016)\citenamefont {Walter}, \citenamefont {Gross},\ and\ \citenamefont {Eisert}}]{Walter2016}%
  \BibitemOpen
  \bibfield  {author} {\bibinfo {author} {\bibfnamefont {M.}~\bibnamefont {Walter}}, \bibinfo {author} {\bibfnamefont {D.}~\bibnamefont {Gross}},\ and\ \bibinfo {author} {\bibfnamefont {J.}~\bibnamefont {Eisert}},\ }\href@noop {} {\bibinfo {title} {Multi-partite entanglement}} (\bibinfo {year} {2016}),\ \bibinfo {note} {arXiv:1612.02437 [quant-ph]},\ \Eprint {https://arxiv.org/abs/1612.02437} {arXiv:1612.02437 [quant-ph]} \BibitemShut {NoStop}%
\bibitem [{\citenamefont {Hein}\ \emph {et~al.}(2003)\citenamefont {Hein}, \citenamefont {Eisert},\ and\ \citenamefont {Briegel}}]{Hein2003}%
  \BibitemOpen
  \bibfield  {author} {\bibinfo {author} {\bibfnamefont {M.}~\bibnamefont {Hein}}, \bibinfo {author} {\bibfnamefont {J.}~\bibnamefont {Eisert}},\ and\ \bibinfo {author} {\bibfnamefont {H.~J.}\ \bibnamefont {Briegel}},\ }\bibfield  {title} {\bibinfo {title} {Multi-party entanglement in graph states},\ }\bibfield  {journal} {\bibinfo  {journal} {Phys. Rev. A 69, 062311 (2004)}\ }\href {https://doi.org/10.1103/PhysRevA.69.062311} {10.1103/PhysRevA.69.062311} (\bibinfo {year} {2003}),\ \Eprint {https://arxiv.org/abs/quant-ph/0307130} {arXiv:quant-ph/0307130 [quant-ph]} \BibitemShut {NoStop}%
\bibitem [{\citenamefont {Hein}\ \emph {et~al.}(2006)\citenamefont {Hein}, \citenamefont {D\"ur}, \citenamefont {Eisert}, \citenamefont {Raussendorf}, \citenamefont {Van~den Nest},\ and\ \citenamefont {Briegel}}]{HeinDurEisertBriegel2006}%
  \BibitemOpen
  \bibfield  {author} {\bibinfo {author} {\bibfnamefont {M.}~\bibnamefont {Hein}}, \bibinfo {author} {\bibfnamefont {W.}~\bibnamefont {D\"ur}}, \bibinfo {author} {\bibfnamefont {J.}~\bibnamefont {Eisert}}, \bibinfo {author} {\bibfnamefont {R.}~\bibnamefont {Raussendorf}}, \bibinfo {author} {\bibfnamefont {M.}~\bibnamefont {Van~den Nest}},\ and\ \bibinfo {author} {\bibfnamefont {H.~J.}\ \bibnamefont {Briegel}},\ }\bibfield  {title} {\bibinfo {title} {Entanglement in graph states and its applications},\ }in\ \href {https://doi.org/10.3254/978-1-61499-018-5-115} {\emph {\bibinfo {booktitle} {Quantum Computers, Algorithms and Chaos}}},\ \bibinfo {editor} {edited by\ \bibinfo {editor} {\bibfnamefont {G.}~\bibnamefont {Casati}}\ and\ \bibinfo {editor} {\bibfnamefont {B.~V.}\ \bibnamefont {Chirikov}}}\ (\bibinfo  {publisher} {International School of Physics ``Enrico Fermi''},\ \bibinfo {year} {2006})\ pp.\ \bibinfo {pages} {115--218},\ \Eprint {https://arxiv.org/abs/quant-ph/0602096} {arXiv:quant-ph/0602096}
  \BibitemShut {NoStop}%
\bibitem [{\citenamefont {Burchardt}\ and\ \citenamefont {Hahn}(2025)}]{Burchardt:2023odi}%
  \BibitemOpen
  \bibfield  {author} {\bibinfo {author} {\bibfnamefont {A.}~\bibnamefont {Burchardt}}\ and\ \bibinfo {author} {\bibfnamefont {F.}~\bibnamefont {Hahn}},\ }\bibfield  {title} {\bibinfo {title} {{The Foliage Partition: An Easy-to-Compute LC-Invariant for Graph States}},\ }\href {https://doi.org/10.22331/q-2025-04-24-1720} {\bibfield  {journal} {\bibinfo  {journal} {Quantum}\ }\textbf {\bibinfo {volume} {9}},\ \bibinfo {pages} {1720} (\bibinfo {year} {2025})},\ \Eprint {https://arxiv.org/abs/2305.07645} {arXiv:2305.07645 [quant-ph]} \BibitemShut {NoStop}%
\bibitem [{\citenamefont {Gong}\ \emph {et~al.}(2021)\citenamefont {Gong}, \citenamefont {Wang}, \citenamefont {Zha}, \citenamefont {Chen}, \citenamefont {Huang}, \citenamefont {Wu}, \citenamefont {Zhu}, \citenamefont {Zhao}, \citenamefont {Li}, \citenamefont {Guo} \emph {et~al.}}]{gong2021quantum}%
  \BibitemOpen
  \bibfield  {author} {\bibinfo {author} {\bibfnamefont {M.}~\bibnamefont {Gong}}, \bibinfo {author} {\bibfnamefont {S.}~\bibnamefont {Wang}}, \bibinfo {author} {\bibfnamefont {C.}~\bibnamefont {Zha}}, \bibinfo {author} {\bibfnamefont {M.-C.}\ \bibnamefont {Chen}}, \bibinfo {author} {\bibfnamefont {H.-L.}\ \bibnamefont {Huang}}, \bibinfo {author} {\bibfnamefont {Y.}~\bibnamefont {Wu}}, \bibinfo {author} {\bibfnamefont {Q.}~\bibnamefont {Zhu}}, \bibinfo {author} {\bibfnamefont {Y.}~\bibnamefont {Zhao}}, \bibinfo {author} {\bibfnamefont {S.}~\bibnamefont {Li}}, \bibinfo {author} {\bibfnamefont {S.}~\bibnamefont {Guo}}, \emph {et~al.},\ }\bibfield  {title} {\bibinfo {title} {Quantum walks on a programmable two-dimensional 62-qubit superconducting processor},\ }\href@noop {} {\bibfield  {journal} {\bibinfo  {journal} {Science}\ }\textbf {\bibinfo {volume} {372}},\ \bibinfo {pages} {948} (\bibinfo {year} {2021})}\BibitemShut {NoStop}%
\bibitem [{\citenamefont {Zhong}\ \emph {et~al.}(2019)\citenamefont {Zhong}, \citenamefont {Chang}, \citenamefont {Satzinger}, \citenamefont {Chou}, \citenamefont {Bienfait}, \citenamefont {Conner}, \citenamefont {Dumur}, \citenamefont {Grebel}, \citenamefont {Peairs}, \citenamefont {Povey}, \citenamefont {Schuster},\ and\ \citenamefont {Cleland}}]{Zhong2019}%
  \BibitemOpen
  \bibfield  {author} {\bibinfo {author} {\bibfnamefont {Y.~P.}\ \bibnamefont {Zhong}}, \bibinfo {author} {\bibfnamefont {H.-S.}\ \bibnamefont {Chang}}, \bibinfo {author} {\bibfnamefont {K.~J.}\ \bibnamefont {Satzinger}}, \bibinfo {author} {\bibfnamefont {M.-H.}\ \bibnamefont {Chou}}, \bibinfo {author} {\bibfnamefont {A.}~\bibnamefont {Bienfait}}, \bibinfo {author} {\bibfnamefont {C.~R.}\ \bibnamefont {Conner}}, \bibinfo {author} {\bibfnamefont {{\'E}.}~\bibnamefont {Dumur}}, \bibinfo {author} {\bibfnamefont {J.}~\bibnamefont {Grebel}}, \bibinfo {author} {\bibfnamefont {G.~A.}\ \bibnamefont {Peairs}}, \bibinfo {author} {\bibfnamefont {R.~G.}\ \bibnamefont {Povey}}, \bibinfo {author} {\bibfnamefont {D.~I.}\ \bibnamefont {Schuster}},\ and\ \bibinfo {author} {\bibfnamefont {A.~N.}\ \bibnamefont {Cleland}},\ }\bibfield  {title} {\bibinfo {title} {Violating {B}ell's inequality with remotely connected superconducting qubits},\ }\href {https://doi.org/10.1038/s41567-019-0507-7} {\bibfield  {journal} {\bibinfo
  {journal} {Nature Physics}\ }\textbf {\bibinfo {volume} {15}},\ \bibinfo {pages} {741} (\bibinfo {year} {2019})}\BibitemShut {NoStop}%
\bibitem [{\citenamefont {Zhong}\ \emph {et~al.}(2021)\citenamefont {Zhong}, \citenamefont {Chang}, \citenamefont {Bienfait}, \citenamefont {Dumur}, \citenamefont {Chou}, \citenamefont {Conner}, \citenamefont {Grebel}, \citenamefont {Povey}, \citenamefont {Yan}, \citenamefont {Schuster},\ and\ \citenamefont {Cleland}}]{Zhong2021}%
  \BibitemOpen
  \bibfield  {author} {\bibinfo {author} {\bibfnamefont {Y.}~\bibnamefont {Zhong}}, \bibinfo {author} {\bibfnamefont {H.-S.}\ \bibnamefont {Chang}}, \bibinfo {author} {\bibfnamefont {A.}~\bibnamefont {Bienfait}}, \bibinfo {author} {\bibfnamefont {{\'E}.}~\bibnamefont {Dumur}}, \bibinfo {author} {\bibfnamefont {M.-H.}\ \bibnamefont {Chou}}, \bibinfo {author} {\bibfnamefont {C.~R.}\ \bibnamefont {Conner}}, \bibinfo {author} {\bibfnamefont {J.}~\bibnamefont {Grebel}}, \bibinfo {author} {\bibfnamefont {R.~G.}\ \bibnamefont {Povey}}, \bibinfo {author} {\bibfnamefont {H.}~\bibnamefont {Yan}}, \bibinfo {author} {\bibfnamefont {D.~I.}\ \bibnamefont {Schuster}},\ and\ \bibinfo {author} {\bibfnamefont {A.~N.}\ \bibnamefont {Cleland}},\ }\bibfield  {title} {\bibinfo {title} {Deterministic multi-qubit entanglement in a quantum network},\ }\href {https://doi.org/10.1038/s41586-021-03288-7} {\bibfield  {journal} {\bibinfo  {journal} {Nature}\ }\textbf {\bibinfo {volume} {590}},\ \bibinfo {pages} {571} (\bibinfo {year}
  {2021})}\BibitemShut {NoStop}%
\bibitem [{\citenamefont {Kurpiers}\ \emph {et~al.}(2018)\citenamefont {Kurpiers}, \citenamefont {Magnard}, \citenamefont {Walter}, \citenamefont {Royer}, \citenamefont {Pechal}, \citenamefont {Heinsoo}, \citenamefont {Salath{\'e}}, \citenamefont {Akin}, \citenamefont {Storz}, \citenamefont {Besse}, \citenamefont {Gasparinetti}, \citenamefont {Blais},\ and\ \citenamefont {Wallraff}}]{Kurpiers2018}%
  \BibitemOpen
  \bibfield  {author} {\bibinfo {author} {\bibfnamefont {P.}~\bibnamefont {Kurpiers}}, \bibinfo {author} {\bibfnamefont {P.}~\bibnamefont {Magnard}}, \bibinfo {author} {\bibfnamefont {T.}~\bibnamefont {Walter}}, \bibinfo {author} {\bibfnamefont {B.}~\bibnamefont {Royer}}, \bibinfo {author} {\bibfnamefont {M.}~\bibnamefont {Pechal}}, \bibinfo {author} {\bibfnamefont {J.}~\bibnamefont {Heinsoo}}, \bibinfo {author} {\bibfnamefont {Y.}~\bibnamefont {Salath{\'e}}}, \bibinfo {author} {\bibfnamefont {A.}~\bibnamefont {Akin}}, \bibinfo {author} {\bibfnamefont {S.}~\bibnamefont {Storz}}, \bibinfo {author} {\bibfnamefont {J.-C.}\ \bibnamefont {Besse}}, \bibinfo {author} {\bibfnamefont {S.}~\bibnamefont {Gasparinetti}}, \bibinfo {author} {\bibfnamefont {A.}~\bibnamefont {Blais}},\ and\ \bibinfo {author} {\bibfnamefont {A.}~\bibnamefont {Wallraff}},\ }\bibfield  {title} {\bibinfo {title} {Deterministic quantum state transfer and remote entanglement using microwave photons},\ }\href
  {https://doi.org/10.1038/s41586-018-0195-y} {\bibfield  {journal} {\bibinfo  {journal} {Nature}\ }\textbf {\bibinfo {volume} {558}},\ \bibinfo {pages} {264} (\bibinfo {year} {2018})}\BibitemShut {NoStop}%
\bibitem [{\citenamefont {Kannan}\ \emph {et~al.}(2023)\citenamefont {Kannan}, \citenamefont {Almanakly}, \citenamefont {Sung}, \citenamefont {Di~Paolo}, \citenamefont {Rower}, \citenamefont {Braum{\"u}ller}, \citenamefont {Melville}, \citenamefont {Niedzielski}, \citenamefont {Karamlou}, \citenamefont {Serniak}, \citenamefont {Veps{\"a}l{\"a}inen}, \citenamefont {Schwartz}, \citenamefont {Yoder}, \citenamefont {Winik}, \citenamefont {Wang}, \citenamefont {Orlando}, \citenamefont {Gustavsson}, \citenamefont {Grover},\ and\ \citenamefont {Oliver}}]{kannan2023}%
  \BibitemOpen
  \bibfield  {author} {\bibinfo {author} {\bibfnamefont {B.}~\bibnamefont {Kannan}}, \bibinfo {author} {\bibfnamefont {A.}~\bibnamefont {Almanakly}}, \bibinfo {author} {\bibfnamefont {Y.}~\bibnamefont {Sung}}, \bibinfo {author} {\bibfnamefont {A.}~\bibnamefont {Di~Paolo}}, \bibinfo {author} {\bibfnamefont {D.~A.}\ \bibnamefont {Rower}}, \bibinfo {author} {\bibfnamefont {J.}~\bibnamefont {Braum{\"u}ller}}, \bibinfo {author} {\bibfnamefont {A.}~\bibnamefont {Melville}}, \bibinfo {author} {\bibfnamefont {B.~M.}\ \bibnamefont {Niedzielski}}, \bibinfo {author} {\bibfnamefont {A.}~\bibnamefont {Karamlou}}, \bibinfo {author} {\bibfnamefont {K.}~\bibnamefont {Serniak}}, \bibinfo {author} {\bibfnamefont {A.}~\bibnamefont {Veps{\"a}l{\"a}inen}}, \bibinfo {author} {\bibfnamefont {M.~E.}\ \bibnamefont {Schwartz}}, \bibinfo {author} {\bibfnamefont {J.~L.}\ \bibnamefont {Yoder}}, \bibinfo {author} {\bibfnamefont {R.}~\bibnamefont {Winik}}, \bibinfo {author} {\bibfnamefont {J.~I.-J.}\ \bibnamefont {Wang}}, \bibinfo {author}
  {\bibfnamefont {T.~P.}\ \bibnamefont {Orlando}}, \bibinfo {author} {\bibfnamefont {S.}~\bibnamefont {Gustavsson}}, \bibinfo {author} {\bibfnamefont {J.~A.}\ \bibnamefont {Grover}},\ and\ \bibinfo {author} {\bibfnamefont {W.~D.}\ \bibnamefont {Oliver}},\ }\bibfield  {title} {\bibinfo {title} {On-demand directional microwave photon emission using waveguide quantum electrodynamics},\ }\href {https://doi.org/10.1038/s41567-022-01869-5} {\bibfield  {journal} {\bibinfo  {journal} {Nature Physics}\ }\textbf {\bibinfo {volume} {19}},\ \bibinfo {pages} {394} (\bibinfo {year} {2023})}\BibitemShut {NoStop}%
\bibitem [{\citenamefont {Negrin}\ \emph {et~al.}(2025)\citenamefont {Negrin}, \citenamefont {Dirnegger}, \citenamefont {Munizzi}, \citenamefont {Talukdar},\ and\ \citenamefont {Narang}}]{Negrin:2024tyj}%
  \BibitemOpen
  \bibfield  {author} {\bibinfo {author} {\bibfnamefont {R.}~\bibnamefont {Negrin}}, \bibinfo {author} {\bibfnamefont {N.}~\bibnamefont {Dirnegger}}, \bibinfo {author} {\bibfnamefont {W.}~\bibnamefont {Munizzi}}, \bibinfo {author} {\bibfnamefont {J.}~\bibnamefont {Talukdar}},\ and\ \bibinfo {author} {\bibfnamefont {P.}~\bibnamefont {Narang}},\ }\bibfield  {title} {\bibinfo {title} {{Efficient multiparty entanglement distribution with DODAG-X protocol}},\ }\href {https://doi.org/10.1117/12.3041486} {\bibfield  {journal} {\bibinfo  {journal} {Proc. SPIE Int. Soc. Opt. Eng.}\ }\textbf {\bibinfo {volume} {13391}},\ \bibinfo {pages} {1339109} (\bibinfo {year} {2025})},\ \Eprint {https://arxiv.org/abs/2408.07118} {arXiv:2408.07118 [quant-ph]} \BibitemShut {NoStop}%
\bibitem [{\citenamefont {Almanakly}\ \emph {et~al.}(2025)\citenamefont {Almanakly}, \citenamefont {Yankelevich}, \citenamefont {Hays}, \citenamefont {Kannan}, \citenamefont {Assouly}, \citenamefont {Greene}, \citenamefont {Gingras}, \citenamefont {Niedzielski}, \citenamefont {Stickler}, \citenamefont {Schwartz}, \citenamefont {Serniak}, \citenamefont {Wang}, \citenamefont {Orlando}, \citenamefont {Gustavsson}, \citenamefont {Grover},\ and\ \citenamefont {Oliver}}]{Almanakly2025}%
  \BibitemOpen
  \bibfield  {author} {\bibinfo {author} {\bibfnamefont {A.}~\bibnamefont {Almanakly}}, \bibinfo {author} {\bibfnamefont {B.}~\bibnamefont {Yankelevich}}, \bibinfo {author} {\bibfnamefont {M.}~\bibnamefont {Hays}}, \bibinfo {author} {\bibfnamefont {B.}~\bibnamefont {Kannan}}, \bibinfo {author} {\bibfnamefont {R.}~\bibnamefont {Assouly}}, \bibinfo {author} {\bibfnamefont {A.}~\bibnamefont {Greene}}, \bibinfo {author} {\bibfnamefont {M.}~\bibnamefont {Gingras}}, \bibinfo {author} {\bibfnamefont {B.~M.}\ \bibnamefont {Niedzielski}}, \bibinfo {author} {\bibfnamefont {H.}~\bibnamefont {Stickler}}, \bibinfo {author} {\bibfnamefont {M.~E.}\ \bibnamefont {Schwartz}}, \bibinfo {author} {\bibfnamefont {K.}~\bibnamefont {Serniak}}, \bibinfo {author} {\bibfnamefont {J.~I.-J.}\ \bibnamefont {Wang}}, \bibinfo {author} {\bibfnamefont {T.~P.}\ \bibnamefont {Orlando}}, \bibinfo {author} {\bibfnamefont {S.}~\bibnamefont {Gustavsson}}, \bibinfo {author} {\bibfnamefont {J.~A.}\ \bibnamefont {Grover}},\ and\ \bibinfo {author}
  {\bibfnamefont {W.~D.}\ \bibnamefont {Oliver}},\ }\bibfield  {title} {\bibinfo {title} {Deterministic remote entanglement using a chiral quantum interconnect},\ }\href {https://doi.org/10.1038/s41567-025-02811-1} {\bibfield  {journal} {\bibinfo  {journal} {Nature Physics}\ }\textbf {\bibinfo {volume} {21}},\ \bibinfo {pages} {825} (\bibinfo {year} {2025})}\BibitemShut {NoStop}%
\bibitem [{\citenamefont {Lalumi\`ere}\ \emph {et~al.}(2013)\citenamefont {Lalumi\`ere}, \citenamefont {Sanders}, \citenamefont {van Loo}, \citenamefont {Fedorov}, \citenamefont {Wallraff},\ and\ \citenamefont {Blais}}]{Lalumiere2013}%
  \BibitemOpen
  \bibfield  {author} {\bibinfo {author} {\bibfnamefont {K.}~\bibnamefont {Lalumi\`ere}}, \bibinfo {author} {\bibfnamefont {B.~C.}\ \bibnamefont {Sanders}}, \bibinfo {author} {\bibfnamefont {A.~F.}\ \bibnamefont {van Loo}}, \bibinfo {author} {\bibfnamefont {A.}~\bibnamefont {Fedorov}}, \bibinfo {author} {\bibfnamefont {A.}~\bibnamefont {Wallraff}},\ and\ \bibinfo {author} {\bibfnamefont {A.}~\bibnamefont {Blais}},\ }\bibfield  {title} {\bibinfo {title} {Input-output theory for waveguide {QED} with an ensemble of inhomogeneous atoms},\ }\href {https://doi.org/10.1103/PhysRevA.88.043806} {\bibfield  {journal} {\bibinfo  {journal} {Phys. Rev. A}\ }\textbf {\bibinfo {volume} {88}},\ \bibinfo {pages} {043806} (\bibinfo {year} {2013})}\BibitemShut {NoStop}%
\bibitem [{\citenamefont {Pichler}\ \emph {et~al.}(2015)\citenamefont {Pichler}, \citenamefont {Ramos}, \citenamefont {Daley},\ and\ \citenamefont {Zoller}}]{Pichler2015}%
  \BibitemOpen
  \bibfield  {author} {\bibinfo {author} {\bibfnamefont {H.}~\bibnamefont {Pichler}}, \bibinfo {author} {\bibfnamefont {T.}~\bibnamefont {Ramos}}, \bibinfo {author} {\bibfnamefont {A.~J.}\ \bibnamefont {Daley}},\ and\ \bibinfo {author} {\bibfnamefont {P.}~\bibnamefont {Zoller}},\ }\bibfield  {title} {\bibinfo {title} {Quantum optics of chiral spin networks},\ }\href {https://doi.org/10.1103/PhysRevA.91.042116} {\bibfield  {journal} {\bibinfo  {journal} {Phys. Rev. A}\ }\textbf {\bibinfo {volume} {91}},\ \bibinfo {pages} {042116} (\bibinfo {year} {2015})}\BibitemShut {NoStop}%
\bibitem [{\citenamefont {Lodahl}\ \emph {et~al.}(2017)\citenamefont {Lodahl}, \citenamefont {Mahmoodian}, \citenamefont {Stobbe}, \citenamefont {Rauschenbeutel}, \citenamefont {Schneeweiss}, \citenamefont {Volz}, \citenamefont {Pichler},\ and\ \citenamefont {Zoller}}]{Lodahl2017}%
  \BibitemOpen
  \bibfield  {author} {\bibinfo {author} {\bibfnamefont {P.}~\bibnamefont {Lodahl}}, \bibinfo {author} {\bibfnamefont {S.}~\bibnamefont {Mahmoodian}}, \bibinfo {author} {\bibfnamefont {S.}~\bibnamefont {Stobbe}}, \bibinfo {author} {\bibfnamefont {A.}~\bibnamefont {Rauschenbeutel}}, \bibinfo {author} {\bibfnamefont {P.}~\bibnamefont {Schneeweiss}}, \bibinfo {author} {\bibfnamefont {J.}~\bibnamefont {Volz}}, \bibinfo {author} {\bibfnamefont {H.}~\bibnamefont {Pichler}},\ and\ \bibinfo {author} {\bibfnamefont {P.}~\bibnamefont {Zoller}},\ }\bibfield  {title} {\bibinfo {title} {Chiral quantum optics},\ }\href {https://doi.org/10.1038/nature21037} {\bibfield  {journal} {\bibinfo  {journal} {Nature}\ }\textbf {\bibinfo {volume} {541}},\ \bibinfo {pages} {473} (\bibinfo {year} {2017})}\BibitemShut {NoStop}%
\bibitem [{\citenamefont {Hughes}\ \emph {et~al.}(2025)\citenamefont {Hughes}, \citenamefont {Munizzi},\ and\ \citenamefont {Narang}}]{Hughes:2025jwa}%
  \BibitemOpen
  \bibfield  {author} {\bibinfo {author} {\bibfnamefont {E.}~\bibnamefont {Hughes}}, \bibinfo {author} {\bibfnamefont {W.}~\bibnamefont {Munizzi}},\ and\ \bibinfo {author} {\bibfnamefont {P.}~\bibnamefont {Narang}},\ }\href@noop {} {\bibinfo {title} {{Improved Simulation of Asynchronous Entanglement Distribution in Noisy Quantum Networks}}} (\bibinfo {year} {2025}),\ \bibinfo {note} {arXiv:2507.22992 [quant-ph]},\ \Eprint {https://arxiv.org/abs/2507.22992} {arXiv:2507.22992 [quant-ph]} \BibitemShut {NoStop}%
\bibitem [{\citenamefont {DiAdamo}\ \emph {et~al.}(2023)\citenamefont {DiAdamo}, \citenamefont {Mandil}, \citenamefont {Qi}, \citenamefont {Miller}, \citenamefont {Kompella},\ and\ \citenamefont {Shabani}}]{DiAdamo:2023tqg}%
  \BibitemOpen
  \bibfield  {author} {\bibinfo {author} {\bibfnamefont {S.}~\bibnamefont {DiAdamo}}, \bibinfo {author} {\bibfnamefont {R.}~\bibnamefont {Mandil}}, \bibinfo {author} {\bibfnamefont {B.}~\bibnamefont {Qi}}, \bibinfo {author} {\bibfnamefont {G.}~\bibnamefont {Miller}}, \bibinfo {author} {\bibfnamefont {R.}~\bibnamefont {Kompella}},\ and\ \bibinfo {author} {\bibfnamefont {A.}~\bibnamefont {Shabani}},\ }\bibfield  {title} {\bibinfo {title} {{Packet-switching in quantum communication: opportunities and challenges}},\ }\href {https://doi.org/10.1117/12.2655710} {\bibfield  {journal} {\bibinfo  {journal} {Proc. SPIE Int. Soc. Opt. Eng.}\ }\textbf {\bibinfo {volume} {12446}},\ \bibinfo {pages} {124460O} (\bibinfo {year} {2023})}\BibitemShut {NoStop}%
\bibitem [{\citenamefont {von Neumann}(1927)}]{vonNeumann1927}%
  \BibitemOpen
  \bibfield  {author} {\bibinfo {author} {\bibfnamefont {J.}~\bibnamefont {von Neumann}},\ }\bibfield  {title} {\bibinfo {title} {Thermodynamik quantenmechanischer gesamtheiten},\ }\href@noop {} {\bibfield  {journal} {\bibinfo  {journal} {Nachrichten von der Gesellschaft der Wissenschaften zu G\"ottingen, Mathematisch-Physikalische Klasse}\ ,\ \bibinfo {pages} {273}} (\bibinfo {year} {1927})}\BibitemShut {NoStop}%
\bibitem [{\citenamefont {Nielsen}\ and\ \citenamefont {Chuang}(2010)}]{nielsen2010quantum}%
  \BibitemOpen
  \bibfield  {author} {\bibinfo {author} {\bibfnamefont {M.~A.}\ \bibnamefont {Nielsen}}\ and\ \bibinfo {author} {\bibfnamefont {I.~L.}\ \bibnamefont {Chuang}},\ }\href@noop {} {\emph {\bibinfo {title} {Quantum Computation and Quantum Information}}}\ (\bibinfo  {publisher} {Cambridge University Press},\ \bibinfo {year} {2010})\BibitemShut {NoStop}%
\bibitem [{\citenamefont {R{\'e}nyi}(1961)}]{Rnyi1961OnMO}%
  \BibitemOpen
  \bibfield  {author} {\bibinfo {author} {\bibfnamefont {A.}~\bibnamefont {R{\'e}nyi}},\ }\bibfield  {title} {\bibinfo {title} {On measures of entropy and information},\ }in\ \href {https://api.semanticscholar.org/CorpusID:123056571} {\emph {\bibinfo {booktitle} {Proceedings of the Fourth Berkeley Symposium on Mathematical Statistics and Probability, Volume 1: Contributions to the Theory of Statistics}}}\ (\bibinfo  {publisher} {University of California Press},\ \bibinfo {year} {1961})\ pp.\ \bibinfo {pages} {547--561}\BibitemShut {NoStop}%
\bibitem [{\citenamefont {Gottesman}(1998)}]{gottesman1998heisenberg}%
  \BibitemOpen
  \bibfield  {author} {\bibinfo {author} {\bibfnamefont {D.}~\bibnamefont {Gottesman}},\ }\bibfield  {title} {\bibinfo {title} {The heisenberg representation of quantum computers},\ }\href@noop {} {\bibfield  {journal} {\bibinfo  {journal} {arXiv preprint quant-ph/9807006}\ } (\bibinfo {year} {1998})}\BibitemShut {NoStop}%
\bibitem [{\citenamefont {Aaronson}\ and\ \citenamefont {Gottesman}(2004)}]{aaronson2004improved}%
  \BibitemOpen
  \bibfield  {author} {\bibinfo {author} {\bibfnamefont {S.}~\bibnamefont {Aaronson}}\ and\ \bibinfo {author} {\bibfnamefont {D.}~\bibnamefont {Gottesman}},\ }\bibfield  {title} {\bibinfo {title} {Improved simulation of stabilizer circuits},\ }\href@noop {} {\bibfield  {journal} {\bibinfo  {journal} {Physical Review A}\ }\textbf {\bibinfo {volume} {70}},\ \bibinfo {pages} {052328} (\bibinfo {year} {2004})}\BibitemShut {NoStop}%
\bibitem [{\citenamefont {Bravyi}\ and\ \citenamefont {Kitaev}(2005)}]{bravyi2005universal}%
  \BibitemOpen
  \bibfield  {author} {\bibinfo {author} {\bibfnamefont {S.}~\bibnamefont {Bravyi}}\ and\ \bibinfo {author} {\bibfnamefont {A.}~\bibnamefont {Kitaev}},\ }\bibfield  {title} {\bibinfo {title} {Universal quantum computation with ideal clifford gates and noisy ancillas},\ }\href@noop {} {\bibfield  {journal} {\bibinfo  {journal} {Physical Review A}\ }\textbf {\bibinfo {volume} {71}},\ \bibinfo {pages} {022316} (\bibinfo {year} {2005})}\BibitemShut {NoStop}%
\bibitem [{\citenamefont {Howard}\ and\ \citenamefont {Campbell}(2017)}]{howard2017application}%
  \BibitemOpen
  \bibfield  {author} {\bibinfo {author} {\bibfnamefont {M.}~\bibnamefont {Howard}}\ and\ \bibinfo {author} {\bibfnamefont {E.~T.}\ \bibnamefont {Campbell}},\ }\bibfield  {title} {\bibinfo {title} {Application of a resource theory for magic states to fault-tolerant quantum computing},\ }\href@noop {} {\bibfield  {journal} {\bibinfo  {journal} {Physical Review Letters}\ }\textbf {\bibinfo {volume} {118}},\ \bibinfo {pages} {090501} (\bibinfo {year} {2017})}\BibitemShut {NoStop}%
\bibitem [{\citenamefont {Leone}\ \emph {et~al.}(2022)\citenamefont {Leone}, \citenamefont {Oliviero},\ and\ \citenamefont {Hamma}}]{Leone:2021rzd}%
  \BibitemOpen
  \bibfield  {author} {\bibinfo {author} {\bibfnamefont {L.}~\bibnamefont {Leone}}, \bibinfo {author} {\bibfnamefont {S.~F.~E.}\ \bibnamefont {Oliviero}},\ and\ \bibinfo {author} {\bibfnamefont {A.}~\bibnamefont {Hamma}},\ }\bibfield  {title} {\bibinfo {title} {{Stabilizer R{\'e}nyi Entropy}},\ }\href {https://doi.org/10.1103/PhysRevLett.128.050402} {\bibfield  {journal} {\bibinfo  {journal} {Phys. Rev. Lett.}\ }\textbf {\bibinfo {volume} {128}},\ \bibinfo {pages} {050402} (\bibinfo {year} {2022})},\ \Eprint {https://arxiv.org/abs/2106.12587} {arXiv:2106.12587 [quant-ph]} \BibitemShut {NoStop}%
\bibitem [{\citenamefont {Cao}\ \emph {et~al.}(2025{\natexlab{a}})\citenamefont {Cao}, \citenamefont {Cheng}, \citenamefont {Hamma}, \citenamefont {Leone}, \citenamefont {Munizzi},\ and\ \citenamefont {Oliviero}}]{cao2025gravitational}%
  \BibitemOpen
  \bibfield  {author} {\bibinfo {author} {\bibfnamefont {C.}~\bibnamefont {Cao}}, \bibinfo {author} {\bibfnamefont {G.}~\bibnamefont {Cheng}}, \bibinfo {author} {\bibfnamefont {A.}~\bibnamefont {Hamma}}, \bibinfo {author} {\bibfnamefont {L.}~\bibnamefont {Leone}}, \bibinfo {author} {\bibfnamefont {W.}~\bibnamefont {Munizzi}},\ and\ \bibinfo {author} {\bibfnamefont {S.~F.}\ \bibnamefont {Oliviero}},\ }\bibfield  {title} {\bibinfo {title} {Gravitational backreaction is magical},\ }\href@noop {} {\bibfield  {journal} {\bibinfo  {journal} {PRX Quantum}\ }\textbf {\bibinfo {volume} {6}},\ \bibinfo {pages} {040375} (\bibinfo {year} {2025}{\natexlab{a}})}\BibitemShut {NoStop}%
\bibitem [{\citenamefont {Andreadakis}\ and\ \citenamefont {Zanardi}(2026)}]{Andreadakis:2025mfw}%
  \BibitemOpen
  \bibfield  {author} {\bibinfo {author} {\bibfnamefont {F.}~\bibnamefont {Andreadakis}}\ and\ \bibinfo {author} {\bibfnamefont {P.}~\bibnamefont {Zanardi}},\ }\bibfield  {title} {\bibinfo {title} {{Exact link between nonlocal nonstabilizerness and operator entanglement}},\ }\href {https://doi.org/10.1103/9x56-2b45} {\bibfield  {journal} {\bibinfo  {journal} {Phys. Rev. A}\ }\textbf {\bibinfo {volume} {113}},\ \bibinfo {pages} {L010404} (\bibinfo {year} {2026})},\ \Eprint {https://arxiv.org/abs/2504.09360} {arXiv:2504.09360 [quant-ph]} \BibitemShut {NoStop}%
\bibitem [{\citenamefont {Qian}\ and\ \citenamefont {Wang}(2025)}]{Qian:2025oit}%
  \BibitemOpen
  \bibfield  {author} {\bibinfo {author} {\bibfnamefont {D.}~\bibnamefont {Qian}}\ and\ \bibinfo {author} {\bibfnamefont {J.}~\bibnamefont {Wang}},\ }\bibfield  {title} {\bibinfo {title} {{Quantum nonlocal nonstabilizerness}},\ }\href {https://doi.org/10.1103/PhysRevA.111.052443} {\bibfield  {journal} {\bibinfo  {journal} {Phys. Rev. A}\ }\textbf {\bibinfo {volume} {111}},\ \bibinfo {pages} {052443} (\bibinfo {year} {2025})},\ \Eprint {https://arxiv.org/abs/2502.06393} {arXiv:2502.06393 [quant-ph]} \BibitemShut {NoStop}%
\bibitem [{\citenamefont {Munizzi}\ and\ \citenamefont {Schnitzer}(2025)}]{Munizzi:2025suf}%
  \BibitemOpen
  \bibfield  {author} {\bibinfo {author} {\bibfnamefont {W.}~\bibnamefont {Munizzi}}\ and\ \bibinfo {author} {\bibfnamefont {H.~J.}\ \bibnamefont {Schnitzer}},\ }\href@noop {} {\bibinfo {title} {{Topological Preparation of Non-Stabilizer States and Clifford Evolution in $SU(2)_1$ Chern-Simons Theory}}} (\bibinfo {year} {2025}),\ \bibinfo {note} {arXiv:2510.15067 [hep-th]},\ \Eprint {https://arxiv.org/abs/2510.15067} {arXiv:2510.15067 [hep-th]} \BibitemShut {NoStop}%
\bibitem [{\citenamefont {Munizzi}\ and\ \citenamefont {Schnitzer}(2026)}]{Munizzi:2026qkv}%
  \BibitemOpen
  \bibfield  {author} {\bibinfo {author} {\bibfnamefont {W.}~\bibnamefont {Munizzi}}\ and\ \bibinfo {author} {\bibfnamefont {H.~J.}\ \bibnamefont {Schnitzer}},\ }\href@noop {} {\bibinfo {title} {{Magic and Non-Clifford Gates in Topological Quantum Field Theory}}} (\bibinfo {year} {2026}),\ \bibinfo {note} {arXiv:2604.14271 [hep-th]},\ \Eprint {https://arxiv.org/abs/2604.14271} {arXiv:2604.14271 [hep-th]} \BibitemShut {NoStop}%
\bibitem [{\citenamefont {Devra}\ \emph {et~al.}(2024)\citenamefont {Devra}, \citenamefont {Glaser}, \citenamefont {Huber},\ and\ \citenamefont {Glaser}}]{devra2024wigner}%
  \BibitemOpen
  \bibfield  {author} {\bibinfo {author} {\bibfnamefont {A.}~\bibnamefont {Devra}}, \bibinfo {author} {\bibfnamefont {N.~J.}\ \bibnamefont {Glaser}}, \bibinfo {author} {\bibfnamefont {D.}~\bibnamefont {Huber}},\ and\ \bibinfo {author} {\bibfnamefont {S.~J.}\ \bibnamefont {Glaser}},\ }\bibfield  {title} {\bibinfo {title} {Wigner state and process tomography on near-term quantum devices: A. devra et al.},\ }\href@noop {} {\bibfield  {journal} {\bibinfo  {journal} {Quantum Information Processing}\ }\textbf {\bibinfo {volume} {23}},\ \bibinfo {pages} {359} (\bibinfo {year} {2024})}\BibitemShut {NoStop}%
\bibitem [{\citenamefont {Andersen}\ \emph {et~al.}(2019)\citenamefont {Andersen}, \citenamefont {Remm}, \citenamefont {Lazar}, \citenamefont {Krinner}, \citenamefont {Heinsoo}, \citenamefont {Besse}, \citenamefont {Gabureac}, \citenamefont {Wallraff},\ and\ \citenamefont {Eichler}}]{andersen2019entanglement}%
  \BibitemOpen
  \bibfield  {author} {\bibinfo {author} {\bibfnamefont {C.~K.}\ \bibnamefont {Andersen}}, \bibinfo {author} {\bibfnamefont {A.}~\bibnamefont {Remm}}, \bibinfo {author} {\bibfnamefont {S.}~\bibnamefont {Lazar}}, \bibinfo {author} {\bibfnamefont {S.}~\bibnamefont {Krinner}}, \bibinfo {author} {\bibfnamefont {J.}~\bibnamefont {Heinsoo}}, \bibinfo {author} {\bibfnamefont {J.-C.}\ \bibnamefont {Besse}}, \bibinfo {author} {\bibfnamefont {M.}~\bibnamefont {Gabureac}}, \bibinfo {author} {\bibfnamefont {A.}~\bibnamefont {Wallraff}},\ and\ \bibinfo {author} {\bibfnamefont {C.}~\bibnamefont {Eichler}},\ }\bibfield  {title} {\bibinfo {title} {Entanglement stabilization using ancilla-based parity detection and real-time feedback in superconducting circuits},\ }\href@noop {} {\bibfield  {journal} {\bibinfo  {journal} {npj Quantum Information}\ }\textbf {\bibinfo {volume} {5}},\ \bibinfo {pages} {69} (\bibinfo {year} {2019})}\BibitemShut {NoStop}%
\bibitem [{\citenamefont {Rosenfeld}\ \emph {et~al.}(2025)\citenamefont {Rosenfeld}, \citenamefont {Gidney}, \citenamefont {Roberts}, \citenamefont {Morvan}, \citenamefont {Lacroix}, \citenamefont {Kafri}, \citenamefont {Marshall}, \citenamefont {Li}, \citenamefont {Sivak}, \citenamefont {Abanin}, \citenamefont {Abbas}, \citenamefont {Acharya}, \citenamefont {Beni}, \citenamefont {Aigeldinger}, \citenamefont {Alcaraz}, \citenamefont {Alcaraz}, \citenamefont {Andersen}, \citenamefont {Ansmann}, \citenamefont {Arute}, \citenamefont {Arya}, \citenamefont {Askew}, \citenamefont {Astrakhantsev}, \citenamefont {Atalaya}, \citenamefont {Babbush}, \citenamefont {Ballard}, \citenamefont {Bardin}, \citenamefont {Bates}, \citenamefont {Bengtsson}, \citenamefont {Karimi}, \citenamefont {Bilmes}, \citenamefont {Bilodeau}, \citenamefont {Borjans}, \citenamefont {Bovaird}, \citenamefont {Bowers}, \citenamefont {Brill}, \citenamefont {Brooks}, \citenamefont {Broughton}, \citenamefont {Browne}, \citenamefont {Buchea},
  \citenamefont {Buckley}, \citenamefont {Burger}, \citenamefont {Burkett}, \citenamefont {Bushnell}, \citenamefont {Busnaina}, \citenamefont {Cabrera}, \citenamefont {Campero}, \citenamefont {Chang}, \citenamefont {Chen}, \citenamefont {Chen}, \citenamefont {Chiaro}, \citenamefont {Chih}, \citenamefont {Cleland}, \citenamefont {Cochrane}, \citenamefont {Cockrell}, \citenamefont {Cogan}, \citenamefont {Conner}, \citenamefont {Cook}, \citenamefont {nas}, \citenamefont {Courtney}, \citenamefont {Crook}, \citenamefont {Curtin}, \citenamefont {Damyanov}, \citenamefont {Das}, \citenamefont {Debroy}, \citenamefont {Demura}, \citenamefont {Donohoe}, \citenamefont {Drozdov}, \citenamefont {Dunsworth}, \citenamefont {Ehimhen}, \citenamefont {Eickbusch}, \citenamefont {Elbag}, \citenamefont {Ella}, \citenamefont {Elzouka}, \citenamefont {Enriquez}, \citenamefont {Erickson}, \citenamefont {Faoro}, \citenamefont {Ferreira}, \citenamefont {Flores}, \citenamefont {Burgos}, \citenamefont {Fontes}, \citenamefont {Forati},
  \citenamefont {Ford}, \citenamefont {Foxen}, \citenamefont {Fukami}, \citenamefont {Fung}, \citenamefont {Fuste}, \citenamefont {Ganjam}, \citenamefont {Garcia}, \citenamefont {Garrick}, \citenamefont {Gasca}, \citenamefont {Gehring}, \citenamefont {Geiger}, \citenamefont {Genois}, \citenamefont {Giang}, \citenamefont {Gilboa}, \citenamefont {Goeders}, \citenamefont {Gonzales}, \citenamefont {Gosula}, \citenamefont {de~Graaf}, \citenamefont {Dau}, \citenamefont {Graumann}, \citenamefont {Grebel}, \citenamefont {Greene}, \citenamefont {Gross}, \citenamefont {Guerrero}, \citenamefont {Guevel}, \citenamefont {Ha}, \citenamefont {Habegger}, \citenamefont {Hadick}, \citenamefont {Hadjikhani}, \citenamefont {Hamilton}, \citenamefont {Hansen}, \citenamefont {Harrigan}, \citenamefont {Harrington}, \citenamefont {Hartshorn}, \citenamefont {Heslin}, \citenamefont {Heu}, \citenamefont {Higgott}, \citenamefont {Hiltermann}, \citenamefont {Hilton}, \citenamefont {Huang}, \citenamefont {Hucka}, \citenamefont {Hudspeth},
  \citenamefont {Huff}, \citenamefont {Huggins}, \citenamefont {Ioffe}, \citenamefont {Jeffrey}, \citenamefont {Jevons}, \citenamefont {Jiang}, \citenamefont {Jin}, \citenamefont {Joshi}, \citenamefont {Juhas}, \citenamefont {Kabel}, \citenamefont {Kang}, \citenamefont {Kang}, \citenamefont {Karamlou}, \citenamefont {Kaufman}, \citenamefont {Kechedzhi}, \citenamefont {Khattar}, \citenamefont {Khezri}, \citenamefont {Kim}, \citenamefont {Klimov}, \citenamefont {Knaut}, \citenamefont {Kobrin}, \citenamefont {Korotkov}, \citenamefont {Kostritsa}, \citenamefont {Kreikebaum}, \citenamefont {Kudo}, \citenamefont {Kueffler}, \citenamefont {Kumar}, \citenamefont {Kurilovich}, \citenamefont {Kutsko}, \citenamefont {Lange-Dei}, \citenamefont {Langley}, \citenamefont {Laptev}, \citenamefont {Lau}, \citenamefont {Leavell}, \citenamefont {Ledford}, \citenamefont {Lee}, \citenamefont {Lee}, \citenamefont {Lester}, \citenamefont {Leung}, \citenamefont {Li}, \citenamefont {Li}, \citenamefont {Lill}, \citenamefont
  {Livingston}, \citenamefont {Lloyd}, \citenamefont {Locharla}, \citenamefont {Lorenzo}, \citenamefont {Lucero}, \citenamefont {Lundahl}, \citenamefont {Lunt}, \citenamefont {Madhuk}, \citenamefont {Maiti}, \citenamefont {Maloney}, \citenamefont {Mandr\`a}, \citenamefont {Martin}, \citenamefont {Martin}, \citenamefont {Mascot}, \citenamefont {Das}, \citenamefont {Maslov}, \citenamefont {Mathews}, \citenamefont {Maxfield}, \citenamefont {McClean}, \citenamefont {McEwen}, \citenamefont {Meeks}, \citenamefont {Megrant}, \citenamefont {Miao}, \citenamefont {Minev}, \citenamefont {Molavi}, \citenamefont {Molina}, \citenamefont {Montazeri}, \citenamefont {Neill}, \citenamefont {Newman}, \citenamefont {Nguyen}, \citenamefont {Nguyen}, \citenamefont {Ni}, \citenamefont {Niu}, \citenamefont {Noll}, \citenamefont {Oas}, \citenamefont {Oliver}, \citenamefont {Orosco}, \citenamefont {Ottosson}, \citenamefont {Pagano}, \citenamefont {Paolo}, \citenamefont {Peek}, \citenamefont {Peterson}, \citenamefont {Pizzuto},
  \citenamefont {Portoles}, \citenamefont {Potter}, \citenamefont {Pritchard}, \citenamefont {Qian}, \citenamefont {Quintana}, \citenamefont {Ramachandran}, \citenamefont {Ranadive}, \citenamefont {Reagor}, \citenamefont {Resnick}, \citenamefont {Rhodes}, \citenamefont {Riley}, \citenamefont {Rodriguez}, \citenamefont {Ropes}, \citenamefont {Rose}, \citenamefont {Rosenberg}, \citenamefont {Rosenstock}, \citenamefont {Rossi}, \citenamefont {Roushan}, \citenamefont {Rower}, \citenamefont {Salazar}, \citenamefont {Sankaragomathi}, \citenamefont {Sarihan}, \citenamefont {Schaefer}, \citenamefont {Schroeder}, \citenamefont {Schurkus}, \citenamefont {Shahingohar}, \citenamefont {Shearn}, \citenamefont {Shorter}, \citenamefont {Shutty}, \citenamefont {Shvarts}, \citenamefont {Small}, \citenamefont {Smith}, \citenamefont {Sobel}, \citenamefont {Spells}, \citenamefont {Springer}, \citenamefont {Sterling}, \citenamefont {Suchard}, \citenamefont {Szasz}, \citenamefont {Sztein}, \citenamefont {Taylor}, \citenamefont
  {Thiruraman}, \citenamefont {Thor}, \citenamefont {Timucin}, \citenamefont {Tomita}, \citenamefont {Torres}, \citenamefont {Torunbalci}, \citenamefont {Tran}, \citenamefont {Vaishnav}, \citenamefont {Vargas}, \citenamefont {Vdovichev}, \citenamefont {Vidal}, \citenamefont {Villalonga}, \citenamefont {Heidweiller}, \citenamefont {Voorhees}, \citenamefont {Waltman}, \citenamefont {Waltz}, \citenamefont {Wang}, \citenamefont {Wang}, \citenamefont {Ware}, \citenamefont {Watson}, \citenamefont {Wei}, \citenamefont {Weidel}, \citenamefont {White}, \citenamefont {Wong}, \citenamefont {Woo}, \citenamefont {Wood}, \citenamefont {Woodson}, \citenamefont {Xing}, \citenamefont {Yao}, \citenamefont {Yeh}, \citenamefont {Ying}, \citenamefont {Yoo}, \citenamefont {Yosri}, \citenamefont {Young}, \citenamefont {Young}, \citenamefont {Zalcman}, \citenamefont {Zhang}, \citenamefont {Zhang}, \citenamefont {Zhu}, \citenamefont {Zobrist}, \citenamefont {Zou}, \citenamefont {Neven}, \citenamefont {Boixo}, \citenamefont {Jones},
  \citenamefont {Kelly}, \citenamefont {Bourassa},\ and\ \citenamefont {Satzinger}}]{rosenfeld2025}%
  \BibitemOpen
  \bibfield  {author} {\bibinfo {author} {\bibfnamefont {E.}~\bibnamefont {Rosenfeld}}, \bibinfo {author} {\bibfnamefont {C.}~\bibnamefont {Gidney}}, \bibinfo {author} {\bibfnamefont {G.}~\bibnamefont {Roberts}}, \bibinfo {author} {\bibfnamefont {A.}~\bibnamefont {Morvan}}, \bibinfo {author} {\bibfnamefont {N.}~\bibnamefont {Lacroix}}, \bibinfo {author} {\bibfnamefont {D.}~\bibnamefont {Kafri}}, \bibinfo {author} {\bibfnamefont {J.}~\bibnamefont {Marshall}}, \bibinfo {author} {\bibfnamefont {M.}~\bibnamefont {Li}}, \bibinfo {author} {\bibfnamefont {V.}~\bibnamefont {Sivak}}, \bibinfo {author} {\bibfnamefont {D.}~\bibnamefont {Abanin}}, \bibinfo {author} {\bibfnamefont {A.}~\bibnamefont {Abbas}}, \bibinfo {author} {\bibfnamefont {R.}~\bibnamefont {Acharya}}, \bibinfo {author} {\bibfnamefont {L.~A.}\ \bibnamefont {Beni}}, \bibinfo {author} {\bibfnamefont {G.}~\bibnamefont {Aigeldinger}}, \bibinfo {author} {\bibfnamefont {R.}~\bibnamefont {Alcaraz}}, \bibinfo {author} {\bibfnamefont {S.}~\bibnamefont {Alcaraz}},
  \bibinfo {author} {\bibfnamefont {T.~I.}\ \bibnamefont {Andersen}}, \bibinfo {author} {\bibfnamefont {M.}~\bibnamefont {Ansmann}}, \bibinfo {author} {\bibfnamefont {F.}~\bibnamefont {Arute}}, \bibinfo {author} {\bibfnamefont {K.}~\bibnamefont {Arya}}, \bibinfo {author} {\bibfnamefont {W.}~\bibnamefont {Askew}}, \bibinfo {author} {\bibfnamefont {N.}~\bibnamefont {Astrakhantsev}}, \bibinfo {author} {\bibfnamefont {J.}~\bibnamefont {Atalaya}}, \bibinfo {author} {\bibfnamefont {R.}~\bibnamefont {Babbush}}, \bibinfo {author} {\bibfnamefont {B.}~\bibnamefont {Ballard}}, \bibinfo {author} {\bibfnamefont {J.~C.}\ \bibnamefont {Bardin}}, \bibinfo {author} {\bibfnamefont {H.}~\bibnamefont {Bates}}, \bibinfo {author} {\bibfnamefont {A.}~\bibnamefont {Bengtsson}}, \bibinfo {author} {\bibfnamefont {M.~B.}\ \bibnamefont {Karimi}}, \bibinfo {author} {\bibfnamefont {A.}~\bibnamefont {Bilmes}}, \bibinfo {author} {\bibfnamefont {S.}~\bibnamefont {Bilodeau}}, \bibinfo {author} {\bibfnamefont {F.}~\bibnamefont {Borjans}},
  \bibinfo {author} {\bibfnamefont {J.}~\bibnamefont {Bovaird}}, \bibinfo {author} {\bibfnamefont {D.}~\bibnamefont {Bowers}}, \bibinfo {author} {\bibfnamefont {L.}~\bibnamefont {Brill}}, \bibinfo {author} {\bibfnamefont {P.}~\bibnamefont {Brooks}}, \bibinfo {author} {\bibfnamefont {M.}~\bibnamefont {Broughton}}, \bibinfo {author} {\bibfnamefont {D.~A.}\ \bibnamefont {Browne}}, \bibinfo {author} {\bibfnamefont {B.}~\bibnamefont {Buchea}}, \bibinfo {author} {\bibfnamefont {B.~B.}\ \bibnamefont {Buckley}}, \bibinfo {author} {\bibfnamefont {T.}~\bibnamefont {Burger}}, \bibinfo {author} {\bibfnamefont {B.}~\bibnamefont {Burkett}}, \bibinfo {author} {\bibfnamefont {N.}~\bibnamefont {Bushnell}}, \bibinfo {author} {\bibfnamefont {J.}~\bibnamefont {Busnaina}}, \bibinfo {author} {\bibfnamefont {A.}~\bibnamefont {Cabrera}}, \bibinfo {author} {\bibfnamefont {J.}~\bibnamefont {Campero}}, \bibinfo {author} {\bibfnamefont {H.-S.}\ \bibnamefont {Chang}}, \bibinfo {author} {\bibfnamefont {S.}~\bibnamefont {Chen}}, \bibinfo
  {author} {\bibfnamefont {Z.}~\bibnamefont {Chen}}, \bibinfo {author} {\bibfnamefont {B.}~\bibnamefont {Chiaro}}, \bibinfo {author} {\bibfnamefont {L.-Y.}\ \bibnamefont {Chih}}, \bibinfo {author} {\bibfnamefont {A.~Y.}\ \bibnamefont {Cleland}}, \bibinfo {author} {\bibfnamefont {B.}~\bibnamefont {Cochrane}}, \bibinfo {author} {\bibfnamefont {M.}~\bibnamefont {Cockrell}}, \bibinfo {author} {\bibfnamefont {J.}~\bibnamefont {Cogan}}, \bibinfo {author} {\bibfnamefont {P.}~\bibnamefont {Conner}}, \bibinfo {author} {\bibfnamefont {H.}~\bibnamefont {Cook}}, \bibinfo {author} {\bibfnamefont {R.~G.~C.}\ \bibnamefont {nas}}, \bibinfo {author} {\bibfnamefont {W.}~\bibnamefont {Courtney}}, \bibinfo {author} {\bibfnamefont {A.~L.}\ \bibnamefont {Crook}}, \bibinfo {author} {\bibfnamefont {B.}~\bibnamefont {Curtin}}, \bibinfo {author} {\bibfnamefont {M.}~\bibnamefont {Damyanov}}, \bibinfo {author} {\bibfnamefont {S.}~\bibnamefont {Das}}, \bibinfo {author} {\bibfnamefont {D.~M.}\ \bibnamefont {Debroy}}, \bibinfo {author}
  {\bibfnamefont {S.}~\bibnamefont {Demura}}, \bibinfo {author} {\bibfnamefont {P.}~\bibnamefont {Donohoe}}, \bibinfo {author} {\bibfnamefont {I.}~\bibnamefont {Drozdov}}, \bibinfo {author} {\bibfnamefont {A.}~\bibnamefont {Dunsworth}}, \bibinfo {author} {\bibfnamefont {V.}~\bibnamefont {Ehimhen}}, \bibinfo {author} {\bibfnamefont {A.}~\bibnamefont {Eickbusch}}, \bibinfo {author} {\bibfnamefont {A.~M.}\ \bibnamefont {Elbag}}, \bibinfo {author} {\bibfnamefont {L.}~\bibnamefont {Ella}}, \bibinfo {author} {\bibfnamefont {M.}~\bibnamefont {Elzouka}}, \bibinfo {author} {\bibfnamefont {D.}~\bibnamefont {Enriquez}}, \bibinfo {author} {\bibfnamefont {C.}~\bibnamefont {Erickson}}, \bibinfo {author} {\bibfnamefont {L.}~\bibnamefont {Faoro}}, \bibinfo {author} {\bibfnamefont {V.~S.}\ \bibnamefont {Ferreira}}, \bibinfo {author} {\bibfnamefont {M.}~\bibnamefont {Flores}}, \bibinfo {author} {\bibfnamefont {L.~F.}\ \bibnamefont {Burgos}}, \bibinfo {author} {\bibfnamefont {S.}~\bibnamefont {Fontes}}, \bibinfo {author}
  {\bibfnamefont {E.}~\bibnamefont {Forati}}, \bibinfo {author} {\bibfnamefont {J.}~\bibnamefont {Ford}}, \bibinfo {author} {\bibfnamefont {B.}~\bibnamefont {Foxen}}, \bibinfo {author} {\bibfnamefont {M.}~\bibnamefont {Fukami}}, \bibinfo {author} {\bibfnamefont {A.~W.~L.}\ \bibnamefont {Fung}}, \bibinfo {author} {\bibfnamefont {L.}~\bibnamefont {Fuste}}, \bibinfo {author} {\bibfnamefont {S.}~\bibnamefont {Ganjam}}, \bibinfo {author} {\bibfnamefont {G.}~\bibnamefont {Garcia}}, \bibinfo {author} {\bibfnamefont {C.}~\bibnamefont {Garrick}}, \bibinfo {author} {\bibfnamefont {R.}~\bibnamefont {Gasca}}, \bibinfo {author} {\bibfnamefont {H.}~\bibnamefont {Gehring}}, \bibinfo {author} {\bibfnamefont {R.}~\bibnamefont {Geiger}}, \bibinfo {author} {\bibfnamefont {E.}~\bibnamefont {Genois}}, \bibinfo {author} {\bibfnamefont {W.}~\bibnamefont {Giang}}, \bibinfo {author} {\bibfnamefont {D.}~\bibnamefont {Gilboa}}, \bibinfo {author} {\bibfnamefont {J.~E.}\ \bibnamefont {Goeders}}, \bibinfo {author} {\bibfnamefont {E.~C.}\
  \bibnamefont {Gonzales}}, \bibinfo {author} {\bibfnamefont {R.}~\bibnamefont {Gosula}}, \bibinfo {author} {\bibfnamefont {S.~J.}\ \bibnamefont {de~Graaf}}, \bibinfo {author} {\bibfnamefont {A.~G.}\ \bibnamefont {Dau}}, \bibinfo {author} {\bibfnamefont {D.}~\bibnamefont {Graumann}}, \bibinfo {author} {\bibfnamefont {J.}~\bibnamefont {Grebel}}, \bibinfo {author} {\bibfnamefont {A.}~\bibnamefont {Greene}}, \bibinfo {author} {\bibfnamefont {J.~A.}\ \bibnamefont {Gross}}, \bibinfo {author} {\bibfnamefont {J.}~\bibnamefont {Guerrero}}, \bibinfo {author} {\bibfnamefont {L.~L.}\ \bibnamefont {Guevel}}, \bibinfo {author} {\bibfnamefont {T.}~\bibnamefont {Ha}}, \bibinfo {author} {\bibfnamefont {S.}~\bibnamefont {Habegger}}, \bibinfo {author} {\bibfnamefont {T.}~\bibnamefont {Hadick}}, \bibinfo {author} {\bibfnamefont {A.}~\bibnamefont {Hadjikhani}}, \bibinfo {author} {\bibfnamefont {M.~C.}\ \bibnamefont {Hamilton}}, \bibinfo {author} {\bibfnamefont {M.}~\bibnamefont {Hansen}}, \bibinfo {author} {\bibfnamefont
  {M.~P.}\ \bibnamefont {Harrigan}}, \bibinfo {author} {\bibfnamefont {S.~D.}\ \bibnamefont {Harrington}}, \bibinfo {author} {\bibfnamefont {J.}~\bibnamefont {Hartshorn}}, \bibinfo {author} {\bibfnamefont {S.}~\bibnamefont {Heslin}}, \bibinfo {author} {\bibfnamefont {P.}~\bibnamefont {Heu}}, \bibinfo {author} {\bibfnamefont {O.}~\bibnamefont {Higgott}}, \bibinfo {author} {\bibfnamefont {R.}~\bibnamefont {Hiltermann}}, \bibinfo {author} {\bibfnamefont {J.}~\bibnamefont {Hilton}}, \bibinfo {author} {\bibfnamefont {H.-Y.}\ \bibnamefont {Huang}}, \bibinfo {author} {\bibfnamefont {M.}~\bibnamefont {Hucka}}, \bibinfo {author} {\bibfnamefont {C.}~\bibnamefont {Hudspeth}}, \bibinfo {author} {\bibfnamefont {A.}~\bibnamefont {Huff}}, \bibinfo {author} {\bibfnamefont {W.~J.}\ \bibnamefont {Huggins}}, \bibinfo {author} {\bibfnamefont {L.~B.}\ \bibnamefont {Ioffe}}, \bibinfo {author} {\bibfnamefont {E.}~\bibnamefont {Jeffrey}}, \bibinfo {author} {\bibfnamefont {S.}~\bibnamefont {Jevons}}, \bibinfo {author} {\bibfnamefont
  {Z.}~\bibnamefont {Jiang}}, \bibinfo {author} {\bibfnamefont {X.}~\bibnamefont {Jin}}, \bibinfo {author} {\bibfnamefont {C.}~\bibnamefont {Joshi}}, \bibinfo {author} {\bibfnamefont {P.}~\bibnamefont {Juhas}}, \bibinfo {author} {\bibfnamefont {A.}~\bibnamefont {Kabel}}, \bibinfo {author} {\bibfnamefont {H.}~\bibnamefont {Kang}}, \bibinfo {author} {\bibfnamefont {K.}~\bibnamefont {Kang}}, \bibinfo {author} {\bibfnamefont {A.~H.}\ \bibnamefont {Karamlou}}, \bibinfo {author} {\bibfnamefont {R.}~\bibnamefont {Kaufman}}, \bibinfo {author} {\bibfnamefont {K.}~\bibnamefont {Kechedzhi}}, \bibinfo {author} {\bibfnamefont {T.}~\bibnamefont {Khattar}}, \bibinfo {author} {\bibfnamefont {M.}~\bibnamefont {Khezri}}, \bibinfo {author} {\bibfnamefont {S.}~\bibnamefont {Kim}}, \bibinfo {author} {\bibfnamefont {P.~V.}\ \bibnamefont {Klimov}}, \bibinfo {author} {\bibfnamefont {C.~M.}\ \bibnamefont {Knaut}}, \bibinfo {author} {\bibfnamefont {B.}~\bibnamefont {Kobrin}}, \bibinfo {author} {\bibfnamefont {A.~N.}\ \bibnamefont
  {Korotkov}}, \bibinfo {author} {\bibfnamefont {F.}~\bibnamefont {Kostritsa}}, \bibinfo {author} {\bibfnamefont {J.~M.}\ \bibnamefont {Kreikebaum}}, \bibinfo {author} {\bibfnamefont {R.}~\bibnamefont {Kudo}}, \bibinfo {author} {\bibfnamefont {B.}~\bibnamefont {Kueffler}}, \bibinfo {author} {\bibfnamefont {A.}~\bibnamefont {Kumar}}, \bibinfo {author} {\bibfnamefont {V.~D.}\ \bibnamefont {Kurilovich}}, \bibinfo {author} {\bibfnamefont {V.}~\bibnamefont {Kutsko}}, \bibinfo {author} {\bibfnamefont {T.}~\bibnamefont {Lange-Dei}}, \bibinfo {author} {\bibfnamefont {B.~W.}\ \bibnamefont {Langley}}, \bibinfo {author} {\bibfnamefont {P.}~\bibnamefont {Laptev}}, \bibinfo {author} {\bibfnamefont {K.-M.}\ \bibnamefont {Lau}}, \bibinfo {author} {\bibfnamefont {E.}~\bibnamefont {Leavell}}, \bibinfo {author} {\bibfnamefont {J.}~\bibnamefont {Ledford}}, \bibinfo {author} {\bibfnamefont {J.}~\bibnamefont {Lee}}, \bibinfo {author} {\bibfnamefont {K.}~\bibnamefont {Lee}}, \bibinfo {author} {\bibfnamefont {B.~J.}\ \bibnamefont
  {Lester}}, \bibinfo {author} {\bibfnamefont {W.}~\bibnamefont {Leung}}, \bibinfo {author} {\bibfnamefont {L.}~\bibnamefont {Li}}, \bibinfo {author} {\bibfnamefont {W.~Y.}\ \bibnamefont {Li}}, \bibinfo {author} {\bibfnamefont {A.~T.}\ \bibnamefont {Lill}}, \bibinfo {author} {\bibfnamefont {W.~P.}\ \bibnamefont {Livingston}}, \bibinfo {author} {\bibfnamefont {M.~T.}\ \bibnamefont {Lloyd}}, \bibinfo {author} {\bibfnamefont {A.}~\bibnamefont {Locharla}}, \bibinfo {author} {\bibfnamefont {L.~D.}\ \bibnamefont {Lorenzo}}, \bibinfo {author} {\bibfnamefont {E.}~\bibnamefont {Lucero}}, \bibinfo {author} {\bibfnamefont {D.}~\bibnamefont {Lundahl}}, \bibinfo {author} {\bibfnamefont {A.}~\bibnamefont {Lunt}}, \bibinfo {author} {\bibfnamefont {S.}~\bibnamefont {Madhuk}}, \bibinfo {author} {\bibfnamefont {A.}~\bibnamefont {Maiti}}, \bibinfo {author} {\bibfnamefont {A.}~\bibnamefont {Maloney}}, \bibinfo {author} {\bibfnamefont {S.}~\bibnamefont {Mandr\`a}}, \bibinfo {author} {\bibfnamefont {L.~S.}\ \bibnamefont {Martin}},
  \bibinfo {author} {\bibfnamefont {O.}~\bibnamefont {Martin}}, \bibinfo {author} {\bibfnamefont {E.}~\bibnamefont {Mascot}}, \bibinfo {author} {\bibfnamefont {P.~M.}\ \bibnamefont {Das}}, \bibinfo {author} {\bibfnamefont {D.}~\bibnamefont {Maslov}}, \bibinfo {author} {\bibfnamefont {M.}~\bibnamefont {Mathews}}, \bibinfo {author} {\bibfnamefont {C.}~\bibnamefont {Maxfield}}, \bibinfo {author} {\bibfnamefont {J.~R.}\ \bibnamefont {McClean}}, \bibinfo {author} {\bibfnamefont {M.}~\bibnamefont {McEwen}}, \bibinfo {author} {\bibfnamefont {S.}~\bibnamefont {Meeks}}, \bibinfo {author} {\bibfnamefont {A.}~\bibnamefont {Megrant}}, \bibinfo {author} {\bibfnamefont {K.~C.}\ \bibnamefont {Miao}}, \bibinfo {author} {\bibfnamefont {Z.~K.}\ \bibnamefont {Minev}}, \bibinfo {author} {\bibfnamefont {R.}~\bibnamefont {Molavi}}, \bibinfo {author} {\bibfnamefont {S.}~\bibnamefont {Molina}}, \bibinfo {author} {\bibfnamefont {S.}~\bibnamefont {Montazeri}}, \bibinfo {author} {\bibfnamefont {C.}~\bibnamefont {Neill}}, \bibinfo
  {author} {\bibfnamefont {M.}~\bibnamefont {Newman}}, \bibinfo {author} {\bibfnamefont {A.}~\bibnamefont {Nguyen}}, \bibinfo {author} {\bibfnamefont {M.}~\bibnamefont {Nguyen}}, \bibinfo {author} {\bibfnamefont {C.-H.}\ \bibnamefont {Ni}}, \bibinfo {author} {\bibfnamefont {M.~Y.}\ \bibnamefont {Niu}}, \bibinfo {author} {\bibfnamefont {N.}~\bibnamefont {Noll}}, \bibinfo {author} {\bibfnamefont {L.}~\bibnamefont {Oas}}, \bibinfo {author} {\bibfnamefont {W.~D.}\ \bibnamefont {Oliver}}, \bibinfo {author} {\bibfnamefont {R.}~\bibnamefont {Orosco}}, \bibinfo {author} {\bibfnamefont {K.}~\bibnamefont {Ottosson}}, \bibinfo {author} {\bibfnamefont {A.}~\bibnamefont {Pagano}}, \bibinfo {author} {\bibfnamefont {A.~D.}\ \bibnamefont {Paolo}}, \bibinfo {author} {\bibfnamefont {S.}~\bibnamefont {Peek}}, \bibinfo {author} {\bibfnamefont {D.}~\bibnamefont {Peterson}}, \bibinfo {author} {\bibfnamefont {A.}~\bibnamefont {Pizzuto}}, \bibinfo {author} {\bibfnamefont {E.}~\bibnamefont {Portoles}}, \bibinfo {author}
  {\bibfnamefont {R.}~\bibnamefont {Potter}}, \bibinfo {author} {\bibfnamefont {O.}~\bibnamefont {Pritchard}}, \bibinfo {author} {\bibfnamefont {M.}~\bibnamefont {Qian}}, \bibinfo {author} {\bibfnamefont {C.}~\bibnamefont {Quintana}}, \bibinfo {author} {\bibfnamefont {G.}~\bibnamefont {Ramachandran}}, \bibinfo {author} {\bibfnamefont {A.}~\bibnamefont {Ranadive}}, \bibinfo {author} {\bibfnamefont {M.~J.}\ \bibnamefont {Reagor}}, \bibinfo {author} {\bibfnamefont {R.}~\bibnamefont {Resnick}}, \bibinfo {author} {\bibfnamefont {D.~M.}\ \bibnamefont {Rhodes}}, \bibinfo {author} {\bibfnamefont {D.}~\bibnamefont {Riley}}, \bibinfo {author} {\bibfnamefont {R.}~\bibnamefont {Rodriguez}}, \bibinfo {author} {\bibfnamefont {E.}~\bibnamefont {Ropes}}, \bibinfo {author} {\bibfnamefont {L.~B.~D.}\ \bibnamefont {Rose}}, \bibinfo {author} {\bibfnamefont {E.}~\bibnamefont {Rosenberg}}, \bibinfo {author} {\bibfnamefont {D.}~\bibnamefont {Rosenstock}}, \bibinfo {author} {\bibfnamefont {E.}~\bibnamefont {Rossi}}, \bibinfo
  {author} {\bibfnamefont {P.}~\bibnamefont {Roushan}}, \bibinfo {author} {\bibfnamefont {D.~A.}\ \bibnamefont {Rower}}, \bibinfo {author} {\bibfnamefont {R.}~\bibnamefont {Salazar}}, \bibinfo {author} {\bibfnamefont {K.}~\bibnamefont {Sankaragomathi}}, \bibinfo {author} {\bibfnamefont {M.~C.}\ \bibnamefont {Sarihan}}, \bibinfo {author} {\bibfnamefont {M.}~\bibnamefont {Schaefer}}, \bibinfo {author} {\bibfnamefont {S.}~\bibnamefont {Schroeder}}, \bibinfo {author} {\bibfnamefont {H.~F.}\ \bibnamefont {Schurkus}}, \bibinfo {author} {\bibfnamefont {A.}~\bibnamefont {Shahingohar}}, \bibinfo {author} {\bibfnamefont {M.~J.}\ \bibnamefont {Shearn}}, \bibinfo {author} {\bibfnamefont {A.}~\bibnamefont {Shorter}}, \bibinfo {author} {\bibfnamefont {N.}~\bibnamefont {Shutty}}, \bibinfo {author} {\bibfnamefont {V.}~\bibnamefont {Shvarts}}, \bibinfo {author} {\bibfnamefont {S.}~\bibnamefont {Small}}, \bibinfo {author} {\bibfnamefont {W.~C.}\ \bibnamefont {Smith}}, \bibinfo {author} {\bibfnamefont {D.~A.}\ \bibnamefont
  {Sobel}}, \bibinfo {author} {\bibfnamefont {B.}~\bibnamefont {Spells}}, \bibinfo {author} {\bibfnamefont {S.}~\bibnamefont {Springer}}, \bibinfo {author} {\bibfnamefont {G.}~\bibnamefont {Sterling}}, \bibinfo {author} {\bibfnamefont {J.}~\bibnamefont {Suchard}}, \bibinfo {author} {\bibfnamefont {A.}~\bibnamefont {Szasz}}, \bibinfo {author} {\bibfnamefont {A.}~\bibnamefont {Sztein}}, \bibinfo {author} {\bibfnamefont {M.}~\bibnamefont {Taylor}}, \bibinfo {author} {\bibfnamefont {J.~P.}\ \bibnamefont {Thiruraman}}, \bibinfo {author} {\bibfnamefont {D.}~\bibnamefont {Thor}}, \bibinfo {author} {\bibfnamefont {D.}~\bibnamefont {Timucin}}, \bibinfo {author} {\bibfnamefont {E.}~\bibnamefont {Tomita}}, \bibinfo {author} {\bibfnamefont {A.}~\bibnamefont {Torres}}, \bibinfo {author} {\bibfnamefont {M.~M.}\ \bibnamefont {Torunbalci}}, \bibinfo {author} {\bibfnamefont {H.}~\bibnamefont {Tran}}, \bibinfo {author} {\bibfnamefont {A.}~\bibnamefont {Vaishnav}}, \bibinfo {author} {\bibfnamefont {J.}~\bibnamefont {Vargas}},
  \bibinfo {author} {\bibfnamefont {S.}~\bibnamefont {Vdovichev}}, \bibinfo {author} {\bibfnamefont {G.}~\bibnamefont {Vidal}}, \bibinfo {author} {\bibfnamefont {B.}~\bibnamefont {Villalonga}}, \bibinfo {author} {\bibfnamefont {C.~V.}\ \bibnamefont {Heidweiller}}, \bibinfo {author} {\bibfnamefont {M.}~\bibnamefont {Voorhees}}, \bibinfo {author} {\bibfnamefont {S.}~\bibnamefont {Waltman}}, \bibinfo {author} {\bibfnamefont {J.}~\bibnamefont {Waltz}}, \bibinfo {author} {\bibfnamefont {S.~X.}\ \bibnamefont {Wang}}, \bibinfo {author} {\bibfnamefont {D.}~\bibnamefont {Wang}}, \bibinfo {author} {\bibfnamefont {B.}~\bibnamefont {Ware}}, \bibinfo {author} {\bibfnamefont {J.~D.}\ \bibnamefont {Watson}}, \bibinfo {author} {\bibfnamefont {Y.}~\bibnamefont {Wei}}, \bibinfo {author} {\bibfnamefont {T.}~\bibnamefont {Weidel}}, \bibinfo {author} {\bibfnamefont {T.}~\bibnamefont {White}}, \bibinfo {author} {\bibfnamefont {K.}~\bibnamefont {Wong}}, \bibinfo {author} {\bibfnamefont {B.~W.~K.}\ \bibnamefont {Woo}}, \bibinfo
  {author} {\bibfnamefont {C.~J.}\ \bibnamefont {Wood}}, \bibinfo {author} {\bibfnamefont {M.}~\bibnamefont {Woodson}}, \bibinfo {author} {\bibfnamefont {C.}~\bibnamefont {Xing}}, \bibinfo {author} {\bibfnamefont {Z.~J.}\ \bibnamefont {Yao}}, \bibinfo {author} {\bibfnamefont {P.}~\bibnamefont {Yeh}}, \bibinfo {author} {\bibfnamefont {B.}~\bibnamefont {Ying}}, \bibinfo {author} {\bibfnamefont {J.}~\bibnamefont {Yoo}}, \bibinfo {author} {\bibfnamefont {N.}~\bibnamefont {Yosri}}, \bibinfo {author} {\bibfnamefont {E.}~\bibnamefont {Young}}, \bibinfo {author} {\bibfnamefont {G.}~\bibnamefont {Young}}, \bibinfo {author} {\bibfnamefont {A.}~\bibnamefont {Zalcman}}, \bibinfo {author} {\bibfnamefont {R.}~\bibnamefont {Zhang}}, \bibinfo {author} {\bibfnamefont {Y.}~\bibnamefont {Zhang}}, \bibinfo {author} {\bibfnamefont {N.}~\bibnamefont {Zhu}}, \bibinfo {author} {\bibfnamefont {N.}~\bibnamefont {Zobrist}}, \bibinfo {author} {\bibfnamefont {Z.}~\bibnamefont {Zou}}, \bibinfo {author} {\bibfnamefont {H.}~\bibnamefont
  {Neven}}, \bibinfo {author} {\bibfnamefont {S.}~\bibnamefont {Boixo}}, \bibinfo {author} {\bibfnamefont {C.}~\bibnamefont {Jones}}, \bibinfo {author} {\bibfnamefont {J.}~\bibnamefont {Kelly}}, \bibinfo {author} {\bibfnamefont {A.}~\bibnamefont {Bourassa}},\ and\ \bibinfo {author} {\bibfnamefont {K.~J.}\ \bibnamefont {Satzinger}},\ }\href {https://arxiv.org/abs/2512.13908} {\bibinfo {title} {Magic state cultivation on a superconducting quantum processor}} (\bibinfo {year} {2025}),\ \bibinfo {note} {arXiv:2512.13908 [quant-ph]},\ \Eprint {https://arxiv.org/abs/2512.13908} {arXiv:2512.13908 [quant-ph]} \BibitemShut {NoStop}%
\bibitem [{\citenamefont {Hashim}\ \emph {et~al.}(2025)\citenamefont {Hashim}, \citenamefont {Yuan}, \citenamefont {Gokhale}, \citenamefont {Chen}, \citenamefont {J{\"u}nger}, \citenamefont {Fruitwala}, \citenamefont {Xu}, \citenamefont {Huang}, \citenamefont {Nowrouzi}, \citenamefont {Jiang} \emph {et~al.}}]{hashim2025efficient}%
  \BibitemOpen
  \bibfield  {author} {\bibinfo {author} {\bibfnamefont {A.}~\bibnamefont {Hashim}}, \bibinfo {author} {\bibfnamefont {M.}~\bibnamefont {Yuan}}, \bibinfo {author} {\bibfnamefont {P.}~\bibnamefont {Gokhale}}, \bibinfo {author} {\bibfnamefont {L.}~\bibnamefont {Chen}}, \bibinfo {author} {\bibfnamefont {C.}~\bibnamefont {J{\"u}nger}}, \bibinfo {author} {\bibfnamefont {N.}~\bibnamefont {Fruitwala}}, \bibinfo {author} {\bibfnamefont {Y.}~\bibnamefont {Xu}}, \bibinfo {author} {\bibfnamefont {G.}~\bibnamefont {Huang}}, \bibinfo {author} {\bibfnamefont {K.}~\bibnamefont {Nowrouzi}}, \bibinfo {author} {\bibfnamefont {L.}~\bibnamefont {Jiang}}, \emph {et~al.},\ }\bibfield  {title} {\bibinfo {title} {Efficient generation of multi-partite entanglement between non-local superconducting qubits using classical feedback},\ }\href@noop {} {\bibfield  {journal} {\bibinfo  {journal} {APL Quantum}\ }\textbf {\bibinfo {volume} {2}} (\bibinfo {year} {2025})}\BibitemShut {NoStop}%
\bibitem [{\citenamefont {Robin}\ and\ \citenamefont {Savage}(2025)}]{Robin:2025ymq}%
  \BibitemOpen
  \bibfield  {author} {\bibinfo {author} {\bibfnamefont {C.~E.~P.}\ \bibnamefont {Robin}}\ and\ \bibinfo {author} {\bibfnamefont {M.~J.}\ \bibnamefont {Savage}},\ }\href@noop {} {\bibinfo {title} {{Anti-Flatness and Non-Local Magic in Two-Particle Scattering Processes}}} (\bibinfo {year} {2025}),\ \bibinfo {note} {arXiv:2510.23426 [quant-ph]},\ \Eprint {https://arxiv.org/abs/2510.23426} {arXiv:2510.23426 [quant-ph]} \BibitemShut {NoStop}%
\bibitem [{\citenamefont {White}\ and\ \citenamefont {White}(2026)}]{White:2024bjp}%
  \BibitemOpen
  \bibfield  {author} {\bibinfo {author} {\bibfnamefont {C.~D.}\ \bibnamefont {White}}\ and\ \bibinfo {author} {\bibfnamefont {M.~J.}\ \bibnamefont {White}},\ }\bibfield  {title} {\bibinfo {title} {{The magic of top quarks}},\ }\href {https://doi.org/10.21468/SciPostPhysProc.18.017} {\bibfield  {journal} {\bibinfo  {journal} {SciPost Phys. Proc.}\ }\textbf {\bibinfo {volume} {18}},\ \bibinfo {pages} {017} (\bibinfo {year} {2026})},\ \Eprint {https://arxiv.org/abs/2412.07479} {arXiv:2412.07479 [hep-ph]} \BibitemShut {NoStop}%
\bibitem [{\citenamefont {White}\ and\ \citenamefont {White}(2024)}]{PhysRevD.110.116016}%
  \BibitemOpen
  \bibfield  {author} {\bibinfo {author} {\bibfnamefont {C.~D.}\ \bibnamefont {White}}\ and\ \bibinfo {author} {\bibfnamefont {M.~J.}\ \bibnamefont {White}},\ }\bibfield  {title} {\bibinfo {title} {Magic states of top quarks},\ }\href {https://doi.org/10.1103/PhysRevD.110.116016} {\bibfield  {journal} {\bibinfo  {journal} {Phys. Rev. D}\ }\textbf {\bibinfo {volume} {110}},\ \bibinfo {pages} {116016} (\bibinfo {year} {2024})}\BibitemShut {NoStop}%
\bibitem [{\citenamefont {Ahmad}\ \emph {et~al.}(2025)\citenamefont {Ahmad}, \citenamefont {Esposito}, \citenamefont {Stasino}, \citenamefont {Odavic}, \citenamefont {Cosenza}, \citenamefont {Sarno}, \citenamefont {Mastrovito}, \citenamefont {Viscardi}, \citenamefont {Cusumano}, \citenamefont {Tafuri} \emph {et~al.}}]{ahmad2025experimental}%
  \BibitemOpen
  \bibfield  {author} {\bibinfo {author} {\bibfnamefont {H.~G.}\ \bibnamefont {Ahmad}}, \bibinfo {author} {\bibfnamefont {G.}~\bibnamefont {Esposito}}, \bibinfo {author} {\bibfnamefont {V.}~\bibnamefont {Stasino}}, \bibinfo {author} {\bibfnamefont {J.}~\bibnamefont {Odavic}}, \bibinfo {author} {\bibfnamefont {C.}~\bibnamefont {Cosenza}}, \bibinfo {author} {\bibfnamefont {A.}~\bibnamefont {Sarno}}, \bibinfo {author} {\bibfnamefont {P.}~\bibnamefont {Mastrovito}}, \bibinfo {author} {\bibfnamefont {M.}~\bibnamefont {Viscardi}}, \bibinfo {author} {\bibfnamefont {S.}~\bibnamefont {Cusumano}}, \bibinfo {author} {\bibfnamefont {F.}~\bibnamefont {Tafuri}}, \emph {et~al.},\ }\bibfield  {title} {\bibinfo {title} {Experimental demonstration of non-local magic in a superconducting quantum processor},\ }\href@noop {} {\bibfield  {journal} {\bibinfo  {journal} {arXiv preprint arXiv:2511.15576}\ } (\bibinfo {year} {2025})}\BibitemShut {NoStop}%
\bibitem [{\citenamefont {Cusumano}\ \emph {et~al.}(2025)\citenamefont {Cusumano}, \citenamefont {Venuti}, \citenamefont {Cepollaro}, \citenamefont {Esposito}, \citenamefont {Iannotti}, \citenamefont {Jasser}, \citenamefont {Viscardi}, \citenamefont {Hamma} \emph {et~al.}}]{cusumano2025non}%
  \BibitemOpen
  \bibfield  {author} {\bibinfo {author} {\bibfnamefont {S.}~\bibnamefont {Cusumano}}, \bibinfo {author} {\bibfnamefont {L.~C.}\ \bibnamefont {Venuti}}, \bibinfo {author} {\bibfnamefont {S.}~\bibnamefont {Cepollaro}}, \bibinfo {author} {\bibfnamefont {G.}~\bibnamefont {Esposito}}, \bibinfo {author} {\bibfnamefont {D.}~\bibnamefont {Iannotti}}, \bibinfo {author} {\bibfnamefont {B.}~\bibnamefont {Jasser}}, \bibinfo {author} {\bibfnamefont {M.}~\bibnamefont {Viscardi}}, \bibinfo {author} {\bibfnamefont {A.}~\bibnamefont {Hamma}}, \emph {et~al.},\ }\bibfield  {title} {\bibinfo {title} {Non-stabilizerness and violations of chsh inequalities},\ }\href@noop {} {\bibfield  {journal} {\bibinfo  {journal} {arXiv preprint arXiv:2504.03351}\ } (\bibinfo {year} {2025})}\BibitemShut {NoStop}%
\bibitem [{\citenamefont {Bukov}\ \emph {et~al.}(2018)\citenamefont {Bukov}, \citenamefont {Day}, \citenamefont {Sels}, \citenamefont {Weinberg}, \citenamefont {Polkovnikov},\ and\ \citenamefont {Mehta}}]{bukov2018reinforcement}%
  \BibitemOpen
  \bibfield  {author} {\bibinfo {author} {\bibfnamefont {M.}~\bibnamefont {Bukov}}, \bibinfo {author} {\bibfnamefont {A.~G.~R.}\ \bibnamefont {Day}}, \bibinfo {author} {\bibfnamefont {D.}~\bibnamefont {Sels}}, \bibinfo {author} {\bibfnamefont {P.}~\bibnamefont {Weinberg}}, \bibinfo {author} {\bibfnamefont {A.}~\bibnamefont {Polkovnikov}},\ and\ \bibinfo {author} {\bibfnamefont {P.}~\bibnamefont {Mehta}},\ }\bibfield  {title} {\bibinfo {title} {Reinforcement learning in different phases of quantum control},\ }\href {https://doi.org/10.1103/PhysRevX.8.031086} {\bibfield  {journal} {\bibinfo  {journal} {Physical Review X}\ }\textbf {\bibinfo {volume} {8}},\ \bibinfo {pages} {031086} (\bibinfo {year} {2018})}\BibitemShut {NoStop}%
\bibitem [{\citenamefont {Niu}\ \emph {et~al.}(2019)\citenamefont {Niu}, \citenamefont {Boixo}, \citenamefont {Smelyanskiy},\ and\ \citenamefont {Neven}}]{niu2019universal}%
  \BibitemOpen
  \bibfield  {author} {\bibinfo {author} {\bibfnamefont {M.~Y.}\ \bibnamefont {Niu}}, \bibinfo {author} {\bibfnamefont {S.}~\bibnamefont {Boixo}}, \bibinfo {author} {\bibfnamefont {V.~N.}\ \bibnamefont {Smelyanskiy}},\ and\ \bibinfo {author} {\bibfnamefont {H.}~\bibnamefont {Neven}},\ }\bibfield  {title} {\bibinfo {title} {Universal quantum control through deep reinforcement learning},\ }\href {https://doi.org/10.1038/s41534-019-0141-3} {\bibfield  {journal} {\bibinfo  {journal} {npj Quantum Information}\ }\textbf {\bibinfo {volume} {5}},\ \bibinfo {pages} {33} (\bibinfo {year} {2019})}\BibitemShut {NoStop}%
\bibitem [{\citenamefont {Baum}\ \emph {et~al.}(2021)\citenamefont {Baum}, \citenamefont {Amico}, \citenamefont {Howell}, \citenamefont {Hush}, \citenamefont {Liuzzi}, \citenamefont {Mundada}, \citenamefont {Merkh}, \citenamefont {Carvalho},\ and\ \citenamefont {Biercuk}}]{baum2021experimental}%
  \BibitemOpen
  \bibfield  {author} {\bibinfo {author} {\bibfnamefont {Y.}~\bibnamefont {Baum}}, \bibinfo {author} {\bibfnamefont {M.}~\bibnamefont {Amico}}, \bibinfo {author} {\bibfnamefont {S.}~\bibnamefont {Howell}}, \bibinfo {author} {\bibfnamefont {M.}~\bibnamefont {Hush}}, \bibinfo {author} {\bibfnamefont {M.}~\bibnamefont {Liuzzi}}, \bibinfo {author} {\bibfnamefont {P.}~\bibnamefont {Mundada}}, \bibinfo {author} {\bibfnamefont {T.}~\bibnamefont {Merkh}}, \bibinfo {author} {\bibfnamefont {A.~R.~R.}\ \bibnamefont {Carvalho}},\ and\ \bibinfo {author} {\bibfnamefont {M.~J.}\ \bibnamefont {Biercuk}},\ }\bibfield  {title} {\bibinfo {title} {Experimental deep reinforcement learning for error-robust gate-set design on a superconducting quantum computer},\ }\href {https://doi.org/10.1103/PRXQuantum.2.040324} {\bibfield  {journal} {\bibinfo  {journal} {PRX Quantum}\ }\textbf {\bibinfo {volume} {2}},\ \bibinfo {pages} {040324} (\bibinfo {year} {2021})}\BibitemShut {NoStop}%
\bibitem [{\citenamefont {Torlai}\ \emph {et~al.}(2018)\citenamefont {Torlai}, \citenamefont {Mazzola}, \citenamefont {Carrasquilla}, \citenamefont {Troyer}, \citenamefont {Melko},\ and\ \citenamefont {Carleo}}]{torlai2018neural}%
  \BibitemOpen
  \bibfield  {author} {\bibinfo {author} {\bibfnamefont {G.}~\bibnamefont {Torlai}}, \bibinfo {author} {\bibfnamefont {G.}~\bibnamefont {Mazzola}}, \bibinfo {author} {\bibfnamefont {J.}~\bibnamefont {Carrasquilla}}, \bibinfo {author} {\bibfnamefont {M.}~\bibnamefont {Troyer}}, \bibinfo {author} {\bibfnamefont {R.}~\bibnamefont {Melko}},\ and\ \bibinfo {author} {\bibfnamefont {G.}~\bibnamefont {Carleo}},\ }\bibfield  {title} {\bibinfo {title} {Neural-network quantum state tomography},\ }\href {https://doi.org/10.1038/s41567-018-0048-5} {\bibfield  {journal} {\bibinfo  {journal} {Nature Physics}\ }\textbf {\bibinfo {volume} {14}},\ \bibinfo {pages} {447} (\bibinfo {year} {2018})}\BibitemShut {NoStop}%
\bibitem [{\citenamefont {Carrasquilla}\ \emph {et~al.}(2019)\citenamefont {Carrasquilla}, \citenamefont {Torlai}, \citenamefont {Melko},\ and\ \citenamefont {Aolita}}]{carrasquilla2019reconstructing}%
  \BibitemOpen
  \bibfield  {author} {\bibinfo {author} {\bibfnamefont {J.}~\bibnamefont {Carrasquilla}}, \bibinfo {author} {\bibfnamefont {G.}~\bibnamefont {Torlai}}, \bibinfo {author} {\bibfnamefont {R.~G.}\ \bibnamefont {Melko}},\ and\ \bibinfo {author} {\bibfnamefont {L.}~\bibnamefont {Aolita}},\ }\bibfield  {title} {\bibinfo {title} {Reconstructing quantum states with generative models},\ }\href {https://doi.org/10.1038/s42256-019-0028-1} {\bibfield  {journal} {\bibinfo  {journal} {Nature Machine Intelligence}\ }\textbf {\bibinfo {volume} {1}},\ \bibinfo {pages} {155} (\bibinfo {year} {2019})}\BibitemShut {NoStop}%
\bibitem [{\citenamefont {Huang}\ \emph {et~al.}(2020)\citenamefont {Huang}, \citenamefont {Kueng},\ and\ \citenamefont {Preskill}}]{huang2020predicting}%
  \BibitemOpen
  \bibfield  {author} {\bibinfo {author} {\bibfnamefont {H.-Y.}\ \bibnamefont {Huang}}, \bibinfo {author} {\bibfnamefont {R.}~\bibnamefont {Kueng}},\ and\ \bibinfo {author} {\bibfnamefont {J.}~\bibnamefont {Preskill}},\ }\bibfield  {title} {\bibinfo {title} {Predicting many properties of a quantum system from very few measurements},\ }\href {https://doi.org/10.1038/s41567-020-0932-7} {\bibfield  {journal} {\bibinfo  {journal} {Nature Physics}\ }\textbf {\bibinfo {volume} {16}},\ \bibinfo {pages} {1050} (\bibinfo {year} {2020})}\BibitemShut {NoStop}%
\bibitem [{\citenamefont {Elben}\ \emph {et~al.}(2023)\citenamefont {Elben}, \citenamefont {Flammia}, \citenamefont {Huang}, \citenamefont {Kueng}, \citenamefont {Preskill}, \citenamefont {Vermersch},\ and\ \citenamefont {Zoller}}]{elben2023randomized}%
  \BibitemOpen
  \bibfield  {author} {\bibinfo {author} {\bibfnamefont {A.}~\bibnamefont {Elben}}, \bibinfo {author} {\bibfnamefont {S.~T.}\ \bibnamefont {Flammia}}, \bibinfo {author} {\bibfnamefont {H.-Y.}\ \bibnamefont {Huang}}, \bibinfo {author} {\bibfnamefont {R.}~\bibnamefont {Kueng}}, \bibinfo {author} {\bibfnamefont {J.}~\bibnamefont {Preskill}}, \bibinfo {author} {\bibfnamefont {B.}~\bibnamefont {Vermersch}},\ and\ \bibinfo {author} {\bibfnamefont {P.}~\bibnamefont {Zoller}},\ }\bibfield  {title} {\bibinfo {title} {The randomized measurement toolbox},\ }\href {https://doi.org/10.1038/s42254-022-00535-2} {\bibfield  {journal} {\bibinfo  {journal} {Nature Reviews Physics}\ }\textbf {\bibinfo {volume} {5}},\ \bibinfo {pages} {9} (\bibinfo {year} {2023})}\BibitemShut {NoStop}%
\bibitem [{\citenamefont {Carleo}\ and\ \citenamefont {Troyer}(2017)}]{carleo2017solving}%
  \BibitemOpen
  \bibfield  {author} {\bibinfo {author} {\bibfnamefont {G.}~\bibnamefont {Carleo}}\ and\ \bibinfo {author} {\bibfnamefont {M.}~\bibnamefont {Troyer}},\ }\bibfield  {title} {\bibinfo {title} {Solving the quantum many-body problem with artificial neural networks},\ }\href {https://doi.org/10.1126/science.aag2302} {\bibfield  {journal} {\bibinfo  {journal} {Science}\ }\textbf {\bibinfo {volume} {355}},\ \bibinfo {pages} {602} (\bibinfo {year} {2017})}\BibitemShut {NoStop}%
\bibitem [{\citenamefont {Sharir}\ \emph {et~al.}(2020)\citenamefont {Sharir}, \citenamefont {Levine}, \citenamefont {Wies}, \citenamefont {Carleo},\ and\ \citenamefont {Shashua}}]{sharir2020deep}%
  \BibitemOpen
  \bibfield  {author} {\bibinfo {author} {\bibfnamefont {O.}~\bibnamefont {Sharir}}, \bibinfo {author} {\bibfnamefont {Y.}~\bibnamefont {Levine}}, \bibinfo {author} {\bibfnamefont {N.}~\bibnamefont {Wies}}, \bibinfo {author} {\bibfnamefont {G.}~\bibnamefont {Carleo}},\ and\ \bibinfo {author} {\bibfnamefont {A.}~\bibnamefont {Shashua}},\ }\bibfield  {title} {\bibinfo {title} {Deep autoregressive models for the efficient variational simulation of many-body quantum systems},\ }\href {https://doi.org/10.1103/PhysRevLett.124.020503} {\bibfield  {journal} {\bibinfo  {journal} {Physical Review Letters}\ }\textbf {\bibinfo {volume} {124}},\ \bibinfo {pages} {020503} (\bibinfo {year} {2020})}\BibitemShut {NoStop}%
\bibitem [{\citenamefont {Carrasquilla}\ and\ \citenamefont {Melko}(2017)}]{carrasquilla2017machine}%
  \BibitemOpen
  \bibfield  {author} {\bibinfo {author} {\bibfnamefont {J.}~\bibnamefont {Carrasquilla}}\ and\ \bibinfo {author} {\bibfnamefont {R.~G.}\ \bibnamefont {Melko}},\ }\bibfield  {title} {\bibinfo {title} {Machine learning phases of matter},\ }\href {https://doi.org/10.1038/nphys4035} {\bibfield  {journal} {\bibinfo  {journal} {Nature Physics}\ }\textbf {\bibinfo {volume} {13}},\ \bibinfo {pages} {431} (\bibinfo {year} {2017})}\BibitemShut {NoStop}%
\bibitem [{\citenamefont {van Nieuwenburg}\ \emph {et~al.}(2017)\citenamefont {van Nieuwenburg}, \citenamefont {Liu},\ and\ \citenamefont {Huber}}]{vannieuwenburg2017learning}%
  \BibitemOpen
  \bibfield  {author} {\bibinfo {author} {\bibfnamefont {E.~P.~L.}\ \bibnamefont {van Nieuwenburg}}, \bibinfo {author} {\bibfnamefont {Y.-H.}\ \bibnamefont {Liu}},\ and\ \bibinfo {author} {\bibfnamefont {S.~D.}\ \bibnamefont {Huber}},\ }\bibfield  {title} {\bibinfo {title} {Learning phase transitions by confusion},\ }\href {https://doi.org/10.1038/nphys4037} {\bibfield  {journal} {\bibinfo  {journal} {Nature Physics}\ }\textbf {\bibinfo {volume} {13}},\ \bibinfo {pages} {435} (\bibinfo {year} {2017})}\BibitemShut {NoStop}%
\bibitem [{\citenamefont {Lipardi}\ \emph {et~al.}(2025)\citenamefont {Lipardi} \emph {et~al.}}]{Lipardi2025SREMachineLearning}%
  \BibitemOpen
  \bibfield  {author} {\bibinfo {author} {\bibfnamefont {V.}~\bibnamefont {Lipardi}} \emph {et~al.},\ }\href@noop {} {\bibinfo {title} {A study on stabilizer {R\'enyi} entropy estimation using machine learning}} (\bibinfo {year} {2025}),\ \bibinfo {note} {arXiv:2509.16799 [quant-ph]},\ \Eprint {https://arxiv.org/abs/2509.16799} {arXiv:2509.16799 [quant-ph]} \BibitemShut {NoStop}%
\bibitem [{\citenamefont {Busoni}\ \emph {et~al.}(2026)\citenamefont {Busoni}, \citenamefont {Gargalionis}, \citenamefont {Wallace},\ and\ \citenamefont {White}}]{Busoni2026NonLocalMagicQudits}%
  \BibitemOpen
  \bibfield  {author} {\bibinfo {author} {\bibfnamefont {G.}~\bibnamefont {Busoni}}, \bibinfo {author} {\bibfnamefont {J.}~\bibnamefont {Gargalionis}}, \bibinfo {author} {\bibfnamefont {E.~N.~V.}\ \bibnamefont {Wallace}},\ and\ \bibinfo {author} {\bibfnamefont {M.~J.}\ \bibnamefont {White}},\ }\href@noop {} {\bibinfo {title} {Analytic formulae for non-local magic in bipartite systems of qutrits and ququints}} (\bibinfo {year} {2026}),\ \bibinfo {note} {arXiv:2603.09155 [quant-ph]},\ \Eprint {https://arxiv.org/abs/2603.09155} {arXiv:2603.09155 [quant-ph]} \BibitemShut {NoStop}%
\bibitem [{\citenamefont {Nguyen}\ \emph {et~al.}(2024)\citenamefont {Nguyen}, \citenamefont {Motzoi}, \citenamefont {Metcalf}, \citenamefont {Whaley}, \citenamefont {Bukov},\ and\ \citenamefont {Schmitt}}]{nam2024reinforcement}%
  \BibitemOpen
  \bibfield  {author} {\bibinfo {author} {\bibfnamefont {H.~N.}\ \bibnamefont {Nguyen}}, \bibinfo {author} {\bibfnamefont {F.}~\bibnamefont {Motzoi}}, \bibinfo {author} {\bibfnamefont {M.}~\bibnamefont {Metcalf}}, \bibinfo {author} {\bibfnamefont {K.~B.}\ \bibnamefont {Whaley}}, \bibinfo {author} {\bibfnamefont {M.}~\bibnamefont {Bukov}},\ and\ \bibinfo {author} {\bibfnamefont {M.}~\bibnamefont {Schmitt}},\ }\bibfield  {title} {\bibinfo {title} {Reinforcement learning pulses for transmon qubit entangling gates},\ }\href@noop {} {\bibfield  {journal} {\bibinfo  {journal} {Machine Learning: Science and Technology}\ }\textbf {\bibinfo {volume} {5}},\ \bibinfo {pages} {025066} (\bibinfo {year} {2024})}\BibitemShut {NoStop}%
\bibitem [{\citenamefont {Genois}\ \emph {et~al.}(2025)\citenamefont {Genois}, \citenamefont {Stevenson}, \citenamefont {Goss}, \citenamefont {Siddiqi},\ and\ \citenamefont {Blais}}]{genois2025quantum}%
  \BibitemOpen
  \bibfield  {author} {\bibinfo {author} {\bibfnamefont {{\'E}.}~\bibnamefont {Genois}}, \bibinfo {author} {\bibfnamefont {N.~J.}\ \bibnamefont {Stevenson}}, \bibinfo {author} {\bibfnamefont {N.}~\bibnamefont {Goss}}, \bibinfo {author} {\bibfnamefont {I.}~\bibnamefont {Siddiqi}},\ and\ \bibinfo {author} {\bibfnamefont {A.}~\bibnamefont {Blais}},\ }\bibfield  {title} {\bibinfo {title} {Quantum optimal control of superconducting qubits based on machine-learning characterization},\ }\href@noop {} {\bibfield  {journal} {\bibinfo  {journal} {Physical Review Applied}\ }\textbf {\bibinfo {volume} {24}},\ \bibinfo {pages} {034073} (\bibinfo {year} {2025})}\BibitemShut {NoStop}%
\bibitem [{\citenamefont {Daraeizadeh}\ \emph {et~al.}(2020)\citenamefont {Daraeizadeh}, \citenamefont {Premaratne}, \citenamefont {Khammassi}, \citenamefont {Song}, \citenamefont {Perkowski},\ and\ \citenamefont {Matsuura}}]{daraeizadeh2020machine}%
  \BibitemOpen
  \bibfield  {author} {\bibinfo {author} {\bibfnamefont {S.}~\bibnamefont {Daraeizadeh}}, \bibinfo {author} {\bibfnamefont {S.~P.}\ \bibnamefont {Premaratne}}, \bibinfo {author} {\bibfnamefont {N.}~\bibnamefont {Khammassi}}, \bibinfo {author} {\bibfnamefont {X.}~\bibnamefont {Song}}, \bibinfo {author} {\bibfnamefont {M.}~\bibnamefont {Perkowski}},\ and\ \bibinfo {author} {\bibfnamefont {A.}~\bibnamefont {Matsuura}},\ }\bibfield  {title} {\bibinfo {title} {Machine-learning-based three-qubit gate design for the toffoli gate and parity check in transmon systems},\ }\href@noop {} {\bibfield  {journal} {\bibinfo  {journal} {Physical Review A}\ }\textbf {\bibinfo {volume} {102}},\ \bibinfo {pages} {012601} (\bibinfo {year} {2020})}\BibitemShut {NoStop}%
\bibitem [{\citenamefont {Wright}\ and\ \citenamefont {De~Sousa}(2023)}]{wright2023fast}%
  \BibitemOpen
  \bibfield  {author} {\bibinfo {author} {\bibfnamefont {E.}~\bibnamefont {Wright}}\ and\ \bibinfo {author} {\bibfnamefont {R.}~\bibnamefont {De~Sousa}},\ }\bibfield  {title} {\bibinfo {title} {Fast quantum gate design with deep reinforcement learning using real-time feedback on readout signals},\ }in\ \href@noop {} {\emph {\bibinfo {booktitle} {2023 IEEE International Conference on Quantum Computing and Engineering (QCE)}}},\ Vol.~\bibinfo {volume} {1}\ (\bibinfo {organization} {IEEE},\ \bibinfo {year} {2023})\ pp.\ \bibinfo {pages} {1295--1303}\BibitemShut {NoStop}%
\bibitem [{\citenamefont {Chatterjee}\ \emph {et~al.}(2025)\citenamefont {Chatterjee}, \citenamefont {Schwinger},\ and\ \citenamefont {Gao}}]{chatterjee2025enhanced}%
  \BibitemOpen
  \bibfield  {author} {\bibinfo {author} {\bibfnamefont {A.}~\bibnamefont {Chatterjee}}, \bibinfo {author} {\bibfnamefont {J.}~\bibnamefont {Schwinger}},\ and\ \bibinfo {author} {\bibfnamefont {Y.~Y.}\ \bibnamefont {Gao}},\ }\bibfield  {title} {\bibinfo {title} {Enhanced qubit readout via reinforcement learning},\ }\href@noop {} {\bibfield  {journal} {\bibinfo  {journal} {Physical Review Applied}\ }\textbf {\bibinfo {volume} {23}},\ \bibinfo {pages} {054057} (\bibinfo {year} {2025})}\BibitemShut {NoStop}%
\bibitem [{\citenamefont {Di~Guglielmo}\ \emph {et~al.}(2025)\citenamefont {Di~Guglielmo}, \citenamefont {Du}, \citenamefont {Campos}, \citenamefont {Boltasseva}, \citenamefont {Dixit}, \citenamefont {Fahim}, \citenamefont {Kudyshev}, \citenamefont {Lopez}, \citenamefont {Ma}, \citenamefont {Perdue} \emph {et~al.}}]{di2025end}%
  \BibitemOpen
  \bibfield  {author} {\bibinfo {author} {\bibfnamefont {G.}~\bibnamefont {Di~Guglielmo}}, \bibinfo {author} {\bibfnamefont {B.}~\bibnamefont {Du}}, \bibinfo {author} {\bibfnamefont {J.}~\bibnamefont {Campos}}, \bibinfo {author} {\bibfnamefont {A.}~\bibnamefont {Boltasseva}}, \bibinfo {author} {\bibfnamefont {A.}~\bibnamefont {Dixit}}, \bibinfo {author} {\bibfnamefont {F.}~\bibnamefont {Fahim}}, \bibinfo {author} {\bibfnamefont {Z.}~\bibnamefont {Kudyshev}}, \bibinfo {author} {\bibfnamefont {S.}~\bibnamefont {Lopez}}, \bibinfo {author} {\bibfnamefont {R.}~\bibnamefont {Ma}}, \bibinfo {author} {\bibfnamefont {G.~N.}\ \bibnamefont {Perdue}}, \emph {et~al.},\ }\bibfield  {title} {\bibinfo {title} {End-to-end workflow for machine learning-based qubit readout with qick and hls4ml},\ }\href@noop {} {\bibfield  {journal} {\bibinfo  {journal} {IEEE Transactions on Quantum Engineering}\ } (\bibinfo {year} {2025})}\BibitemShut {NoStop}%
\bibitem [{\citenamefont {Liu}(2025)}]{liu2025superconducting}%
  \BibitemOpen
  \bibfield  {author} {\bibinfo {author} {\bibfnamefont {Y.}~\bibnamefont {Liu}},\ }\bibfield  {title} {\bibinfo {title} {Superconducting quantum computing optimization based on multi-objective deep reinforcement learning},\ }\href@noop {} {\bibfield  {journal} {\bibinfo  {journal} {Scientific Reports}\ }\textbf {\bibinfo {volume} {15}},\ \bibinfo {pages} {3828} (\bibinfo {year} {2025})}\BibitemShut {NoStop}%
\bibitem [{\citenamefont {Porotti}\ \emph {et~al.}(2022)\citenamefont {Porotti}, \citenamefont {Essig}, \citenamefont {Huard},\ and\ \citenamefont {Marquardt}}]{porotti2022deep}%
  \BibitemOpen
  \bibfield  {author} {\bibinfo {author} {\bibfnamefont {R.}~\bibnamefont {Porotti}}, \bibinfo {author} {\bibfnamefont {A.}~\bibnamefont {Essig}}, \bibinfo {author} {\bibfnamefont {B.}~\bibnamefont {Huard}},\ and\ \bibinfo {author} {\bibfnamefont {F.}~\bibnamefont {Marquardt}},\ }\bibfield  {title} {\bibinfo {title} {Deep reinforcement learning for quantum state preparation with weak nonlinear measurements},\ }\href@noop {} {\bibfield  {journal} {\bibinfo  {journal} {Quantum}\ }\textbf {\bibinfo {volume} {6}},\ \bibinfo {pages} {747} (\bibinfo {year} {2022})}\BibitemShut {NoStop}%
\bibitem [{\citenamefont {Reuer}\ \emph {et~al.}(2023)\citenamefont {Reuer}, \citenamefont {Landgraf}, \citenamefont {F{\"o}sel}, \citenamefont {O'Sullivan}, \citenamefont {Beltr{\'a}n}, \citenamefont {Akin}, \citenamefont {Norris}, \citenamefont {Remm}, \citenamefont {Kerschbaum}, \citenamefont {Besse} \emph {et~al.}}]{reuer2023realizing}%
  \BibitemOpen
  \bibfield  {author} {\bibinfo {author} {\bibfnamefont {K.}~\bibnamefont {Reuer}}, \bibinfo {author} {\bibfnamefont {J.}~\bibnamefont {Landgraf}}, \bibinfo {author} {\bibfnamefont {T.}~\bibnamefont {F{\"o}sel}}, \bibinfo {author} {\bibfnamefont {J.}~\bibnamefont {O'Sullivan}}, \bibinfo {author} {\bibfnamefont {L.}~\bibnamefont {Beltr{\'a}n}}, \bibinfo {author} {\bibfnamefont {A.}~\bibnamefont {Akin}}, \bibinfo {author} {\bibfnamefont {G.~J.}\ \bibnamefont {Norris}}, \bibinfo {author} {\bibfnamefont {A.}~\bibnamefont {Remm}}, \bibinfo {author} {\bibfnamefont {M.}~\bibnamefont {Kerschbaum}}, \bibinfo {author} {\bibfnamefont {J.-C.}\ \bibnamefont {Besse}}, \emph {et~al.},\ }\bibfield  {title} {\bibinfo {title} {Realizing a deep reinforcement learning agent for real-time quantum feedback},\ }\href@noop {} {\bibfield  {journal} {\bibinfo  {journal} {Nature Communications}\ }\textbf {\bibinfo {volume} {14}},\ \bibinfo {pages} {7138} (\bibinfo {year} {2023})}\BibitemShut {NoStop}%
\bibitem [{\citenamefont {Cao}\ \emph {et~al.}(2025{\natexlab{b}})\citenamefont {Cao}, \citenamefont {Zhang}, \citenamefont {Alghadeer}, \citenamefont {Fasciati}, \citenamefont {Piscitelli}, \citenamefont {Bakr}, \citenamefont {Leek},\ and\ \citenamefont {Aspuru-Guzik}}]{cao2025automating}%
  \BibitemOpen
  \bibfield  {author} {\bibinfo {author} {\bibfnamefont {S.}~\bibnamefont {Cao}}, \bibinfo {author} {\bibfnamefont {Z.}~\bibnamefont {Zhang}}, \bibinfo {author} {\bibfnamefont {M.}~\bibnamefont {Alghadeer}}, \bibinfo {author} {\bibfnamefont {S.~D.}\ \bibnamefont {Fasciati}}, \bibinfo {author} {\bibfnamefont {M.}~\bibnamefont {Piscitelli}}, \bibinfo {author} {\bibfnamefont {M.}~\bibnamefont {Bakr}}, \bibinfo {author} {\bibfnamefont {P.}~\bibnamefont {Leek}},\ and\ \bibinfo {author} {\bibfnamefont {A.}~\bibnamefont {Aspuru-Guzik}},\ }\bibfield  {title} {\bibinfo {title} {Automating quantum computing laboratory experiments with an agent-based ai framework},\ }\href@noop {} {\bibfield  {journal} {\bibinfo  {journal} {Patterns}\ }\textbf {\bibinfo {volume} {6}} (\bibinfo {year} {2025}{\natexlab{b}})}\BibitemShut {NoStop}%
\end{thebibliography}%

\clearpage



\end{document}